\documentclass[printer]{aa} 
\usepackage{graphicx}
\usepackage{amsmath}
\usepackage{amssymb}
\usepackage{bm}
\usepackage{epstopdf}
\usepackage{enumerate}
\usepackage{color}
\usepackage{hyperref}
\usepackage{natbib}
\usepackage{setspace}

\newcommand{\LL}{\left(}
\newcommand{\RR}{\right)}

\renewcommand{\P}[2]{\frac{\partial #1}{\partial #2}}

\newcommand{\curl}{\nabla \times}
\renewcommand{\div}{\nabla \cdot}

\newcommand{\pr}{\parallel}
\newcommand{\g}{\gamma}

\newcommand{\dg}{^{\circ}}

\DeclareRobustCommand*{\unit}[1]{\def~{\,}\ensuremath{\mathrm{\,#1}}}
\graphicspath{{all_figures/}}
\bibpunct{(}{)}{;}{a}{}{,} 
\usepackage{txfonts} 
\usepackage{subfig}     
\usepackage{graphicx}

\begin{document} 

\title{Particle acceleration with anomalous pitch angle scattering in 3D separator reconnection}

 \author{A. Borissov \inst{\ref{inst:edinburgh},\ref{inst:standrews}}
    \and 
         T. Neukirch\inst{\ref{inst:standrews}}
    \and 
       E. P. Kontar\inst{\ref{inst:glasgow}}
    \and 
       J. Threlfall\inst{\ref{inst:standrews}}
    \and
       C. E. Parnell\inst{\ref{inst:standrews}}
  } 
  
  \institute{School of Physics and Astronomy, 
             University of Edinburgh, 
             James Clerk Maxwell Building, Peter Guthrie Tait Road,
             Edinburgh EH9 3FD, UK\\ 
             \email{alexei.borissov@ed.ac.uk} 
             \label{inst:edinburgh}
        \and 
             School of Mathematics and Statistics, 
             University of St Andrews,  
             St Andrews KY16 9SS, UK
             \label{inst:standrews}
       \and 
             School of Physics and Astronomy,
             University of Glasgow,
             Glasgow G12 8QQ, UK
             \label{inst:glasgow}
              } 
             

\abstract
{Understanding how the release of stored magnetic energy contributes to the generation of non-thermal high energy particles during solar flares is an important open problem in solar physics. There is a general consensus that magnetic reconnection plays a fundamental role in the energy release and conversion processes taking place during flares. A common approach for investigating how reconnection contributes to particle acceleration is to use test particle calculations in electromagnetic fields derived from numerical magnetohydrodynamic (MHD) simulations of reconnecting magnetic fields. These MHD simulations use anomalous resistivities that are orders of magnitude larger than the Spitzer resistivity that is based on Coulomb collisions. The processes leading to such an enhanced resistivity should also affect the test particles, for example, through pitch angle scattering. This study explores the effect of such a link between the level of resistivity and its impact on particle orbits and builds on a previous study using a 2D MHD simulation of magnetic reconnection.}
{This paper aims to extend the previous investigation to a 3D magnetic reconnection configuration and to study the effect on test particle orbits.}
{We carried out orbit calculations using a 3D MHD simulation of reconnection in a magnetic field with a magnetic separator. The orbit calculations use the relativistic guiding centre approximation but, crucially, they also include pitch angle scattering using stochastic differential equations. The effects of varying the resistivity and the models for pitch angle scattering on particle orbit trajectories, final positions, energy spectra, final pitch angle distribution, and orbit duration are all studied in detail.}
{ Pitch angle scattering widens highly collimated beams of unscattered orbit trajectories, allowing orbits to access previously unaccessible field lines; this causes final positions to spread along other topological structures which could not be accessed without scattering. Scattered orbit energy spectra are found to be predominantly affected by the level of anomalous resistivity, with the pitch angle scattering model only playing a role in specific, isolated cases. This is in contrast to the study involving a 2D MHD simulation of magnetic reconnection, where pitch angle scattering had a more noticeable effect on the energy spectra. Pitch scattering effects are found to play a crucial role in determining the pitch angle and orbit duration distributions. }
{}


 \titlerunning{Particle acceleration in separator reconnection}
 \authorrunning{Borissov et al.}

\maketitle

\section{Introduction}

During solar flares, magnetic energy is converted into kinetic energy (plasma flows), turbulence, thermal energy (plasma heating), and non-thermal energy (high-energy particle populations). The consequences of this energy release can be observed across the electromagnetic spectrum \cite[e.g.][]{benz2017}. In particular, the high-energy (hard X-ray and gamma ray) and radio parts of the observed flare-related radiation are closely linked to the acceleration of charged particles to high energies during solar flares \citep[e.g.][]{pick-vilmer2008,benz2017}.

The large scale dynamics of flares are normally described by magnetohydrodynamics (MHD) and magnetic reconnection is thought to play a key role 
in the release of magnetic energy during solar flares \citep[e.g.][]{priest-forbes2002,shibata-magara2011,janvier2017}. In particular, magnetic reconnection is thought to contribute either directly or indirectly to the generation of the non-thermal particle population \citep[e.g.][]{aschwanden2002,neukirch-et-al2007,zharkova-et-al2011,cargill-et-al2012}. Here, acceleration via the generic reconnection electric field parallel to the magnetic field \citep[e.g.][]{schindler-et-al1988,hesse-schindler1988, schindler-et-al1991,hesse1995} would be classified as direct, whereas other acceleration mechanisms associated with reconnection, such as stochastic acceleration in turbulent reconnection outflows \citep[e.g][]{liu-et-al2008,2017PhRvL.118o5101K,2018ApJ...855...95H,2019PPCF...61a4020V}, collapsing magnetic traps \citep[e.g][]{somov-kosugi1997}, or termination shocks \citep[e.g.][]{mann-warmuth2009}, would be classified as indirect \citep[see e.g.][for a more detailed discussion]{neukirch-et-al2007}.

The main tool for current investigations of particle acceleration associated with magnetic reconnection are still  test particle calculations in given electromagnetic fields \citep[we include in this both particle orbit calculations and solutions of kinetic equations without back reaction onto the fields; see e.g][]{minoshima-et-al2010,minoshima-et-al2011}. It is, however, important to mention that there is a growing body of work that uses kinetic theory to investigate particle acceleration in the context of collisionless magnetic reconnection \citep[e.g][]{hoshino-et-al2001,drake-et-al2006,baumann-et-al2013, 2013PhRvL.110o1101B,dahlin-et-al2014,dahlin-et-al2015,dahlin-et-al2017,munoz-buechner2018}, along with recent attempts to couple the microscopic and macroscopic scales self-consistently \citep[e.g][]{drake-et-al2019,gordovsky-et-al2019}. There is extensive previous work on test particle calculations related to particle acceleration in solar flares, both in analytically prescribed electromagnetic fields \citep[e.g.][]{kliem1994,litvinenko1996,arzner-vlahos2004,vlahos-et-al2004,zharkova-gordovskyy2004,zharkova-gordovskyy2005,wood-neukirch2005, giuliani-et-al2005, dalla-browning2006, dalla-browning2008,grady-et-al2012,eradat_oskoui-et-al2014,stanier-et-al2012,threlfall-et-al2015, borissov-et-al2016, threlfall-et-al2017} and in fields taken from numerical simulations \citep[e.g][]{turkmani-et-al2005,turkmani-et-al2006,karlicky-barta2006, gordovskyy-et-al2010a, gordovskyy-et-al2010b,guo-et-al2010,rosdahl-galsgaard2010,burge-et-al2014, zhou-et-al2015,zhou-et-al2016, threlfall-et-al2016b,borissov-et-al2017, birn-et-al2017,threlfall-et-al2018,xia-zharkova2018,ye-et-al2019}.

For MHD, the reconnection process depends crucially on the resistivity $\eta$. Generically, MHD reconnection occurs when there is a localised break-down of ideal Ohm's law due to a non-vanishing resistivity $\eta$ and the associated development of an electric field component parallel to the magnetic field ($\textbf{E}_\parallel$) \citep[e.g.][]{schindler-et-al1988,hesse-schindler1988}, which is proportional to the parallel component of the current density ($\textbf{j}_\parallel$) with $\textbf{E}_\parallel = \eta \textbf{j}_\parallel$.  The physical process leading to a non-vanishing resistivity is particle scattering \citep[e.g.][]{treumann-baumjohann2015}, with the level of resistivity is determined by the scattering rate. Coulomb scattering leads to the classical Spitzer resistivity, which for the solar corona is found to be far too small to explain the rapid energy release observed during solar flares \citep[e.g.][]{1969ARA&A...7..149S,priest2014}. Therefore, an enhanced, anomalous resistivity based on scattering by other processes such as turbulence \citep[e.g][]{papadopoulos1977,treumann2001,2016ApJ...824...78B} is often used \citep[e.g.][]{gordovskyy-et-al2010a}. Importantly, recent observations \citep[e.g.][]{2014ApJ...780..176K,2017PhRvL.118o5101K, 2018A&A...610A...6M} suggest enhanced non-collisional scattering should be present in solar flares to explain X-ray, radio and extreme ultraviolet (EUV) data.

Although the electromagnetic fields obtained from MHD simulations with an anomalous resistivity imply that scattering processes happen in the diffusion region, test particle simulations typically do not take this into account.  Occasionally the effect of Coulomb collisions is taken into account \citep[e.g.][]{gordovskyy-et-al2013,gordovskyy-et-al2014,burge-et-al2014} but the Coulomb collision cross section decreases with increasing kinetic energy of the particles and, hence,  the effect of Coulomb collisions is primarily important for the thermal particle population rather than for the highly accelerated non-thermal population.  However, due to the enhanced anomalous resistivity assumed in MHD simulations of magnetic reconnection, it should be expected that non-thermal test particles should also experience an enhanced scattering rate \citep[e.g.][]{2014ApJ...780..176K}. 

The effect of such an enhanced scattering rate in the reconnection region was recently explored by \citet{borissov-et-al2017} by computing test particle orbit trajectories and energy spectra in fields obtained from a series of 2D MHD reconnection simulations with different values for the resistivity.  Test particle orbits were integrated both with and without pitch angle scattering as well as with scattering at different rates.  Differences in orbit trajectories due to scattering caused particles to traverse the reconnection region multiple times, resulting in higher energy gains than would be possible in the absence of scattering.  Since the reconnection rate and the strength of the parallel electric field did not vary much for different values of the resistivity, it was not possible to fully examine the relationship between scattering and acceleration within the 2D MHD model.

In this paper we present a natural extension of the 2D work of \citet{borissov-et-al2017} by investigating particle acceleration with resistivity dependent scattering in the context of 3D separator reconnection \citep[e.g][]{parnell-et-al2010a,stevenson-parnell2015a,stevenson-parnell2015b}. In particular, a 3D separator is chosen due to the local structural similarity between the 2D magnetic field with guide field used in \citet{borissov-et-al2017} and a 3D separator field projected onto a plane perpendicular to the separator. Magnetic separators are found to be ubiquitous even in simple magnetic field models of the solar corona \citep[e.g.][]{edwards-et-al2016}, so while the particular setup used in this paper is idealised, it will still provide valuable insight into the particle dynamics in regions of the solar corona with a local magnetic separator topology. Furthermore, it is crucial to gain a good understanding of particle dynamics under the influence of scattering in relatively simple, high resolution simulations before moving on to more complex topologies where the individual effects of scattering and the underlying magnetic fields are much harder to disentangle. This will also provide an opportunity to further investigate the relationship between pitch angle scattering and acceleration that was not fully possible in the 2D case. Therefore, in this paper we perform 3D MHD simulations of separator reconnection, similar to those undertaken by \cite{stevenson-parnell2015a}, with different levels of anomalous resistivity.  We then compute test particle orbit trajectories and energy spectra in the resulting fields, with pitch angle scattering implemented at both constant and resistivity dependent rates. 

The structure of the paper is as follows. In Section \ref{mhd-sim3d}, we describe the setup of the 3D separator reconnection experiments. Section \ref{eom} briefly describes the governing equations of motion for the test particles, including our implementation of pitch angle scattering. We go on to present the results of test particle orbit calculations first examining the behaviour of a small number of individual orbits in Section \ref{single-particle}, followed by an analysis of the energy spectra, pitch angle, and orbit duration distributions for a larger sample of test particles in Section \ref{many-particle}. In Section \ref{eta-comparison}, we compare the effect of differing MHD resistivity on energy spectra and orbit pitch angle and duration distributions in the presence of a resistivity dependent scattering model. A discussion of these results and conclusions follow in Sections \ref{discussion} and \ref{conclusions}.


\section{MHD simulation of 3D separator reconnection}
\label{mhd-sim3d}

To obtain the background electromagnetic fields for our orbit calculations we use the resistive MHD equations in the form of
\begin{equation}\label{MHD1}
\P\rho t + \div \LL \rho \textbf u \RR = 0,
\end{equation}
\begin{equation}\label{MHD2}
\rho \LL \P{\textbf u}{t} + \textbf u \cdot \nabla \textbf u\RR = \textbf j \times \textbf B - \nabla P,
\end{equation}
\begin{equation}\label{MHD3}
\P \epsilon t + \textbf u \cdot \nabla \epsilon = -\frac 1 \rho P\div \textbf u + \frac{\eta j^2}{\rho},
\end{equation}
\begin{equation}\label{MHD4}
\P{\textbf B}{t} = \eta\nabla^2 \textbf B + \curl\LL \textbf u \times \textbf B \RR,
\end{equation}
\begin{equation}\label{MHD5}
\textbf E + \textbf u \times \textbf B = \eta \textbf j,
\end{equation}
\begin{equation}\label{MHD6}
\div \textbf B = 0,
\end{equation}
\begin{equation}\label{MHD7}
\curl \textbf B = \mu_0 \textbf j,
\end{equation}
\begin{equation}\label{MHD8}
P = \rho k_B T/\mu_m,
\end{equation}
where $\rho$ is the plasma density, $\textbf u$ is the plasma velocity, $\textbf B$ the magnetic field, $P$ is the pressure, $\epsilon = P/\LL\rho(\g_g-1) \RR$ the internal energy, $\gamma_g = \frac{C_p}{C_v}$ is the ratio of specific heats (the subscript differentiates it from the Lorentz factor used in the rest of this paper), $\eta$ the resistivity, $\textbf j$ is the current density, and $\mu_m$ the reduced mass (the subscript is used to differentiate it from the magnetic moment noted elsewhere in this paper). 

We use the \textit{Lare3D} code \citep[see e.g.][]{arber-et-al2001} to solve Eqs.\ \ref{MHD1}-\ref{MHD8}) numerically. Our 3D separator reconnection simulations follow the technique described in \cite{stevenson-parnell2015a}. The initial conditions for the simulation are taken from \cite{stevenson-et-al2015}, featuring a magnetic separator that was allowed to relax to a numerical equilibrium by evolving the MHD equations with a non-zero viscosity, but zero resistivity. Magnetic reconnection is induced by specifying a constant value of anomalous resistivity wherever the current density exceeds a threshold value. The resulting fields consist of a separator connecting two nulls located at $x = y = 0 \unit m, z = 0 \unit m$, and $z = 300 \unit m$ , within a computational box that is $x,y\in [-25,25]\unit m, z \in [-100,400]\unit m$. 

Since heat conduction and radiation are not used in these simulations, an elevated density is prescribed to ensure realistic temperatures during reconnection (in this case the average density over the box is $8.59\times 10^{-8}\unit{kg\cdot m^{-3}}$, and the maximum $2.12\times 10^{-7} \unit{kg \cdot m^{-3}}$). This is necessary so that the calculation of the Spitzer resistivity is representative of the value in the solar corona. 

In our simulations, we specify the critical current to be $j_{max} = 7.2 \times 10^3 \unit{A \cdot m^{-2}}$ and perform simulations for anomalous resistivities at the rates $\eta_a = 10^{-3},10^{-4},10^{-5}$, all in normalised units. The normalising scales used in the simulations are $L_0 = 100\unit m$, $B_0 = 0.12 \unit T$ and $\rho_0 = 1.67\times 10^{-7}\unit{kg \cdot m^{-3}}$, resulting in resistivities of $\eta_a = 3.3\times 10^{-2}$ to $\eta_a = 3.3\times 10^{-4} \unit{\Omega \cdot m}$. We refer to the resistivities used by their non-dimensional values. In this work it was necessary to choose a short length scale in comparison to other studies of particle acceleration in coronal environments because the pitch angle scattering rate imposed a mean free path independent of, and far shorter than, this length scale. Using a longer length scale would make the particle orbit calculation at a high scattering rate prohibitively expensive. For the chosen scales the timescale imposed by the{ Alfv\'en} crossing time is $T_A = L_0/v_A = 3.8\times 10^{-4}\unit s$ which is comparable to a few orbit durations but also significantly larger than the orbit durations of the most energised particles (as will be discussed in Section \ref{many-particle}). Furthermore, the particle gyrofrequency is always less than the scattering frequency, which is itself much shorter than $T_A$. A snapshot at the beginning of the simulation is chosen for particles to be injected into it. This is done to ensure that the reconnection rate and, hence, the parallel electric field, is relatively high since it decreases towards the end of the simulation \citep{stevenson-parnell2015a}. 
\begin{figure*}
        \resizebox{0.95\textwidth}{!}{\includegraphics[clip=true, trim=80 80 40 40]{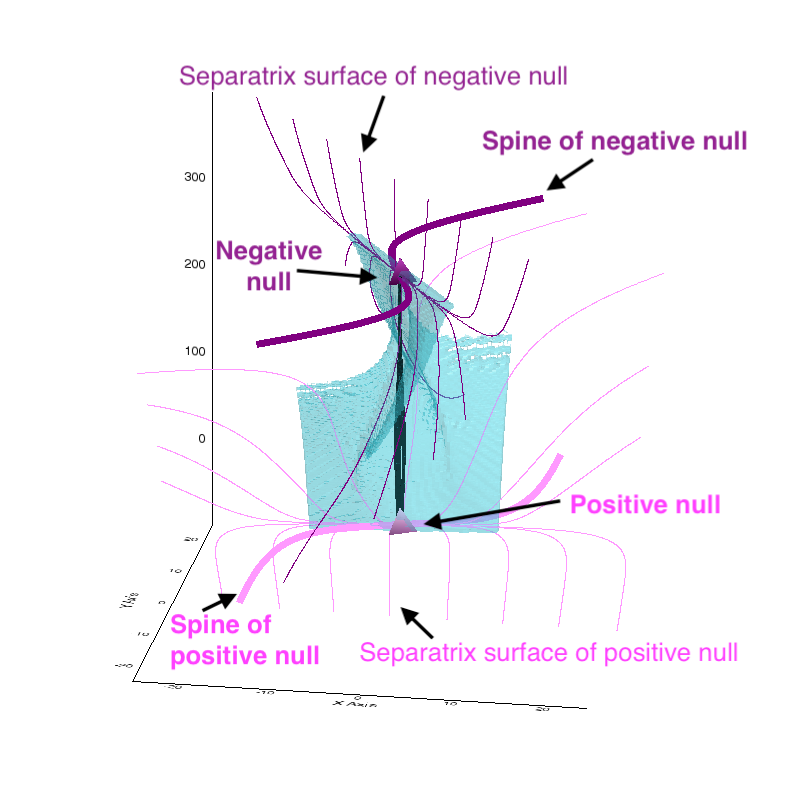}}
        \caption{Topological structure of MHD configuration, with separator (black) marking the intersection of separatrix surfaces of a positive and negative null, overlaid with an isosurface representing reconnection region in light blue (where the current is found above a critical value).}
        \label{f1}
\end{figure*}

The resulting field structure, including the locations of the nulls, is shown in Figure \ref{f1}. The reconnection region, which determines where pitch angle scattering occurs in the test particle simulation (as well as the location of non-zero parallel electric field) can be seen as a blue isosurface in this 3D image, extending out along the separatrix surfaces of each null. A 3D midplane cut through this structure is similar to the shape of the region of non-zero resistivity in the 2D simulations in \cite{borissov-et-al2017}. In contrast to the 2D case, the orientation of the reconnection region twists in successive cuts at different elevations perpendicular to the separator. There is also only a finite extent of the reconnection region along the separator, unlike the 2D case where it can be made to be arbitrarily long. 

\section{Test particle equations of motion}\label{eom}

A frequent approach to studying test particle dynamics is through the use of the guiding centre approximation \citep[e.g.][]{wood-neukirch2005,giuliani-et-al2005,gordovskyy-et-al2010a,gordovskyy-et-al2010b,borissov-et-al2016,threlfall-et-al2016a,birn-et-al2017}. This approach averages over the particle gyrational motion caused by the magnetic field, which reduces the computational load by not requiring that the simulation temporally resolve each particle gyration. The centre of particle gyration, obtained by averaging over the position of the particle over one gyration, is called the guiding centre. The equations of motion for the particle guiding centre, $\textbf R$, in relativistic form are given by:

\begin{align}\label{NRGC1}
\dot{\textbf R}_{\perp} &= \frac{\textbf b}{B} \times \left[ -\textbf E + \frac{\mu}{\gamma q} \nabla B + \frac{mU}{q} \frac{d\textbf b}{dt} + \frac{m\gamma}{q}\frac{d\textbf u_{E}}{dt} \right. \nonumber \\ &\hspace{20pt} \left. + \frac{U}{\gamma}E_{\parallel}\textbf u_{E} + \frac{\mu}{\gamma q} \textbf u_{E} \frac{\partial B}{\partial t} \right],
\end{align}

\begin{equation}\label{NRGC2}
m\frac{dU}{dt} = m\gamma \textbf u_{E} \cdot \frac{d\textbf b}{dt} + q E_{\parallel} - \frac{\mu}{\gamma}\frac{\partial B}{\partial s},
\end{equation}
where $\dot{\textbf R}_{\perp}$ the drift of the guiding centre perpendicular to the magnetic field, $v_\pr$ is the parallel velocity of the guiding centre, and $U = \gamma v_\pr$ is the relativistic parallel velocity. Equations (\ref{NRGC1}) and (\ref{NRGC2}) are the same as equivalent expressions from \cite{northrop1963} up to factors multiplying $\textbf B$ which approach unity in cases where the $\textbf E \times \textbf B$ drift is non-relativistic. Furthermore, \cite{northrop1963} contains a differential equation for the Lorentz factor, $\gamma$, which we omit by simply using the definition:

\begin{equation}\label{NRGC3}
\gamma = \sqrt{1 + \frac{U^2 + u_E^2}{c^2} + \frac{2\mu B}{mc^2}}.
\end{equation}
Since the code tracks relativistic quantities $U$, $u_E$ and $\mu$ rather than their non-relativistic counterparts, it is necessary to use this modified definition of the Lorentz factor compared to the usual one. Here $u_E = \gamma \textbf E \times \textbf B /B^2$ the relativistic $\textbf E \times \textbf B$ drift, the relativistic magnetic moment is given by $\mu = \gamma^2 m v_\perp/2B$, where $v_\perp$ is the particle's velocity perpendicular to the magnetic field. Finally, $m$, $q$, and $c$ are the particle mass, charge and speed of light, respectively. In this paper, we examine the dynamics of electrons only. Where the anomalous resistivity is zero Equations (\ref{NRGC1}) and (\ref{NRGC2}) are solved using an adaptive time-step Runge-Kutta scheme, with the Lorentz factor (and hence the energy) being updated with Equation (\ref{NRGC3}). 

In regions where resistivity is non-zero pitch angle scattering is implemented following \cite{borissov-et-al2017}. This approach involves updating the cosine of the pitch angle $\beta = \cos \theta = v_{\pr}/v_{tot}$ (where $v_{tot} = \sqrt{v_\pr^2 + v_\perp^2}$ is the speed of the particle) at the start of each time-step by numerically integrating the following stochastic differential equation (SDE):
\begin{equation}\label{SDE1}
d\beta = \LL \dot \beta - \nu\beta \RR dt + \sqrt{\nu\LL 1-\beta^2 \RR}dW,
\end{equation}
where $\nu$ is the scattering rate and $W$ the Wiener process. In the stepping algorithm $\Delta W = \zeta \sqrt{\Delta t}$, with $\zeta$ being a random number drawn from a normal distribution with mean zero and standard deviation one \citep[e.g.][]{2014ApJ...787...86J}. The value of $\dot \beta$ is the rate of change of the particle's pitch angle cosine caused by propagation through the given electromagnetic fields. It is obtained by setting the time derivative of the magnetic moment to zero and rearranging the expression \citep[for a derivation see the appendix of][]{borissov-et-al2017}. The resulting expression is: 
\begin{equation}\label{SDE2}
\dot \beta = \left( \frac{1}{U}\frac{dU}{dt} - \frac{1}{2B}\frac{dB}{dt} \right) \beta\left( 1-\beta^2 \right).
\end{equation}

We consider two models for the scattering rate. In both cases we set the scattering rate to be $\nu = \nu_{ei}n'/\kappa$, where $\nu_{ei} = 2.91\times10^{-6} n \log \Lambda T^{-3/2} \unit {s^{-1}}$, using cgs units, with temperature in \unit{eV} \citep{huba1998}. We introduce $n' = 10^{-5}$ to adjust the scattering rate to something which would be closer to scattering rates for number densities typical of coronal conditions, since the number density in the MHD simulations was intentionally elevated to keep the temperatures reasonable. The different scattering models examined are characterised by different values of $\kappa$. We used $\kappa = 10^{-8}$ and 
$\kappa = \eta_{sp}/\eta$, where $\eta_{sp}$ 
is the local Spitzer resistivity. 
Throughout this paper, we refer to the scattering rates as specified by $\kappa$ if necessary.

Test particles are injected into a snapshot from the MHD simulation whereupon particle orbit trajectories and energy gain are computed subject to the equations described above. The trajectories are integrated until the test particle exits the computational box or the simulation time exceeds $10^{-4}\unit s$. 


\section{Individual particle orbits}
\label{single-particle}
\begin{figure*}[t]
\centering
        \subfloat[Orbit trajectories without scattering]{\label{orbits-a}\resizebox{0.49\textwidth}{!}{\includegraphics[clip=true, trim=60 20 65 25]{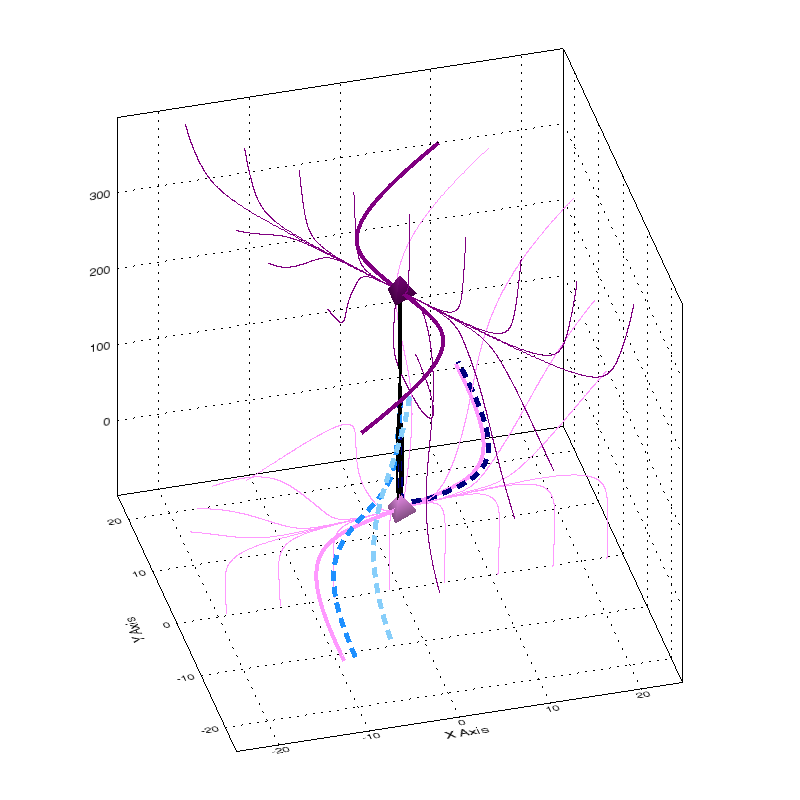}}}
                \subfloat[Orbit trajectories with scattering $\LL\kappa = \eta_{sp}/\eta_a\RR$]{\label{orbits-b}\resizebox{0.49\textwidth}{!}{\includegraphics[clip=true, trim=60 20 65 25]{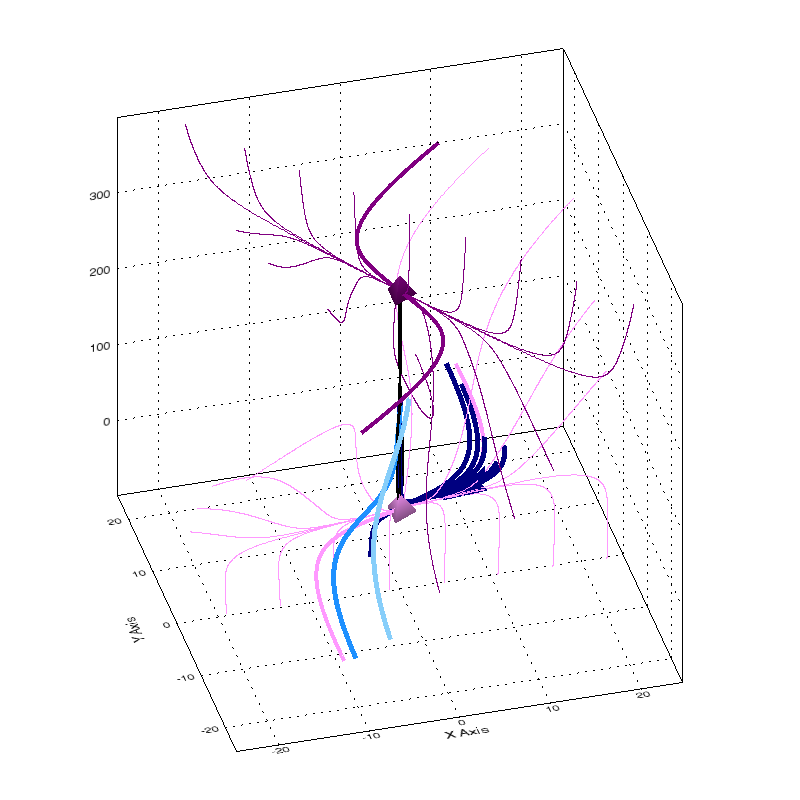}}}\\    
        \subfloat[Kinetic energy evolution]{\label{orbits-c}\resizebox{0.49\textwidth}{!}{\includegraphics[clip=true, trim=30 5 5 25]{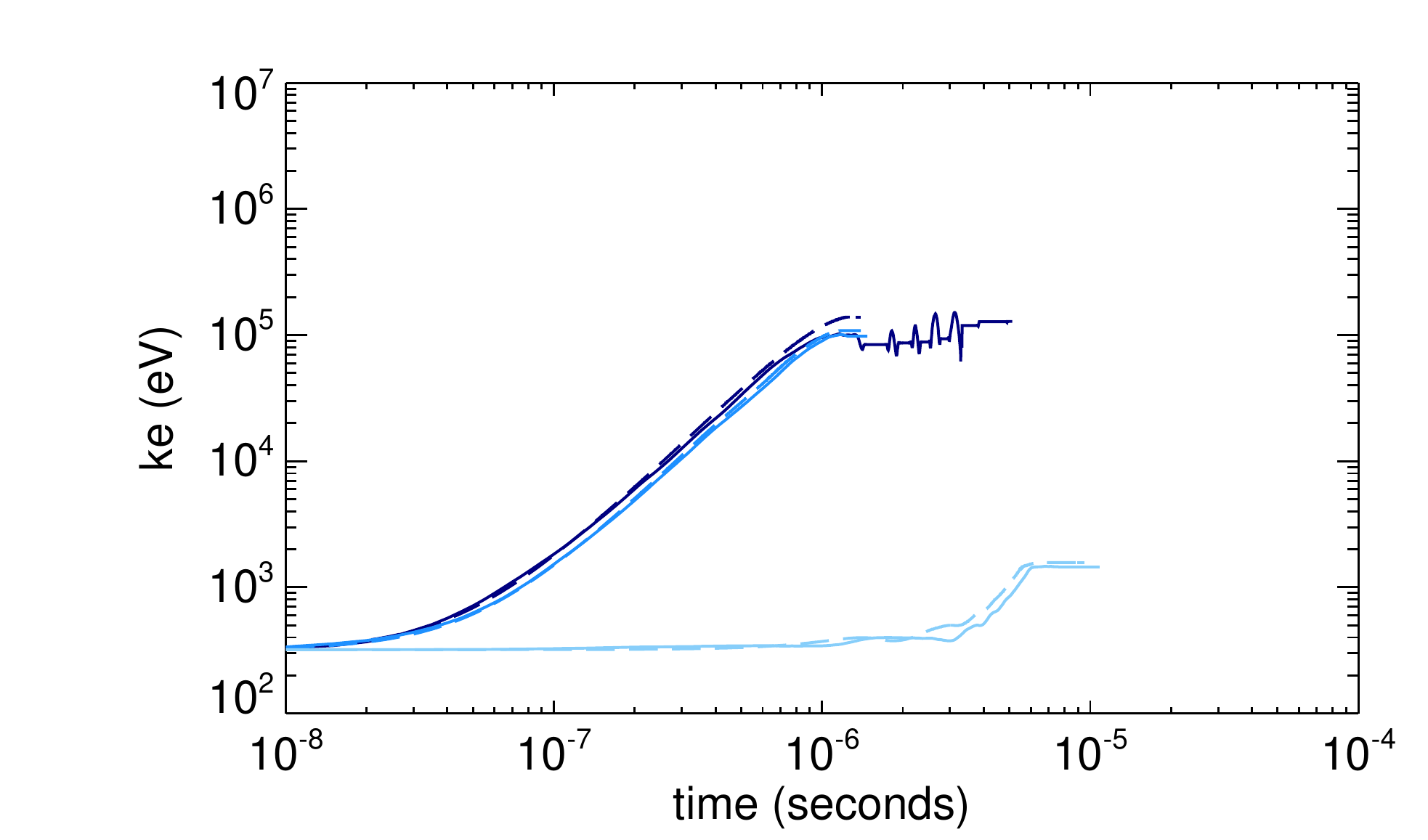}}}
        \subfloat[Pitch angle evolution]{\label{orbits-d}\resizebox{0.49\textwidth}{!}{\includegraphics[clip=true, trim=5 5 5 25]{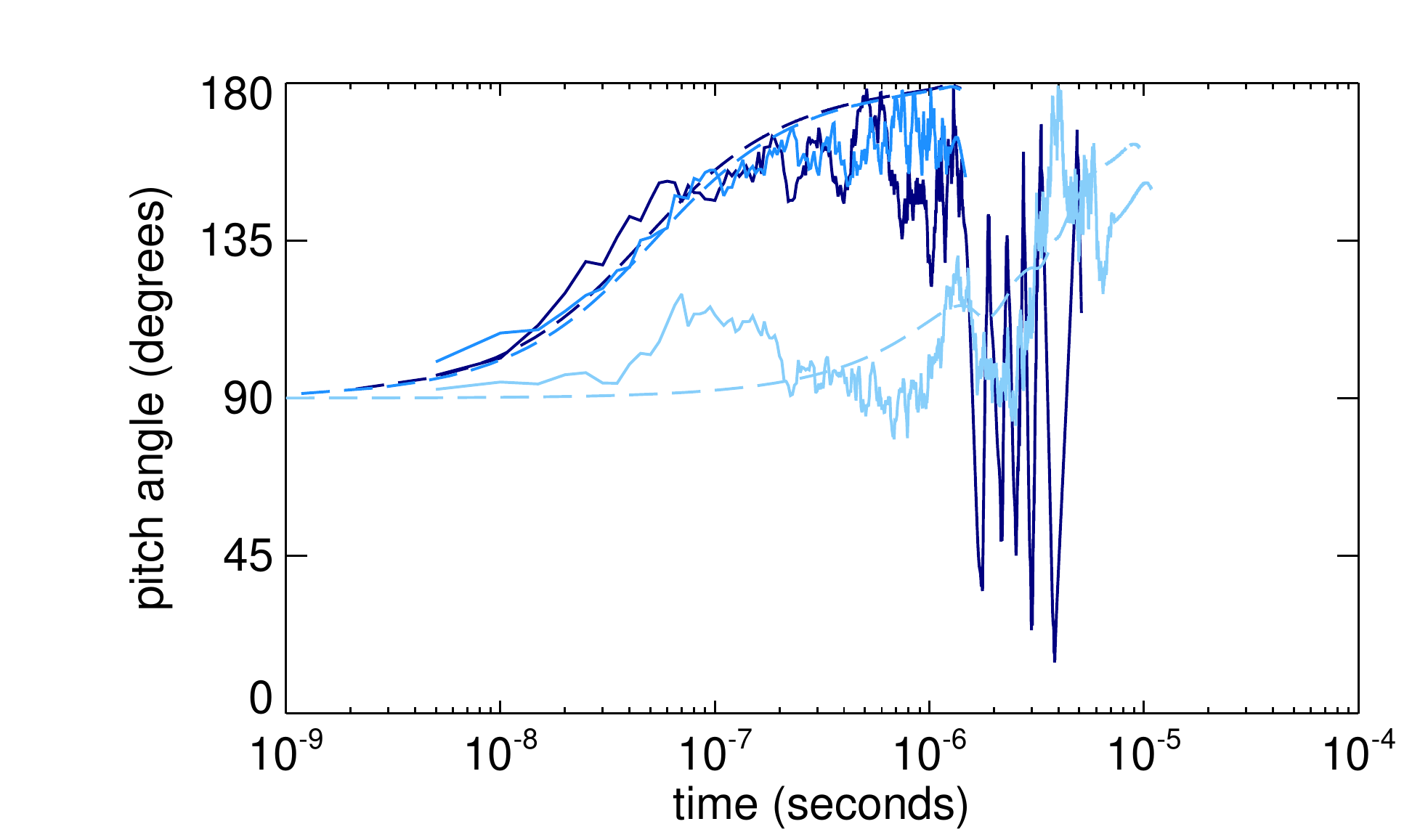}}}
        \caption{Comparison of orbit trajectories and properties when scattering is included or omitted.
        Trajectories which \protect\subref{orbits-a} omit or \protect\subref{orbits-b} include scattering for three cases initialised at different radial distances along the midplane of the separator ($x_0 = 0.01,0.1,1\unit m$ seen as dark blue, blue and light blue lines respectively) are overlaid with topological features of the MHD simulation (for key see Figure~\ref{f1}). \protect\subref{orbits-c} shows the evolution of kinetic energy of each of the orbits, while \protect\subref{orbits-d} shows the pitch angle evolution over time; cases which include scattering are shown as solid lines, while those which omit scattering are shown using dashed lines. }
        \label{orbits}
\end{figure*}

Our initial objective is to gauge the effect of pitch angle scattering on orbit trajectories, energy gains, and, indeed, pitch angle evolution. For the purposes of clarity, we begin by considering specific single particle orbit calculations prior to calculations of large numbers of orbits. Individual test particles are initialised with energy 320 \unit{eV} {and $90\dg$ pitch angles, while the initial distance from the separator is varied in one of three positions ($x_0 = 0.01,0.1,1 \unit m$, while fixing $y_0 = 0\unit m$ and $z_0 = 150\unit m$).}This energy corresponds to the median of a $2.5\times 10^6\unit K$ {Maxwellian} distribution, while the initial pitch angle is chosen so as not to introduce any velocity parallel to the magnetic field. {The orbits are calculated using the MHD simulation with the highest resistivity case ($\eta = 10^{-3}$), to most clearly distinguish between cases with and without scattering.}

{The results of the calculations are summarised in Figure \ref{orbits}. In the absence of scattering, Figure~\ref{orbits-a} shows the particle orbit paths, (blue dashed lines) which closely follow the separator and exit the computational box along field lines close to the spines of the lower null. When scattering is included, the same initial positions yield the orbit trajectories seen in Figure~\ref{orbits-b}. In this case, the orbit initialised closest to the separator has a much more complex trajectory (seen in dark blue in Figure~\ref{orbits-b}), with several strong scattering events causing the particle trajectory to repeatedly and sharply change close to the lower null. Orbits initialised further from the separator (medium and light blue) experience significantly less scattering and, hence, their trajectories are well matched when comparing Figures~\ref{orbits-a} and~\ref{orbits-b}. Viewed from the perspective of energy and pitch angle evolutions over time (Figures~\ref{orbits-c} and~\ref{orbits-d}), 
the orbit initially closest to the separator (dark blue) gains the most energy, while also experiencing the largest swings in pitch angle when scattering is included (compared to both cases initialised further from the separator, seen in medium and light blue, and unscattered cases, seen as dashed lines). We also see that in each of the three initial position cases in Figure~\ref{orbits-d}, the final kinetic energy appears well matched whether scattering is included or not; the solid and dashed lines of each colour ultimately appear to reach very similar final energy levels. The same cannot be said for the pitch angles seen in Figure~\ref{orbits-d} however, where large variations can be seen between final pitch angles when comparing cases which include or omit scattering and which begin at the same location (i.e. comparing solid and dashed lines of the same colour).}
\begin{figure*}[t]
        \centering
        \subfloat[$\eta_a = 10^{-5}$]{\label{to1a}\resizebox{0.49\textwidth}{!}{\includegraphics{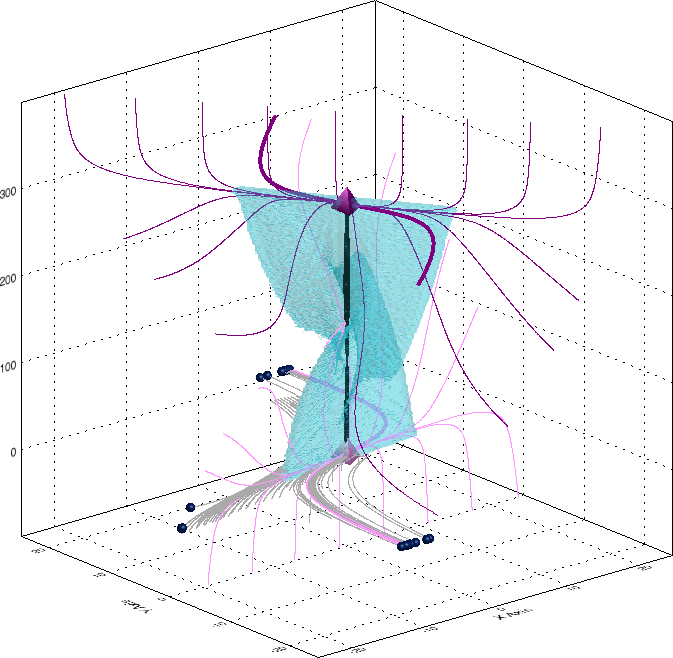}}}
        \subfloat[$\eta_a = 10^{-4}$]{\label{to1b}\resizebox{0.49\textwidth}{!}{\includegraphics{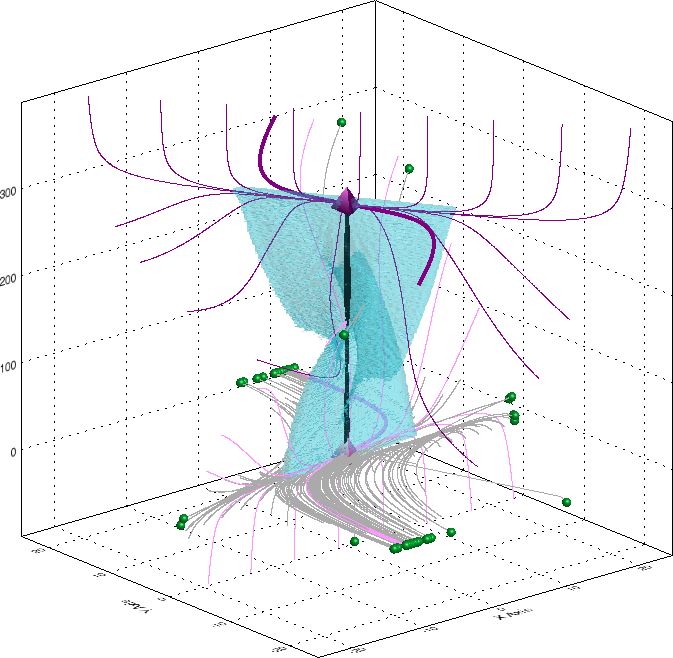}}}\\
        \subfloat[$\eta_a = 10^{-3}$]{\label{to1c}\resizebox{0.49\textwidth}{!}{\includegraphics{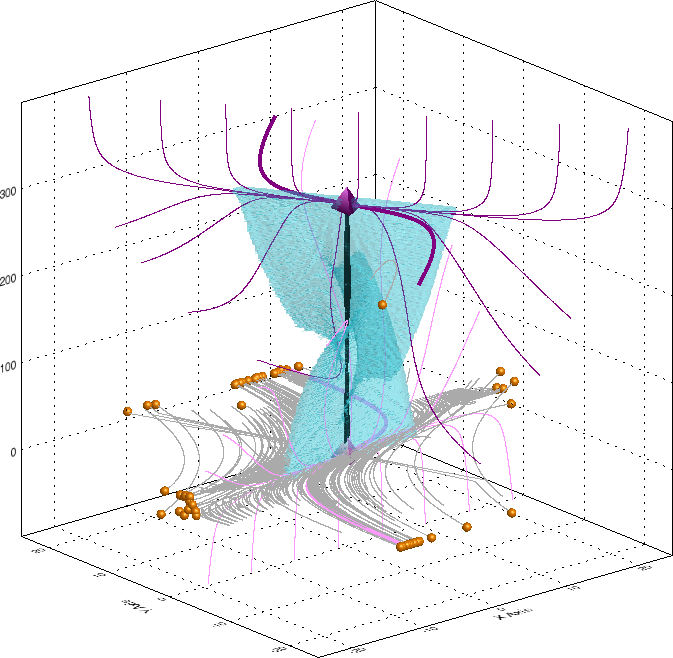}}{\hspace{1ex}}{\resizebox{0.155\textwidth}{!}{\includegraphics{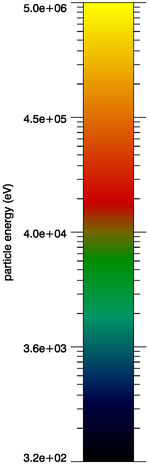} }}}
        \caption{Orbit trajectories, final positions, energies and field structure for 100 orbits with identical initial conditions, but with scattering effects, subject to MHD resistivities \protect\subref{to1a} $\eta_a = 10^{-5}$, \protect\subref{to1b} $\eta_a = 10^{-4}$ and \protect\subref{to1c} $\eta_a = 10^{-3}$. Initial conditions are $[x,y,z] = [0,0,150]\unit m$, pitch angle $90\dg$, kinetic energy $320\unit{eV}$. Coloured orbs represent final orbit positions, coloured according to final kinetic energy (for key see colour bar), grey lines represent orbit paths, while topological features are also overlaid (as defined in Figure~\ref{f1}).}
        \label{to1}
\end{figure*}
\begin{figure*}[t]
  \centering
        \subfloat[z$(t)$, $\eta_a = 10^{-5}$]{\label{to2-a}{\includegraphics[width = 0.43\textwidth]{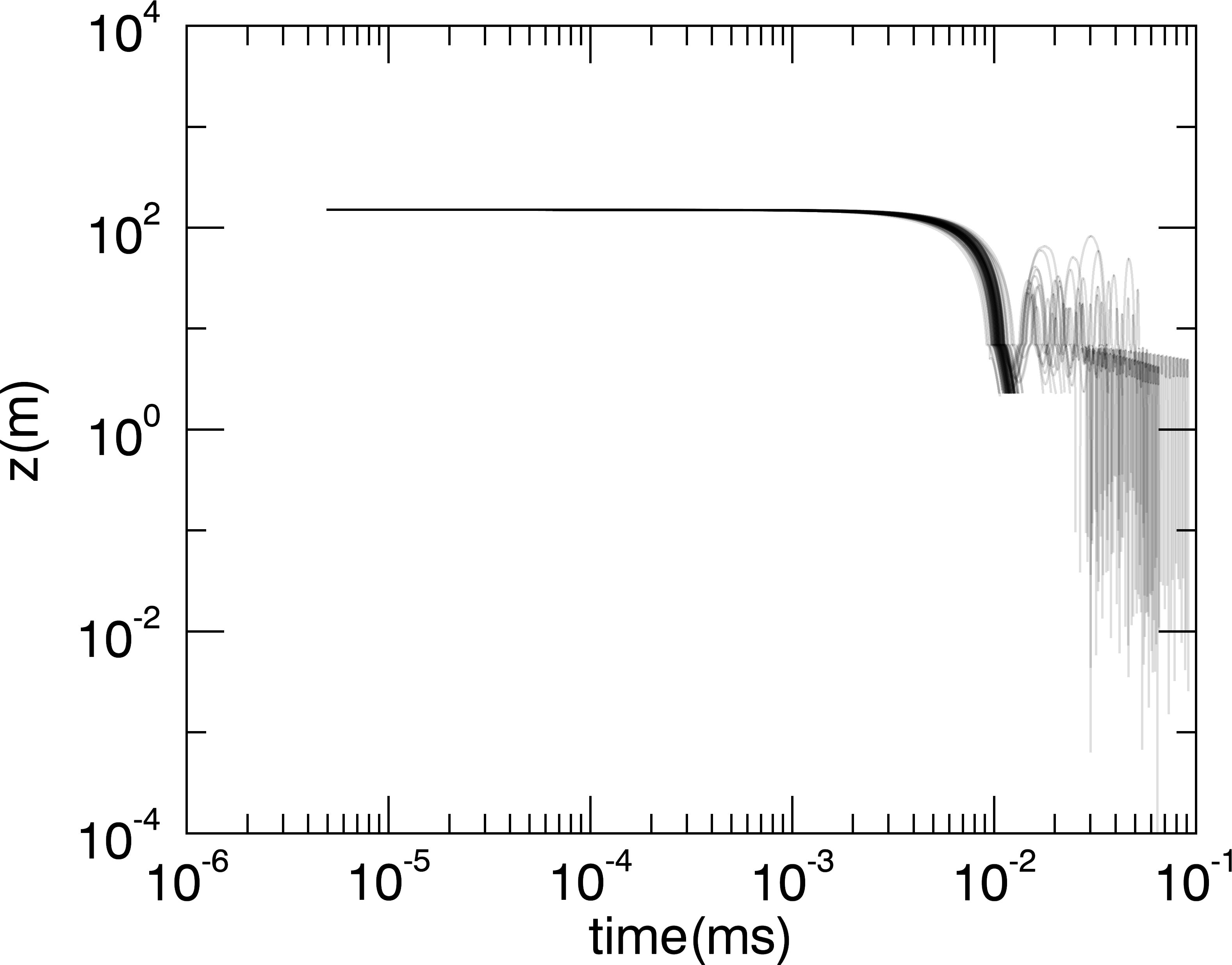}}}
        \hspace{1cm}
        \subfloat[KE$(t)$, $\eta_a = 10^{-5}$]{\label{to2-b}{\includegraphics[width = 0.43\textwidth]{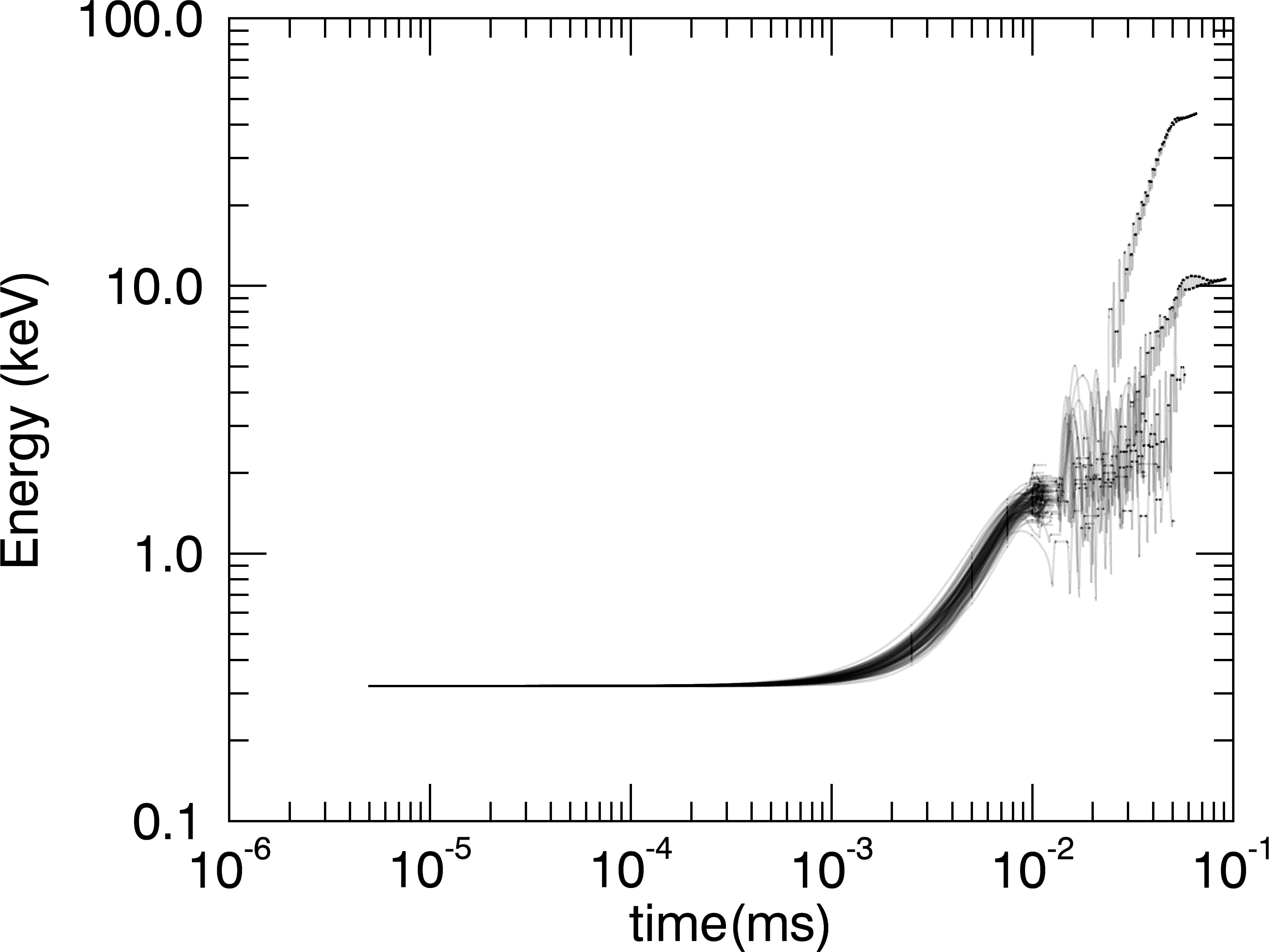}}}\\
        \subfloat[z$(t)$, $\eta_a = 10^{-4}$]{\label{to2-c}{\includegraphics[width = 0.43\textwidth]{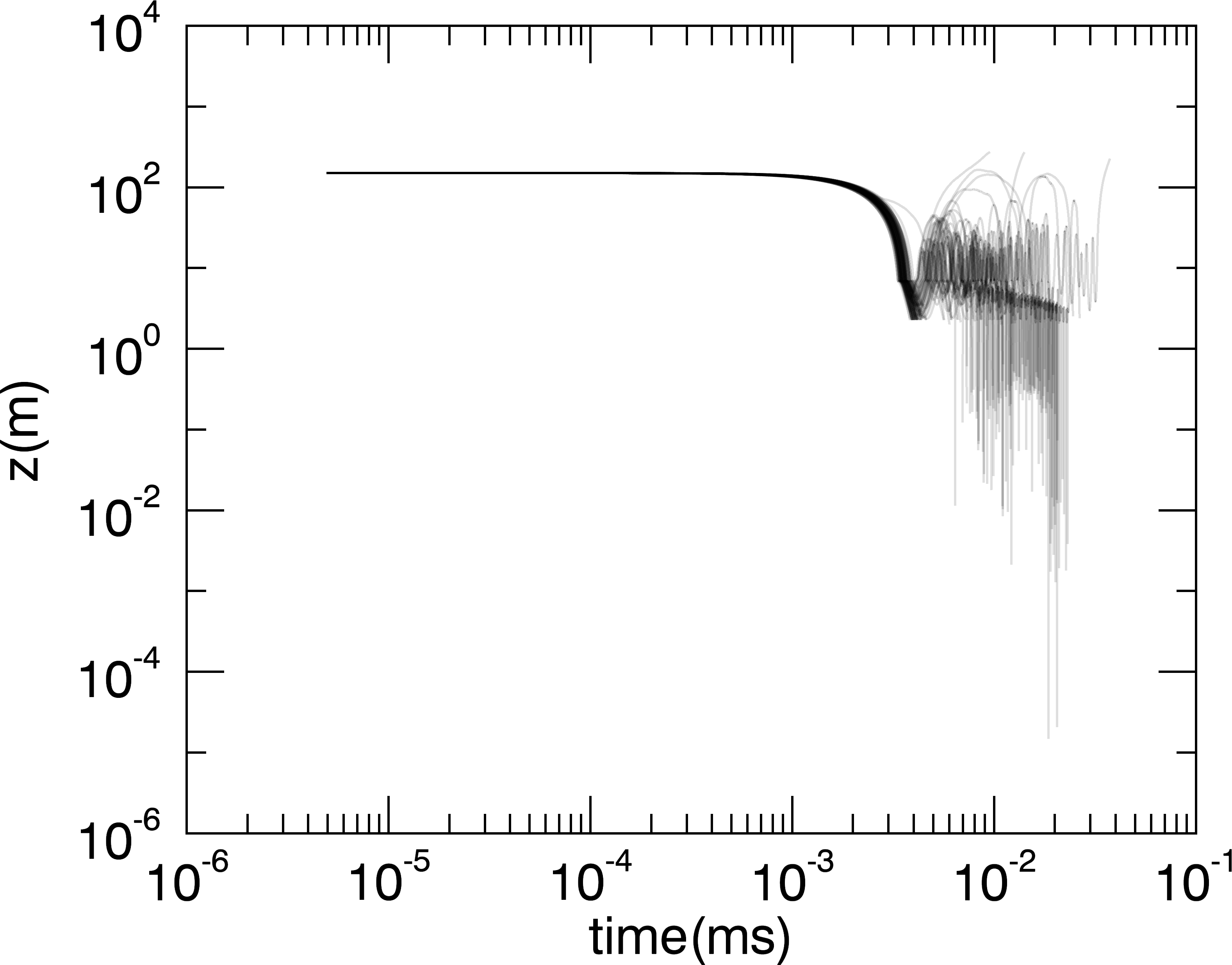}}}
        \hspace{1cm}
        \subfloat[KE$(t)$, $\eta_a = 10^{-4}$]{\label{to2-d}{\includegraphics[width = 0.43\textwidth]{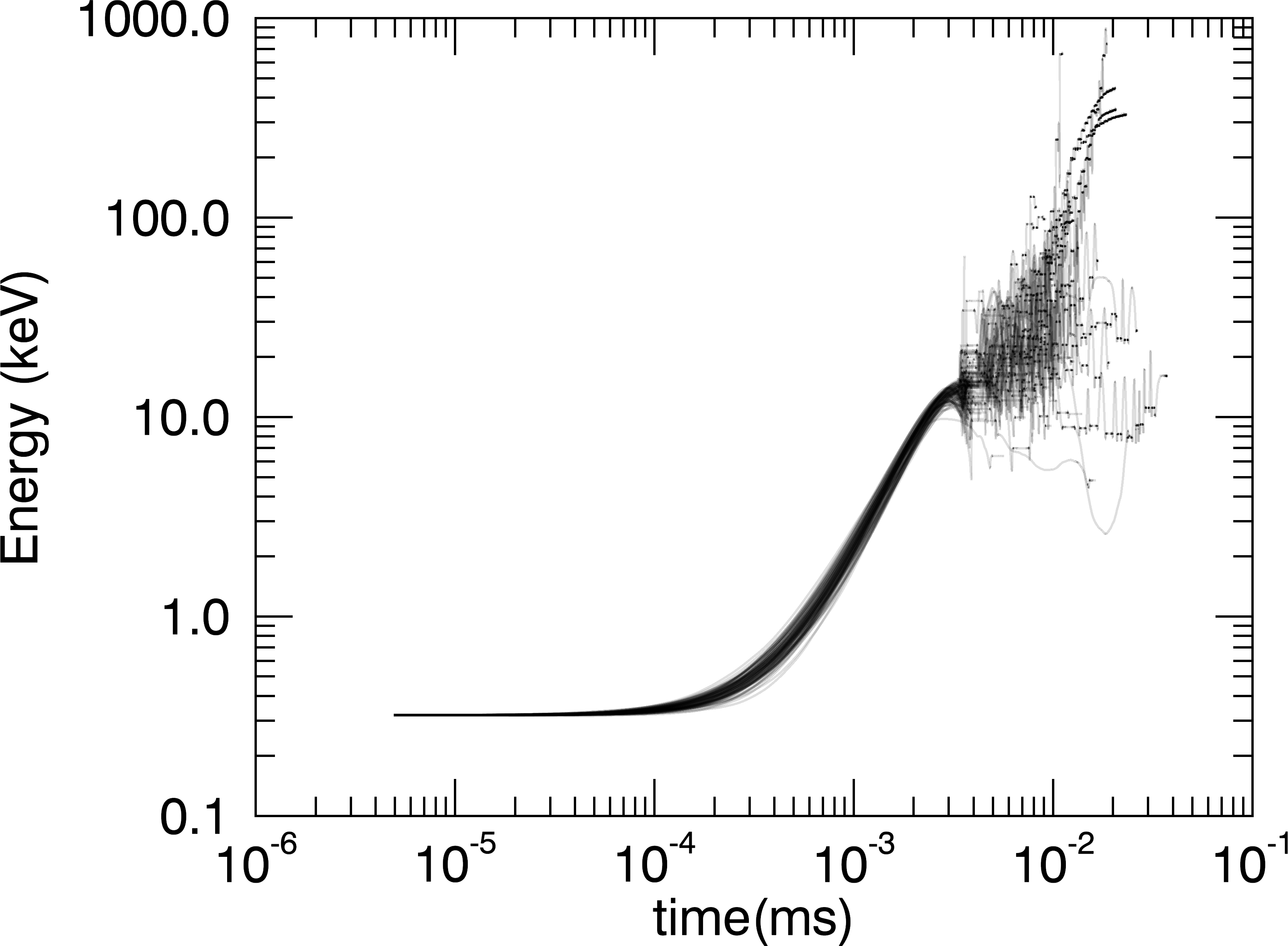}}}\\
        \subfloat[z$(t)$, $\eta_a = 10^{-3}$]{\label{to2-e}{\includegraphics[width = 0.43\textwidth]{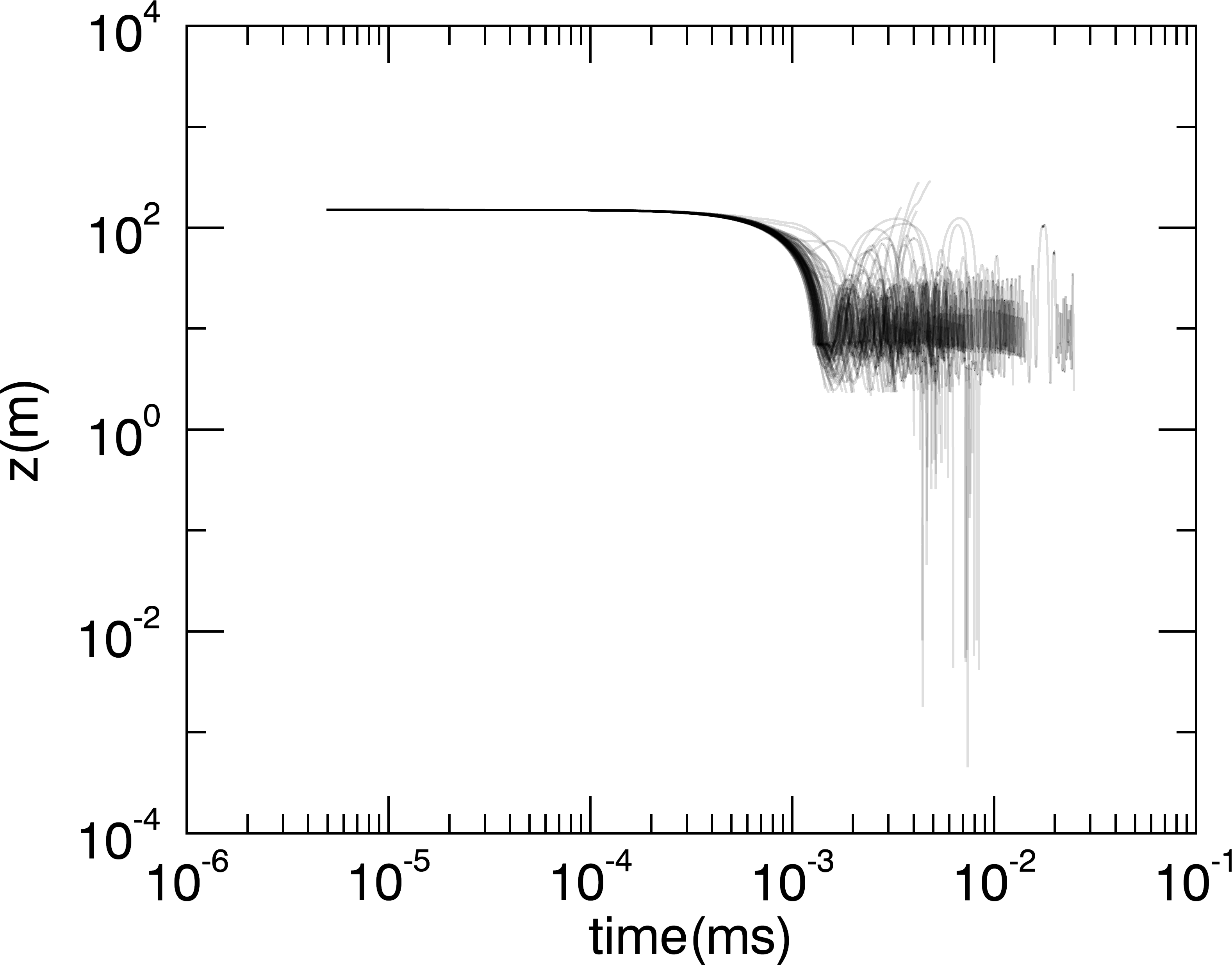}}}
        \hspace{1cm}
        \subfloat[KE$(t)$, $\eta_a = 10^{-3}$]{\label{to2-f}{\includegraphics[width = 0.43\textwidth]{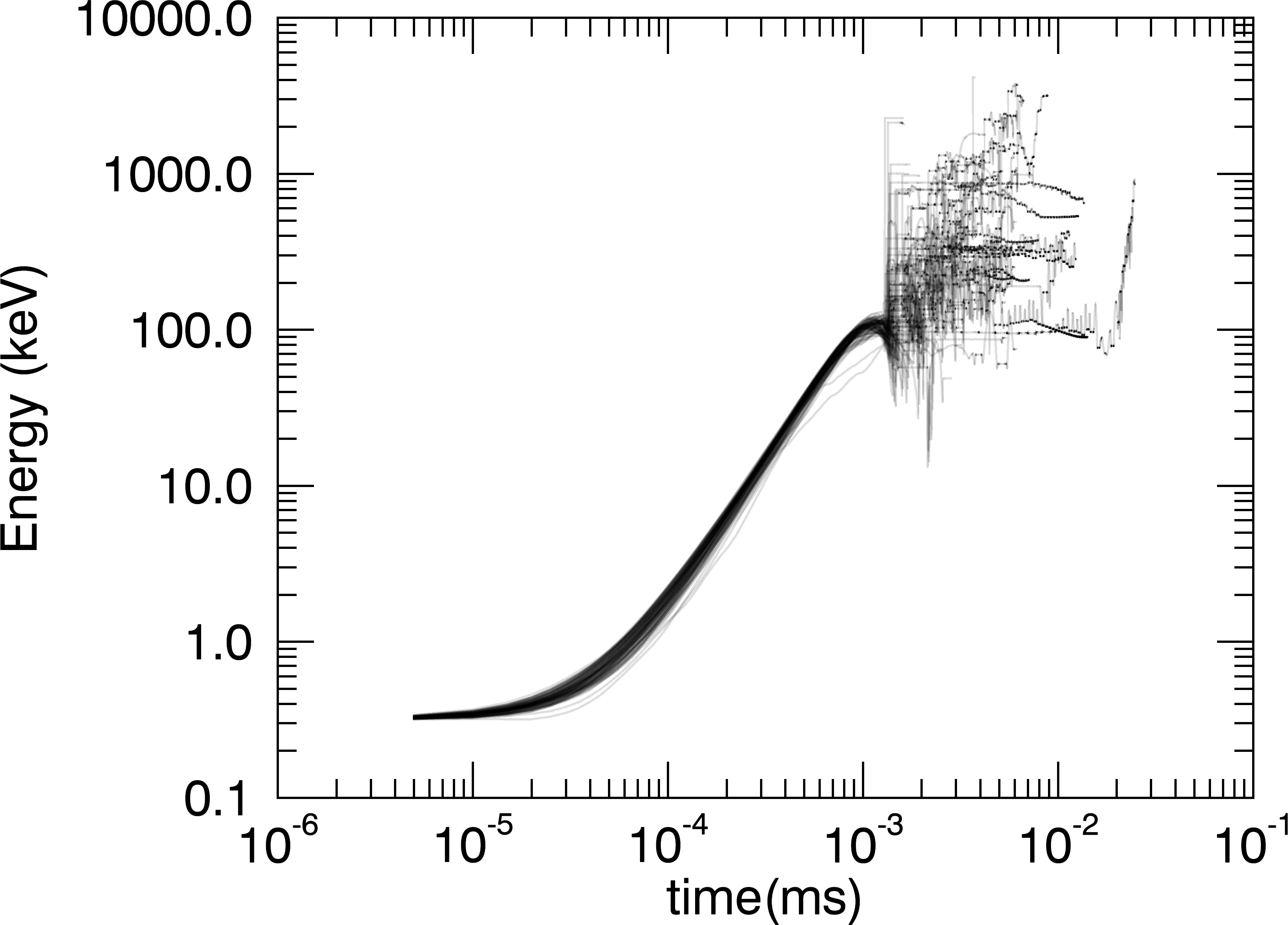}}}
        \caption{Orbit $z$-displacement (left column) and kinetic energy (KE, right column) over time for 100 orbits with identical initial conditions but subject to scattering effects (with trajectories seen in Figure \ref{to1}).}
        \label{to2}
\end{figure*}

Much like any stochastic process, it is difficult to glean much information about the role of pitch angle scattering in only a single example. By repeating the same calculation many times, we are able to better assess the role pitch angle scattering plays in the range of particle orbit behaviour ultimately recovered. We therefore perform 100 particle orbit calculations with identical initial conditions (energy $320\unit{eV}$, pitch angle $90\dg$, {$[x,y,z] = [0,0,150]\unit m$}). {The calculations are repeating using a single snapshot from MHD simulations performed at different resistivites, $\eta = 10^{-3}, 10^{-4}, 10^{-5}$. In Figure \ref{to1} we outline the resulting 3D orbit trajectories at each resistivity, as well as the final kinetic energy of each orbit at the final orbit position and the topology of the magnetic field structure in the MHD simulation snapshot used as the orbit environment. We clearly see that the increasing resistivity not only increases the final kinetic energy of each orbit (seen in the colour of the orbs at the positions where the orbits leave the computational box) but it also allows the orbits to leave the box along different topological features. In the case of the weakest resistivity, $\eta = 10^{-5}$, Figure~\ref{to1a} shows that the majority of the orbits show minimal energy gains and predominantly leave the numerical domain close to the spines of the lower null (solid pink lines). Increasing the resistivity to $\eta = 10^{-4}$ and $\eta = 10^{-3}$ (seen in Figures~\ref{to1b} and~\ref{to1c} respectively) sees the orbits gain more energy, and exhibit much wider ranges of final trajectories, with many orbits now able to leave the domain along the fan plane of the lower null. Such alternative trajectories are only accessible due to pitch angle scattering, which allows orbits to 'escape' along previously inaccessible field lines.

In order to better illustrate the properties (and complexity) of our results for 100 different orbit realisations based on only a single initial position, in Figure~\ref{to2} we display the z-displacement and kinetic energy of each set of orbits, for each case of resistivity considered in this investigation. By comparing the vertical displacements of each resitivity cases (Figures~\ref{to2-a},~\ref{to2-c},~\ref{to2-e}), we see that the resistivity enhancements cause orbits to accelerate faster down the separator, with scattering occuring earlier and more frequently as the resistivity increases.
This too is reflected in the kinetic energy evolutions (seen in Figures~\ref{to2-b},~\ref{to2-d},~\ref{to2-f}), where jumps in kinetic energy are generally earlier, larger and more frequent as resistivity increases, ultimately leading to larger final kinetic energies compared to weaker resistivity values.}

\section{Particle energy spectra and distributions}\label{many-particle}

Having studied specific examples of particle initial conditions, we then examined much larger distributions of particles subject to a broader range of initial conditions. We considered calculations of sets of $10^4$ particle orbits integrated with scattering at rates given by $\kappa = 10^{-8}$, $\kappa=\eta_{sp}/\eta_a$, as well as without scattering. We examined populations of orbits originating in different locations with respect to the separator: in {Case 1} (Section \ref{centred-particles}) we considered a population of particle orbits originating {along} the central (vertical) third of the separator. In {Case 2} (Section \ref{elongated-particles}) orbits originate along the whole length of the separator, while in {Cases 3 \& 4} (Sections \ref{extremity-particles} and~\ref{extremity-particles2}) orbits originate at either the top or bottom of the separator. In all cases the initial pitch angle distribution takes on 100 uniformly distributed values between $10\dg$ and $170\dg$, and the initial energy 25 values between $10\unit{eV}$ and $320\unit{eV}$. This range of initial energies covers over $90\%$ of the Maxwellian distribution at $10^6 \unit K$. The particle orbits are weighted when computing the spectra so that the initial energy distribution is Maxwellian and the initial distribution of the pitch angle cosine is uniform. Unless otherwise specified, we perform these particle calculations in the MHD fields using the $\eta_a = 10^{-3}$ simulation.

\subsection{Case 1: Initial particle positions near middle of separator}\label{centred-particles}

\begin{figure*}
\subfloat[KE: $x,y\in{[-10,10]}\unit{m}$]{\label{dist1a}\resizebox{0.33\textwidth}{!}{\includegraphics[clip=true, trim=5 5 15 15]{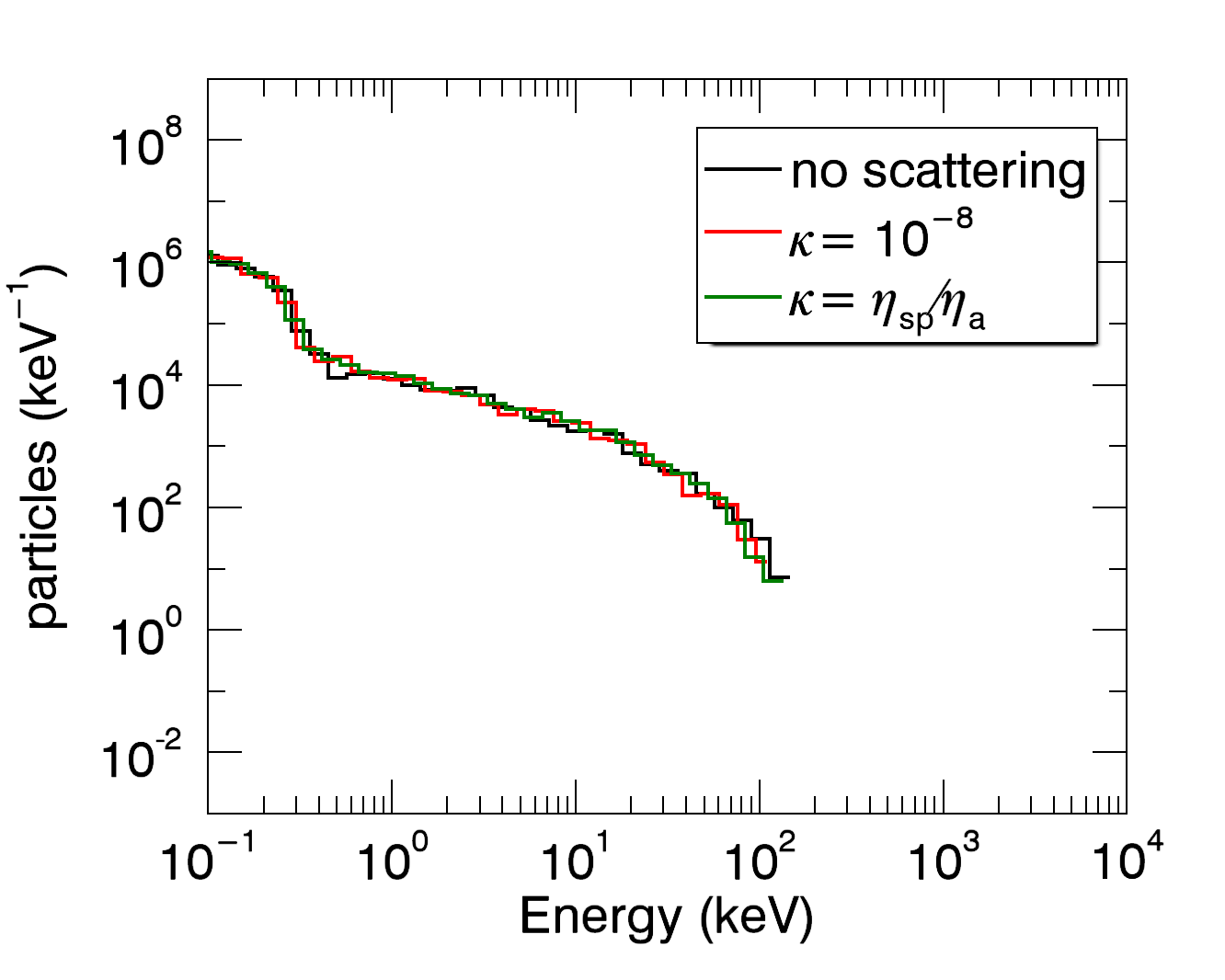}}}
\subfloat[$\theta$: $x,y\in{[-10,10]}\unit{m}$ ]{\label{dist1b}\resizebox{0.33\textwidth}{!}{\includegraphics[clip=true, trim=5 5 20 20]{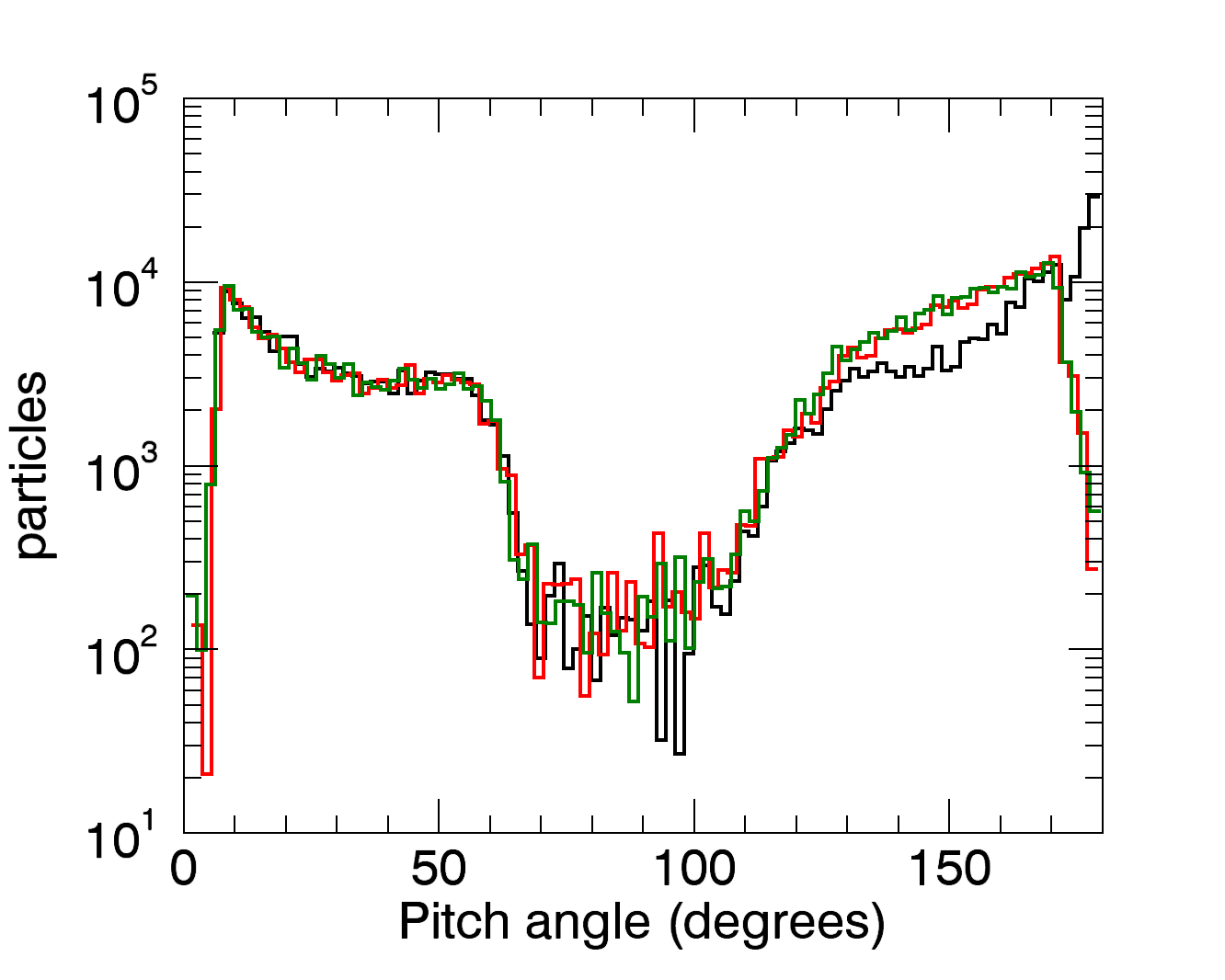}}}
\subfloat[duration: $x,y\in{[-10,10]}\unit{m}$ ]{\label{dist1c}\resizebox{0.33\textwidth}{!}{\includegraphics[clip=true, trim=5 10 12 20]{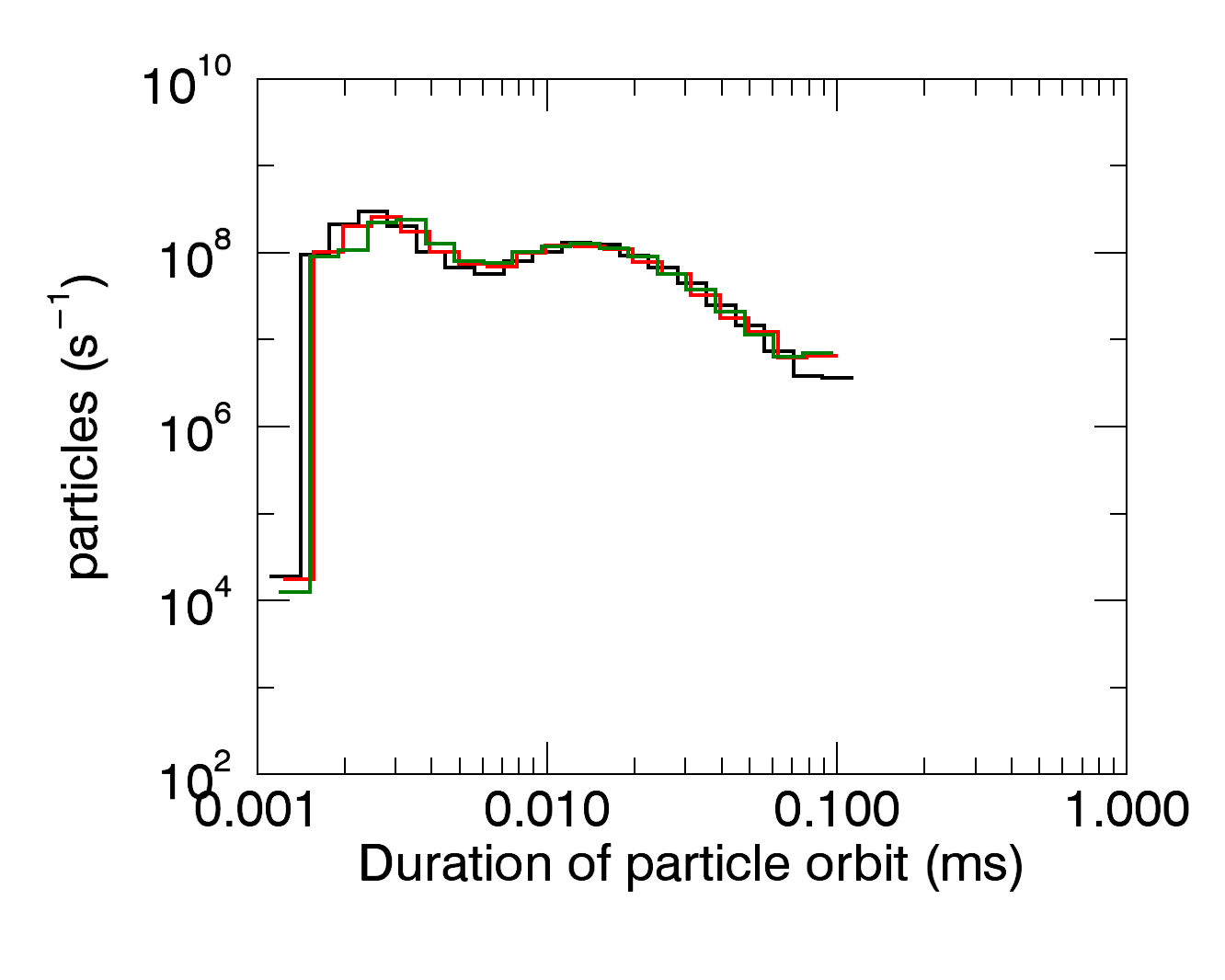}}}\\
\subfloat[KE: $x,y\in{[-1,1]}\unit{m}$]{\label{dist1d}\resizebox{0.33\textwidth}{!}{\includegraphics[clip=true, trim=5 5 15 15]{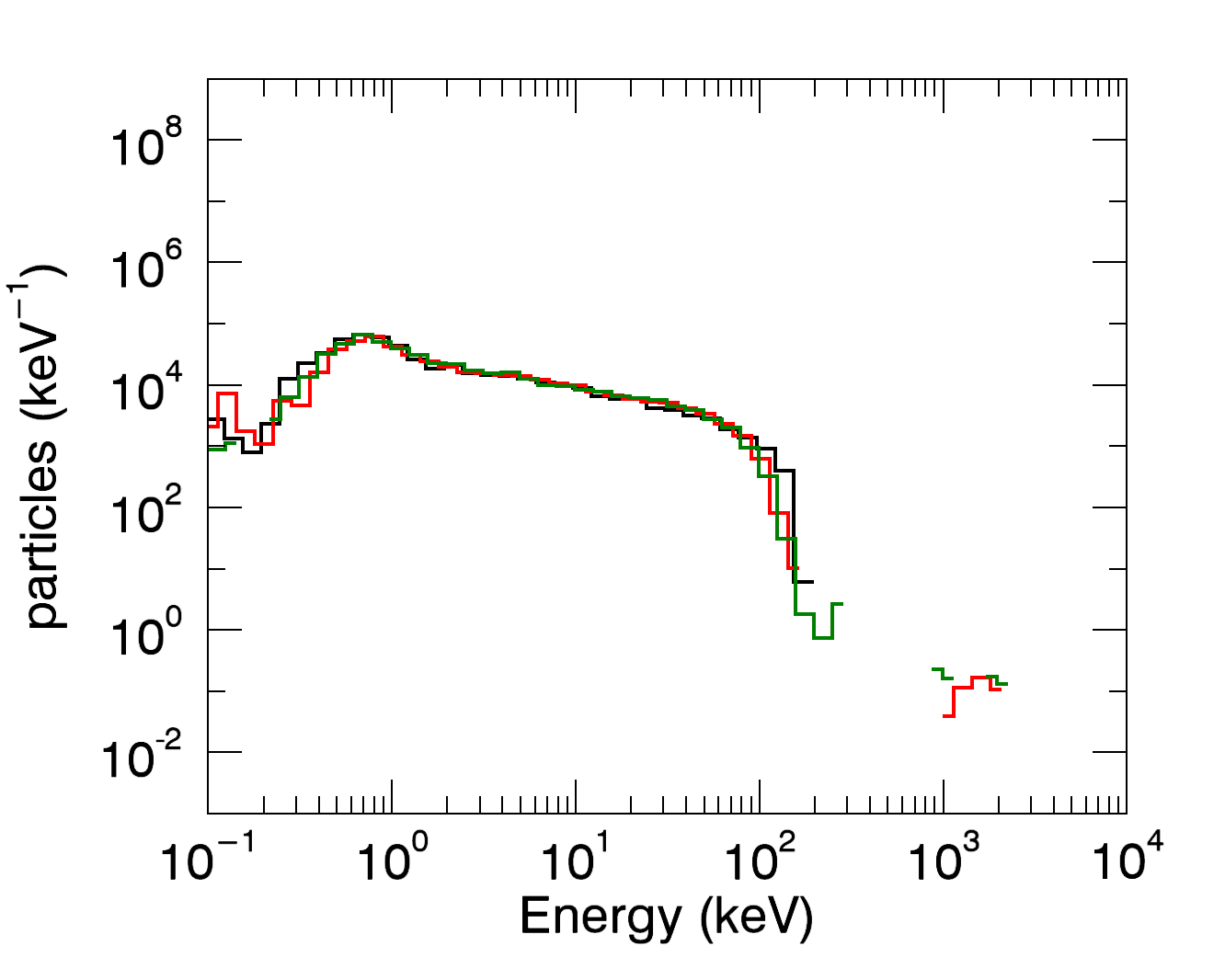}}}
\subfloat[$\theta$: $x,y\in{[-1,1]}\unit{m}$ ]{\label{dist1e}\resizebox{0.33\textwidth}{!}{\includegraphics[clip=true, trim=5 5 20 20]{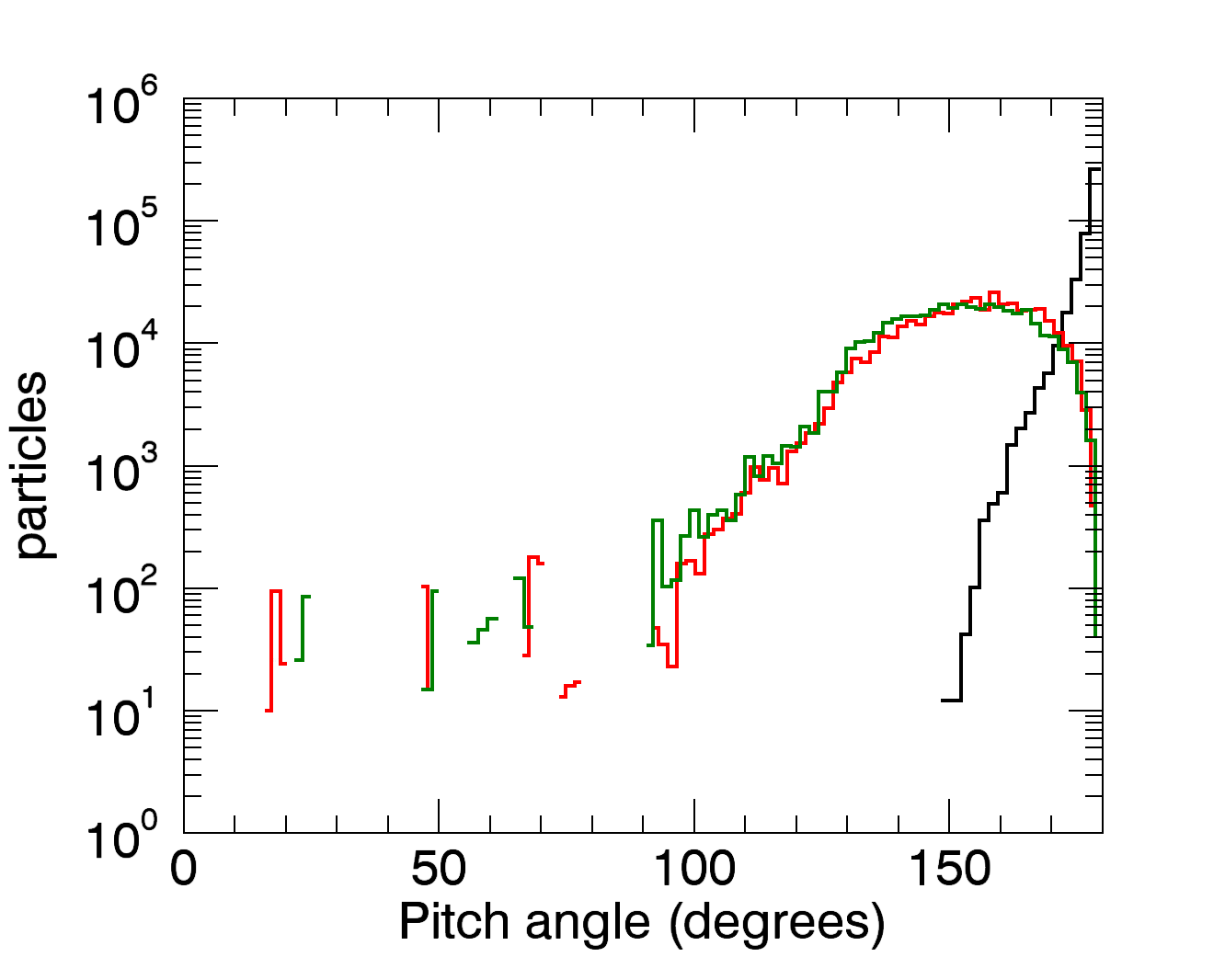}}}
\subfloat[duration: $x,y\in{[-1,1]}\unit{m}$]{\label{dist1f}\resizebox{0.33\textwidth}{!}{\includegraphics[clip=true, trim=5 10 12 20]{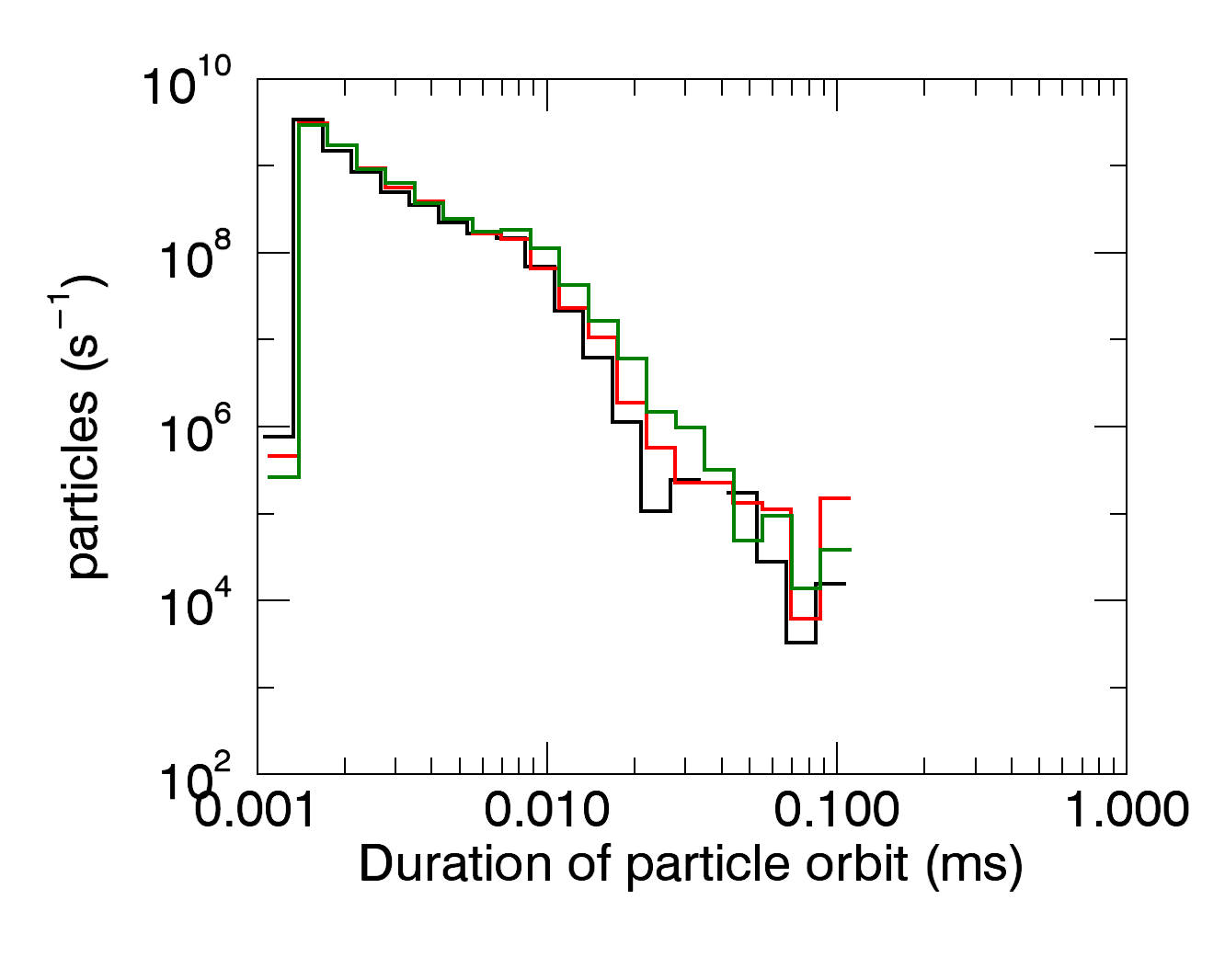}}}\\
\caption{Particle energy spectra, pitch angle and orbit duration distributions for Case 1 orbits: the colours represent orbits without scattering, with scattering at a rate $\kappa = 10^{-8}$, and $\kappa = \eta_{sp}/\eta_a$ respectively (see legend) based upon initial energies $\in[10,320] \unit{eV}$, initial pitch angles $\in[10,170]\dg$ and in fields from MHD simulation with $\eta = 10^{-3}$. All orbits originate from within $z \in [100,200]\unit m$: \protect\subref{dist1a}-\protect\subref{dist1c} are initialised within $x,y \in [-10,10]\unit m$, while \protect\subref{dist1d}-\protect\subref{dist1f} originate within $x,y \in [-1,1]\unit m$.}
\label{dist1}
\end{figure*}

Case 1 concerns a population of test particles with initial positions uniformly distributed in $x,y \in [-10,10]\unit m$ and $z \in [100,200]\unit m$, corresponding to a box encompassing the central part of the separator. The resulting spectra, pitch angle and orbit duration distributions are shown in Figures \ref{dist1a}-\ref{dist1c}. {The energy spectra for particle orbits both with and without scattering are almost identical (see Figure \ref{dist1a}). This is similar to the single orbit results seen in, for example, Fig~\ref{orbits}, where the kinetic energies with and without scattering were well-matched, but directly contrasting the 2D findings of \citet{borissov-et-al2017}. }

The orbit durations (Figure \ref{dist1c}) are also almost identical, however the pitch angle distribution (Figure \ref{dist1b}) shows fewer particles with pitch angles $\theta > 170\dg$ ($23.9\%$ of non-scattered particles achieve these pitch angles, compared to $6.7\%$ and $4.8\%$ for the $\kappa = 10^{-8}$ and $\kappa = \eta_{sp}/\eta_a$ scattering cases respectively). More than half ($64\%$) of all particle orbits have final pitch angles $\theta > 90\dg$ (regardless of scattering model). Scattering ensures a $\sim50\%$ increase of particle orbits in the range $90\dg \leq \theta \leq 170\dg$ compared to orbits without scattering. 

In order to establish why energy spectra in Figure \ref{dist1a} are indistinguishable, we repeat our orbit calculations but modify the initial positions of Case 1 orbits to originate closer to the separator. In Figures \ref{dist1d}-\ref{dist1f} we show the distributions which result, with initial positions uniformly distributed within $x,y \in [-1,1]\unit m$. Unlike the original Case 1 initial positions, all orbits now start within the reconnection region, which modifies the recovered energy spectra (Figure~\ref{dist1d}), reducing the number of orbits with lower energies and increasing the number of orbits with large energies. Furthermore, small differences between the scattering models are now apparent in Figure~\ref{dist1d} at the largest energies. The proportion of particles achieving energy more than $100\unit{keV}$ is $9.4\%$ in the absence of scattering, and $\sim 4\%$ for each of the scattering cases. If this threshold increases to $200\unit{keV}$, a change in behaviour can be observed: while more particles achieve $100\unit{keV}$ in the absence of scattering compared to the scattering cases, more scattered particles reach $200\unit{keV}$ than those without scattering. Specifically, $0.01\%$ of particle orbits reach energies above $200\unit{keV}$ in the absence of scattering, compared to $0.05\%$ for scattering at a rate $\kappa=10^{-8}$ and $0.1\%$ for $\kappa=\eta_{sp}/\eta_a,$ respectively. 

Considering the pitch angle distribution (Figure \ref{dist1e}) clear differences emerge between the scattered and unscattered simulations. Whereas without scattering almost all ($99.95\%$) of particles have final pitch angle $\theta > 145\dg$, only $75\%$ do for scattering with $\kappa = 10^{-8}$ and $72\%$ for scattering with $\kappa = \eta_{sp}/\eta_a$. This implies that scattering can have a strong effect on the final pitch angle distribution without necessarily yielding large differences in the final energy spectra.

Moving the initial orbit positions closer to the separator also starts to bring out smaller, yet notable differences between the scattering models in orbit lifetimes. Figure~\ref{dist1f} highlights that only $3.5\%$ of particles having an orbit duration greater than $0.01\unit{ms}$ in the absence of scattering, compared to $4.9\%$ with $\kappa = 10^{-8}$ and $9.2\%$ for $\kappa = \eta_{sp}/\eta_a$. These results are consistent with the pitch angle distribution containing fewer orbits with pitch angles close to $0\dg/180\dg$, implying on average a smaller parallel velocity and a longer time before the particles reach the computational boundary.

\subsection{Case 2: Initial positions along entire length of separator}\label{elongated-particles}

\begin{figure*}
\subfloat[KE]{\label{dist11g}\resizebox{0.33\textwidth}{!}{\includegraphics[clip=true, trim=5 5 15 15]{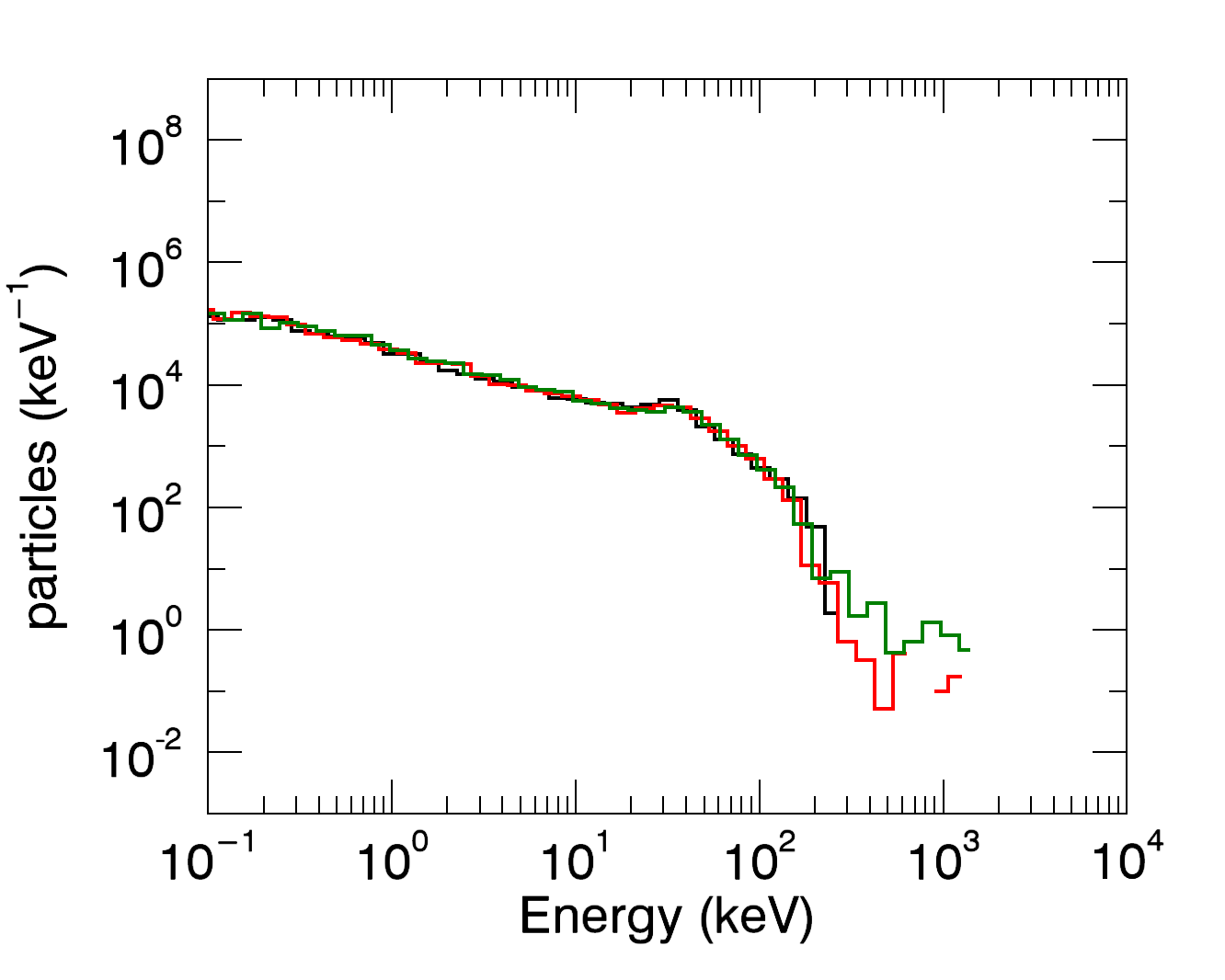}}}
\subfloat[$\theta$]{\label{dist11h}\resizebox{0.33\textwidth}{!}{\includegraphics[clip=true, trim=5 5 20 20]{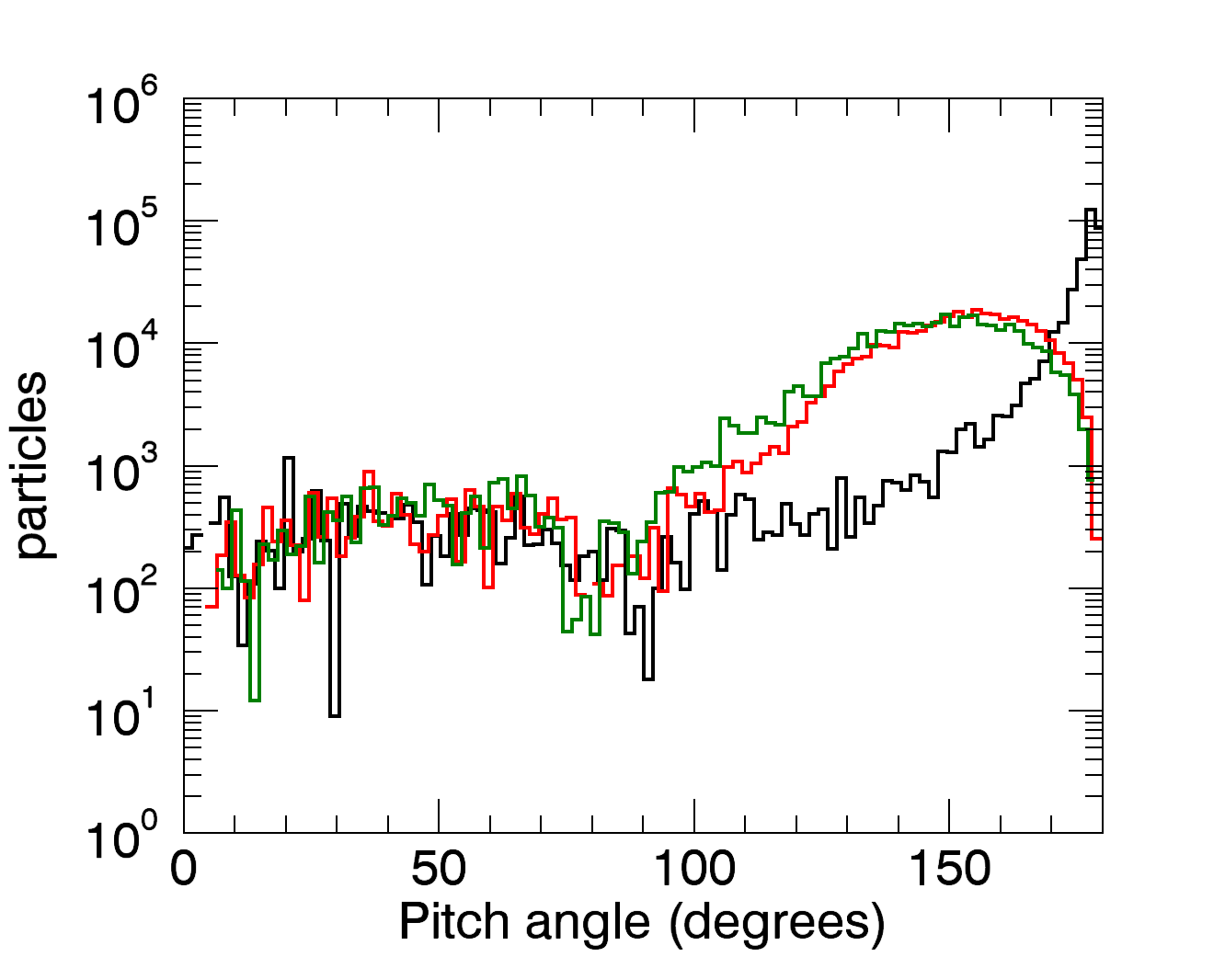}}}
\subfloat[duration]{\label{dist11i}\resizebox{0.33\textwidth}{!}{\includegraphics[clip=true, trim=5 10 12 20]{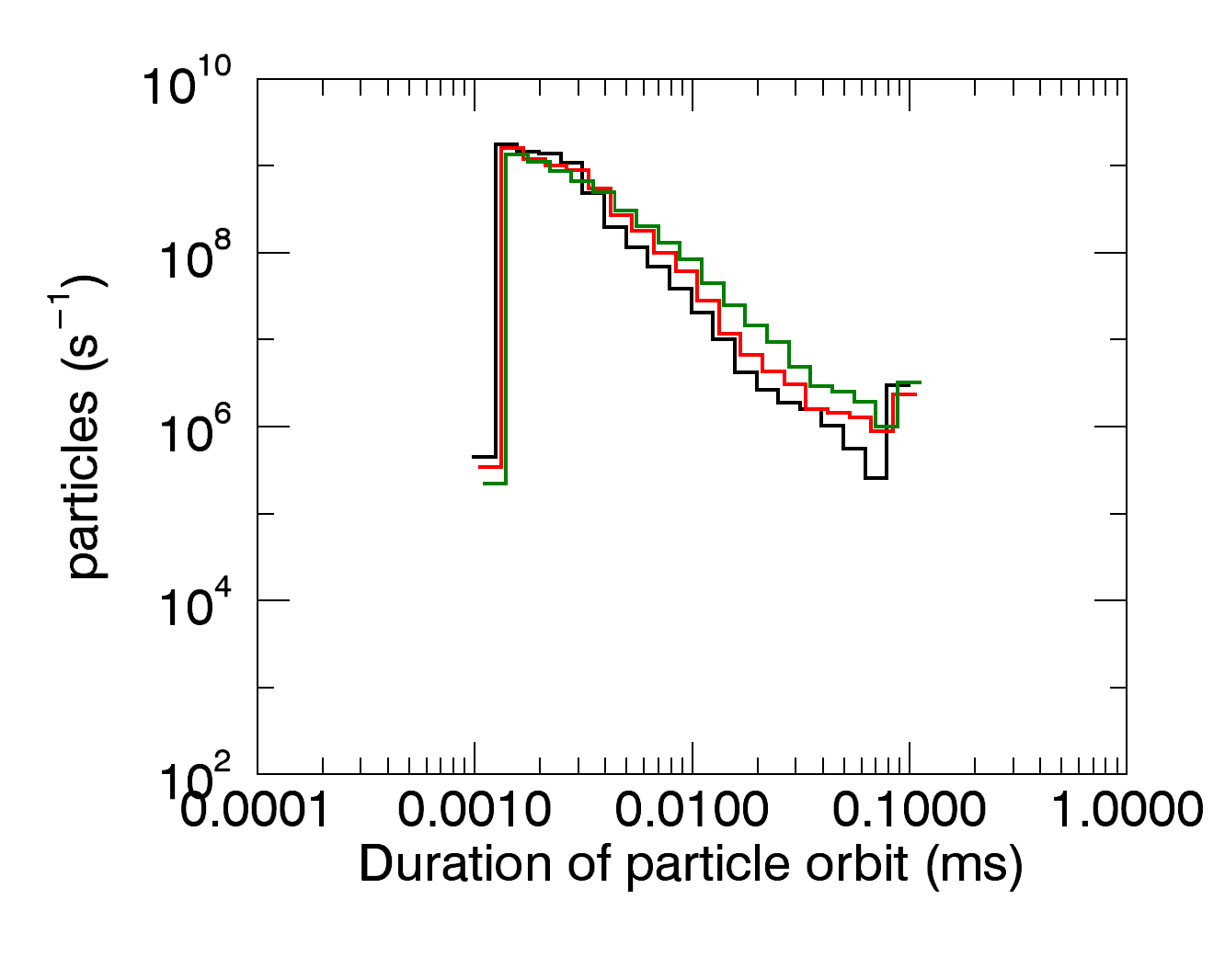}}}
\caption{Particle energy spectra, pitch angle, and orbit duration distributions for Case 2 orbits: the colours represent orbits without scattering, with scattering at a rate $\kappa = 10^{-8}$, and $\kappa = \eta_{sp}/\eta_a$ respectively (see legend), based upon initial energies $\in[10,320] \unit{eV}$, initial pitch angles $\in[10,170]\dg$ and in fields from MHD simulation with $\eta = 10^{-3}$. Orbits are initialised within $x,y \in [-1,1]\unit m$, $z \in [1,299]\unit m$.}
\label{dist11}
\end{figure*}

By extending the range of initial positions in z beyond that in Case 1, Case 2 aims to focus on orbit behaviour in the vicinity of the null points (noting that, of course, the guiding centre approach breaks down if the orbits closely approach the null). The extended range of vertical initial positions used in Case 2 ($x,y \in [-1,1]\unit m$ and $z \in [1,299]\unit m$) yields the distributions seen in Figures \ref{dist11g}-\ref{dist11i}). The percentage of particles achieving more than $100\unit{keV}$ energies in Figure~\ref{dist11g} remains $\sim 6\%$ in all cases. However, the fraction of particles obtaining energies more than $200\unit{keV}$ is $0.6\%$ in the absence of scattering, $0.2\%$ with scattering at $\kappa = 10^{-8}$ and again $0.6\%$ with $\kappa = \eta_{sp}/\eta_a$. Examining particle orbits gaining energies more than $300\unit{keV}$, we find $0.03\%$ in the absence of scattering, $0.09\%$ for $\kappa = 10^{-8}$ and $0.4\%$ for $\kappa = \eta_{sp}/\eta_a$. This implies that scattering can play a role in the energisation of a small number of particles to very high energies. 
The Case 2 pitch angle distribution, seen in Figure~\ref{dist11h}, shows smaller differences between the scattered and unscattered models than in Case 1 (Figure \ref{dist1e}). This is likely due to the wider range of initial $z$ values creating more orbits which are only weakly affected by the electric field, effectively reducing the final pitch angle. As a result, without scattering most ($93\%$) of the particles in Figure~\ref{dist11h} have final pitch angle $\theta > 145\dg$ (compared to the $99.95\%$ in Figure \ref{dist1e}). An extended range of initial $z$ values also results in a more pronounced difference in the pitch angle distributions between the two scattering models. In Case 2, $67\%$ of particle orbits have pitch angle $\theta > 145\dg$ for scattering with $\kappa = 10^{-8}$ and $58\%$ when $\kappa = \eta_{sp}/\eta_a$. Both values are lower than equivalent Case 1 values (seen in Figure \ref{dist1e}). Finally, the orbit duration distribution in Case 2 (Figure~\ref{dist11i}) shows minimal differences between scattered and unscattered particle orbit distributions; $6\%$ of orbits lasting longer than $0.01\unit{ms}$ in the absence of scattering, comparing to $9\%$ with $\kappa = 10^{-8}$ and $16\%$ for $\kappa = \eta_{sp}/\eta_a$.

\subsection{Case 3: Initial positions near bottom of separator}\label{extremity-particles}

\begin{figure*}
\centering
\subfloat[KE: $\eta_a=10^{-5}$]{\label{dist2a}\resizebox{0.33\textwidth}{!}{\includegraphics[clip=true, trim=5 5 15 15]{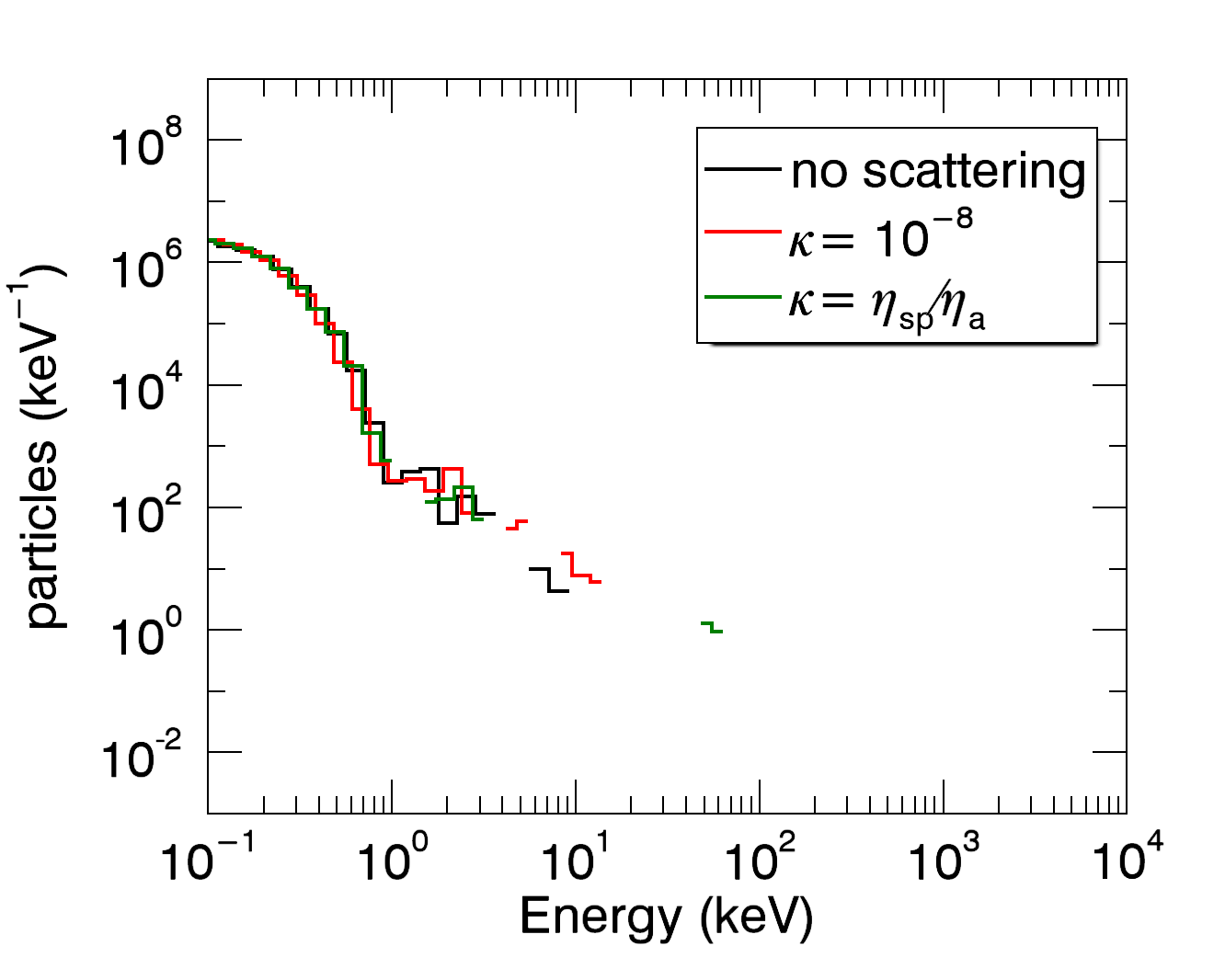}}}
\subfloat[$\theta$: $\eta_a=10^{-5}$]{\label{dist2b}\resizebox{0.33\textwidth}{!}{\includegraphics[clip=true, trim=5 5 20 20]{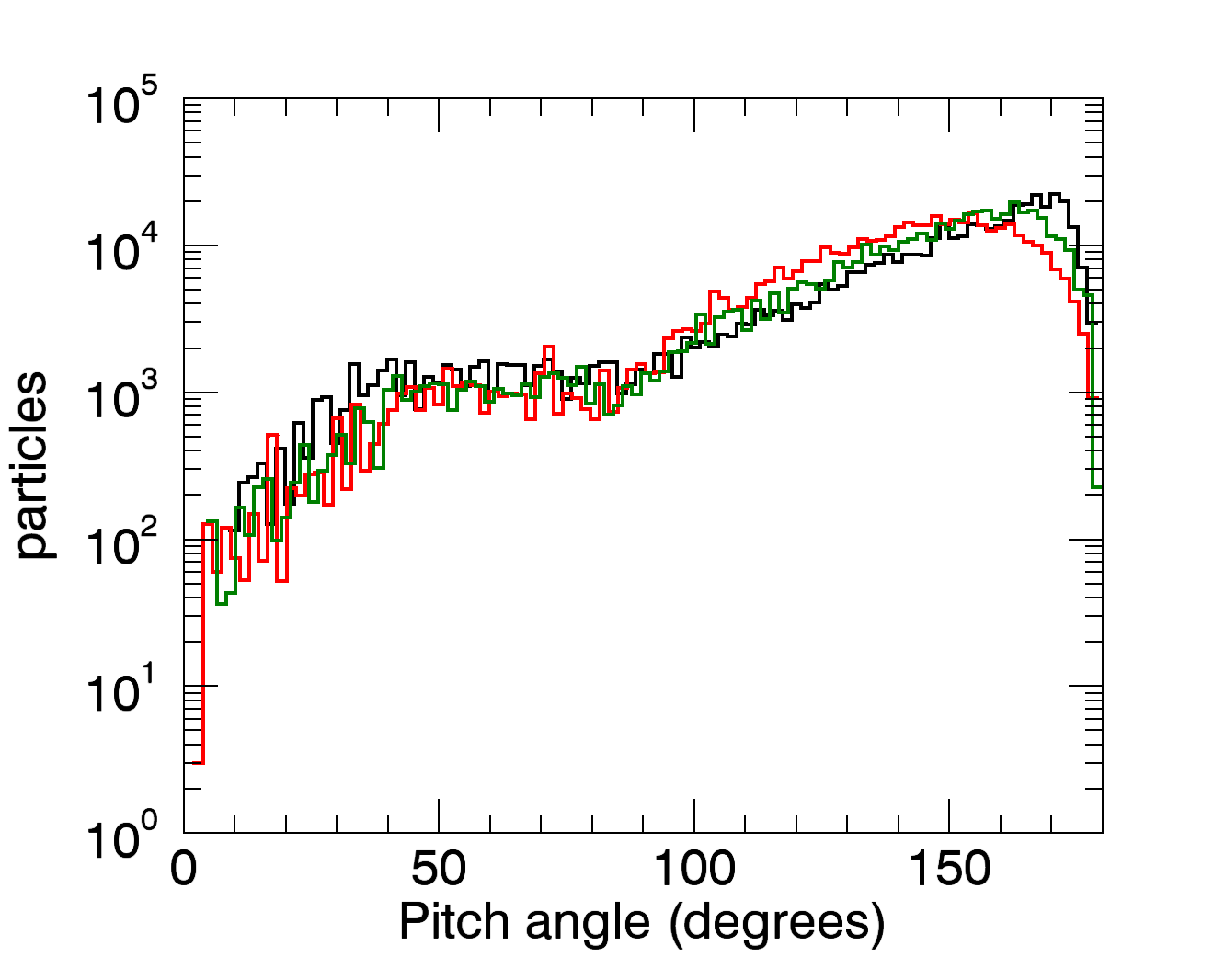}}}
\subfloat[duration: $\eta_a=10^{-5}$]{\label{dist2c}\resizebox{0.33\textwidth}{!}{\includegraphics[clip=true, trim=5 10 12 20]{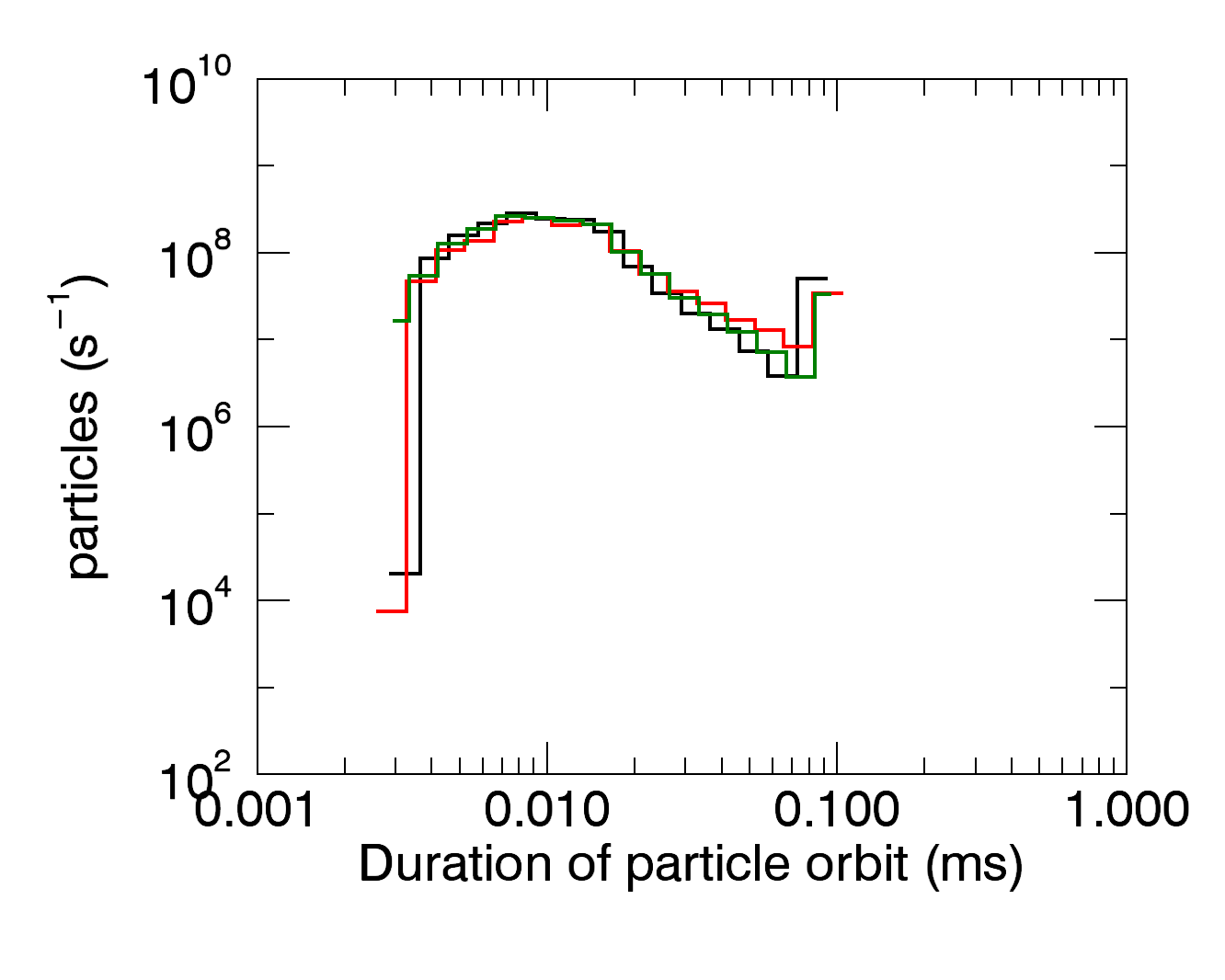}}}\\
\subfloat[KE: $\eta_a=10^{-4}$]{\label{dist2d}\resizebox{0.33\textwidth}{!}{\includegraphics[clip=true, trim=5 5 15 15]{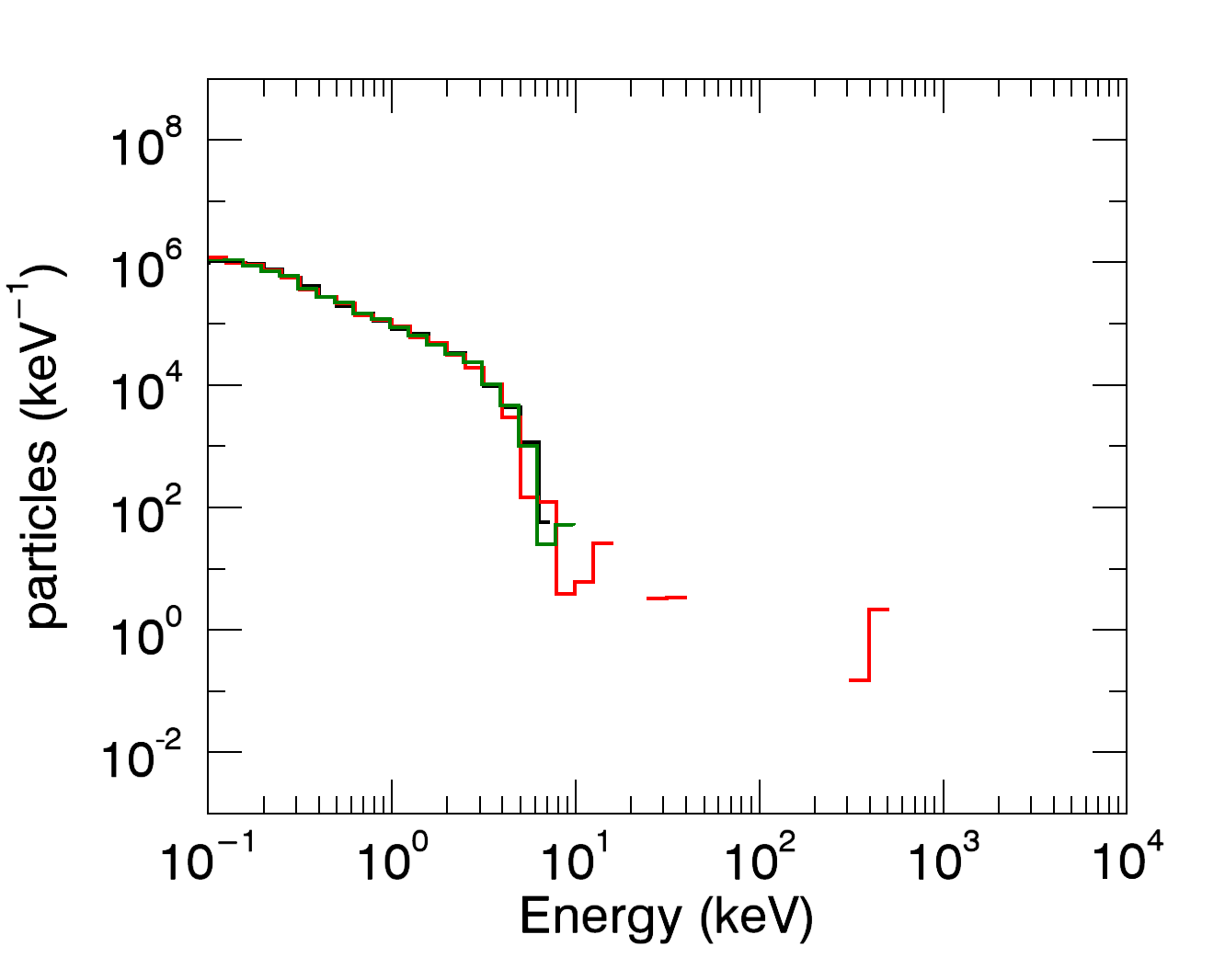}}}
\subfloat[$\theta$: $\eta_a=10^{-4}$]{\label{dist2e}\resizebox{0.33\textwidth}{!}{\includegraphics[clip=true, trim=5 5 20 20]{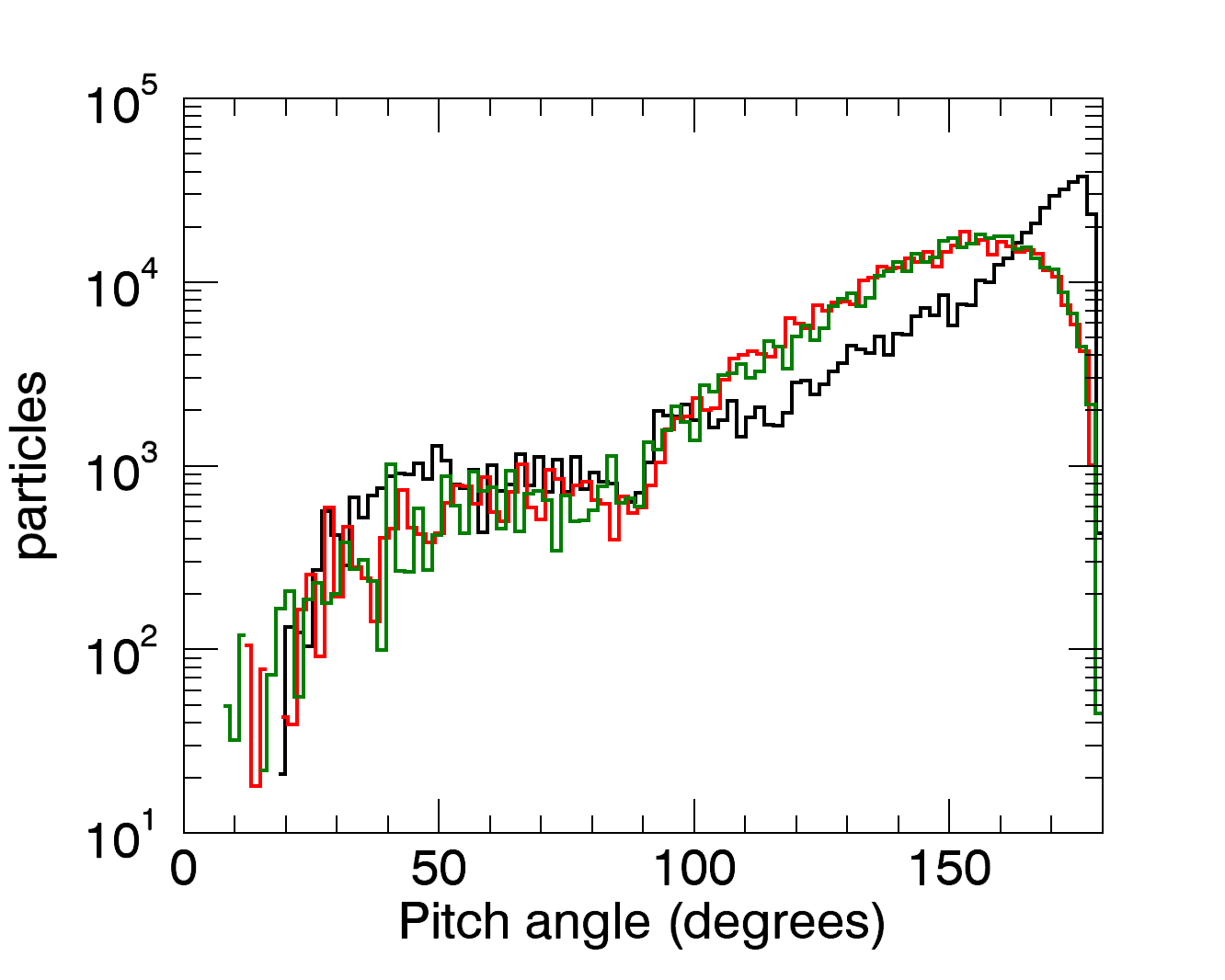}}}
\subfloat[duration: $\eta_a=10^{-4}$]{\label{dist2f}\resizebox{0.33\textwidth}{!}{\includegraphics[clip=true, trim=5 10 12 20]{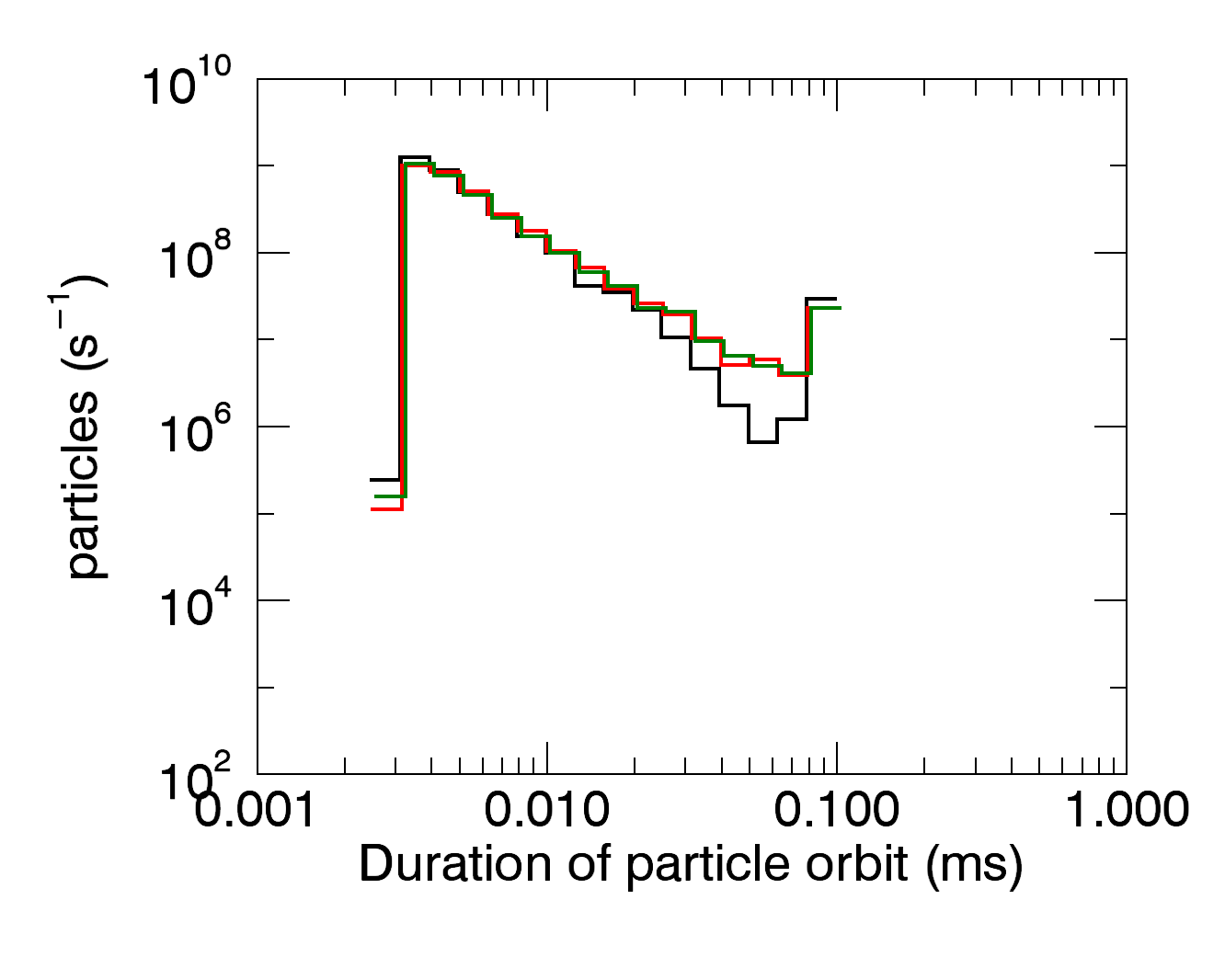}}}\\
\subfloat[KE: $\eta_a=10^{-3}$]{\label{dist2g}\resizebox{0.33\textwidth}{!}{\includegraphics[clip=true, trim=5 5 15 15]{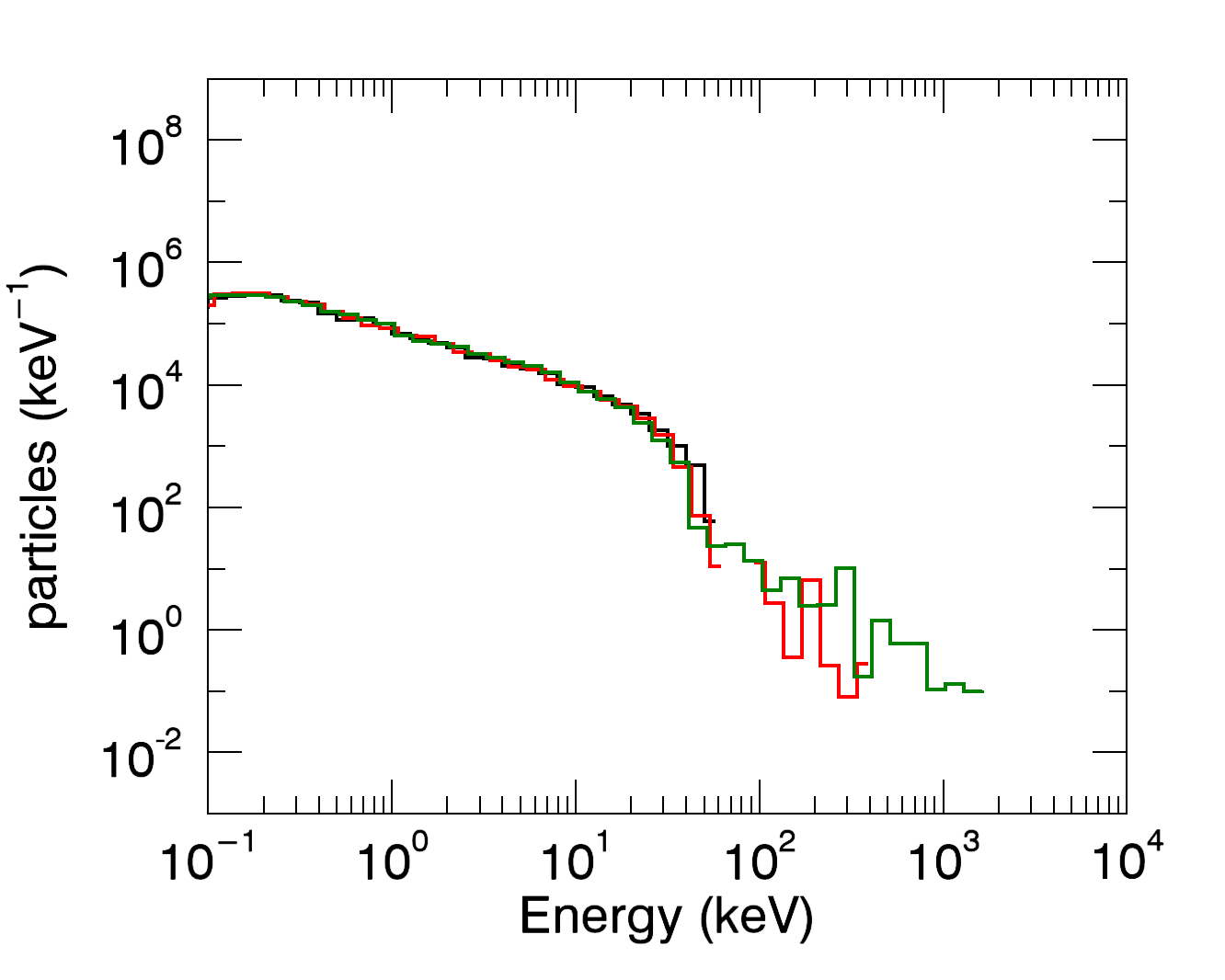}}}
\subfloat[$\theta$: $\eta_a=10^{-3}$]{\label{dist2h}\resizebox{0.33\textwidth}{!}{\includegraphics[clip=true, trim=5 5 20 20]{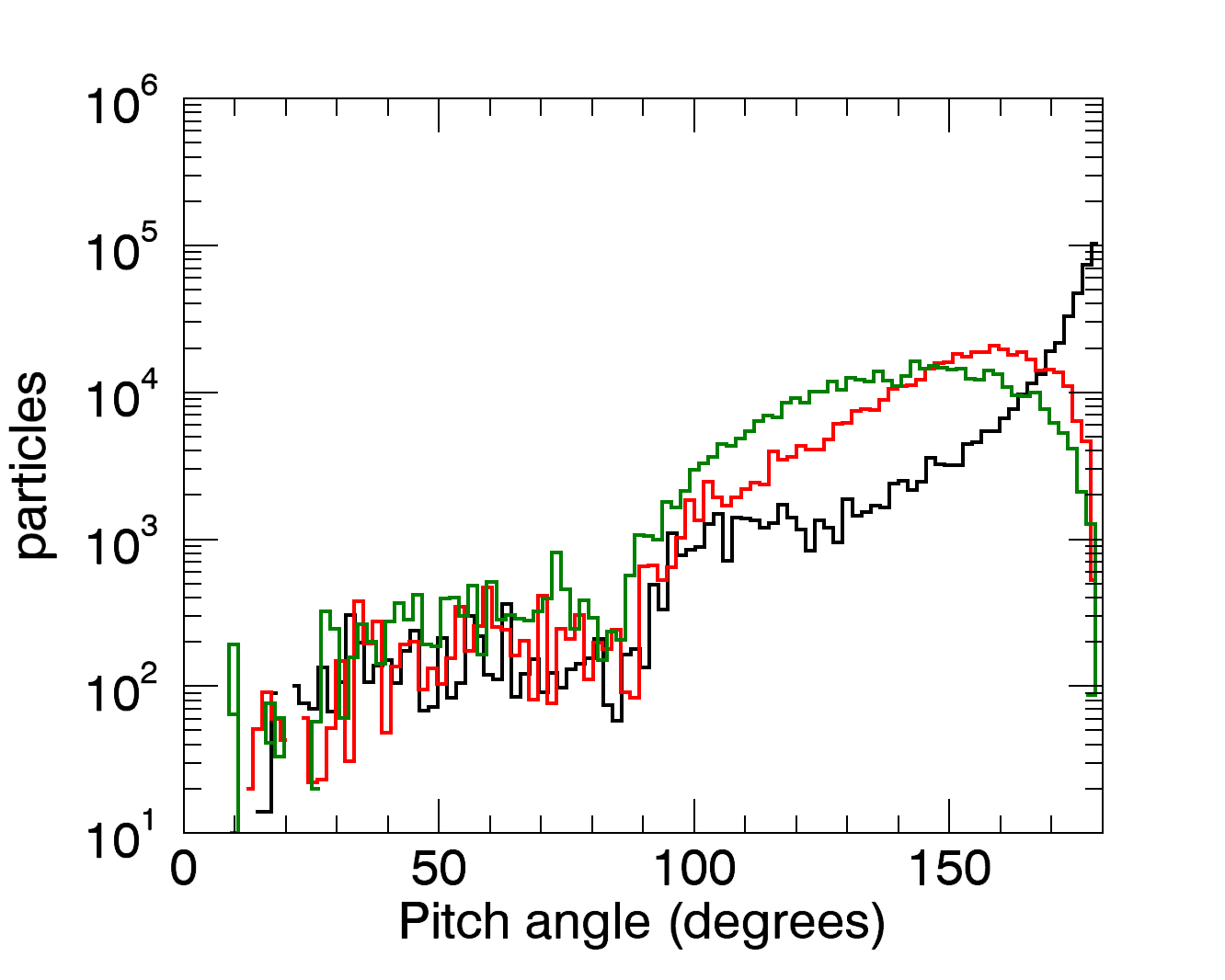}}}
\subfloat[duration: $\eta_a=10^{-3}$]{\label{dist2i}\resizebox{0.33\textwidth}{!}{\includegraphics[clip=true, trim=5 10 12 20]{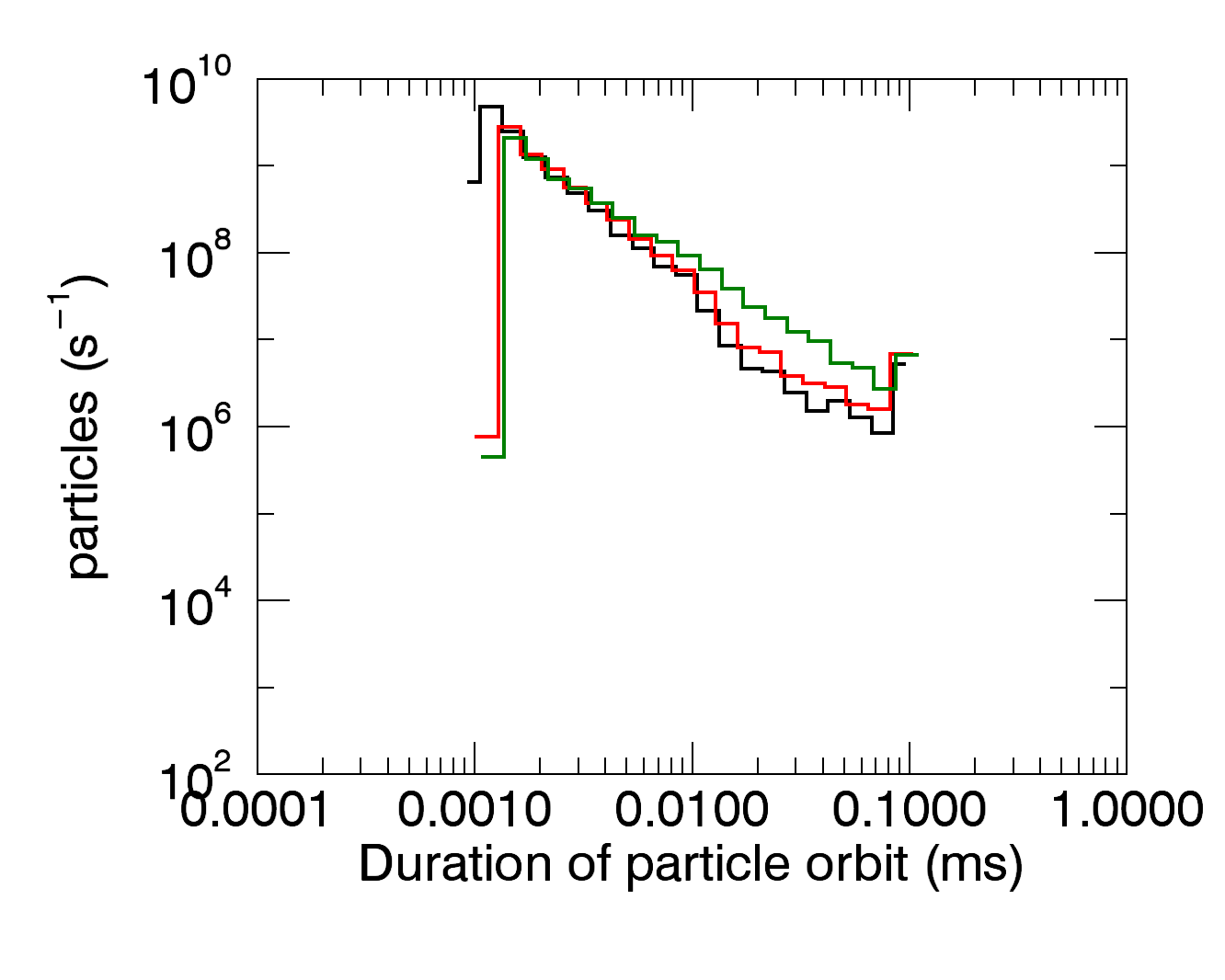}}}
\caption{Particle energy spectra, pitch angle and orbit duration distributions for Case 3 orbits: the colours represent orbits without scattering, with scattering at a rate $\kappa = 10^{-8}$, and $\kappa = \eta_{sp}/\eta_a$ respectively (see legend), based upon initial energies $\in[10,320] \unit{eV}$, initial pitch angles $\in[10,170]\dg$ and originating from $x,y \in [-1,1]\unit m, z \in [1,100]\unit m$. Each row uses different MHD resistivity: top row (Figures~\protect\subref{dist2a}-\protect\subref{dist2c}) uses $\eta_a=10^{-5}$, middle row (Figures~\protect\subref{dist2d}-\protect\subref{dist2f}) uses $\eta_a=10^{-4}$ , and bottom row (Figures~\protect\subref{dist2g}-\protect\subref{dist2i}) uses $\eta_a=10^{-3}$.}
\label{dist2}
\end{figure*}

The results from Cases 1 and 2 (Sections \ref{centred-particles} and \ref{elongated-particles}) imply that the initial vertical position of orbits results in variations in energy, pitch angle, and duration distributions in different scattering models. To examine this further, Case 3 considers initial orbit positions solely near the lower null, with $x,y \in [-1,1]\unit m$ and $z \in [1,100]\unit m$. The results of these orbit calculations are shown in Figure \ref{dist2} using different levels of MHD resistivity.

Firstly (and as expected), as the resistivity is increased in Figure~\ref{dist2}, the orbits become progressively more energised in all scattering model cases. Comparing Figures~\ref{dist2a},~\ref{dist2d}, and \ref{dist2g}, for example, shows the gradual lengthening of a high energy power law component of the energy spectra, with progressively more orbits achieving the highest energies. Furthermore, the pitch angle distributions seen in Figures~\ref{dist2b},~\ref{dist2e} and \ref{dist2h} reveal that increasing resistivity generally leads to more asymmetric pitch angle distributions, again regardless of scattering model. Finally, as one might also expect, resistivity appears to reduce the particle orbit durations, with Figures~\ref{dist2b},~\ref{dist2e} and \ref{dist2h} showing a shift in the peak of the distributions towards shorter orbit durations as resistivity increases, along with a reduction in the number of orbits which reach the maximum lifetime of $0.1$\unit{ms}. This is to be expected, as greater resistivity leads to faster acceleration, meaning that orbits are more likely to rapidly reach the edge of the computational domain than in cases where the resistivity is reduced. 
\begin{table*}
\caption{Percentage of particle orbits per energy band in $\eta_a = 10^{-3}$ MHD simulation.}
\label{energy-table}
\centering
\begin{tabular}{l c c c c c}
\hline\hline
Initial conditions & Scattering model & $5-30\unit{keV}$ & $30-100\unit{keV}$ & $>100\unit{keV}$ & peak energy~(keV) \\
\hline
$z \in [1,299]\unit m$ & no scattering & 28.7 & 34.3 & 12.1 & 258\\
$z \in [1,299]\unit m$ & $\kappa = \eta_{sp}/\eta_a$ & 27.8 & 33.5 & 7.6 & 3809\\
$z \in [1,100]\unit m$ & no scattering & 34.5 & 4.7 & 0 & 67\\
$z \in [1,100]\unit m$ & $\kappa = \eta_{sp}/\eta_a$ & 34.1 & 2.5 & 0.4 & 2973\\
$z \in [200,299]\unit m$ & no scattering & 19.9 & 58.0 & 12.8 & 263\\
$z \in [200,299]\unit m$ & $\kappa = \eta_{sp}/\eta_a$ & 19.8 & 59.8 & 12.0 & 4122
\end{tabular}
\end{table*}

Focussing temporarily on the pitch angles, we note that the scattering appears to play the greatest role in the final pitch angle distributions in Case 3 (minor differences between scattering models are also present in the lifetime distributions, but the scattering models do not appear to play a major role in the energy spectra). For context, in the $\eta_a = 10^{-5}$ case, the percentage of orbits with final pitch angle $\theta > 170\dg$ is approximately $14\%$ in the absence of scattering, and to $4.6\%$ and $6.8\%$ for scattering with $\kappa = 10^{-8}$ and $\kappa = \eta_{sp}/\eta_a$ models respectively. Increasing the resistivity by a factor of ten, we find that the percentage where $\theta > 170\dg$ is $35.6\%$ in the absence of scattering (compared to more than $65\%$ for the $\eta_a = 10^{-3}$ simulation), and $6.6\%$ and $7.7\%$ respectively for the $\kappa = 10^{-8}$ and $\kappa = \eta_{sp}/\eta_a$ simulations. We estimate $89.6\%$ of the orbits achieve $\theta > 145\dg$ in the unscattered model, $68.4\%$ if $\kappa = 10^{-8}$ and $48.1\%$ for $\kappa = \eta_{sp}/\eta_a$ when $\eta_a = 10^{-3}$. We again believe that this is linked to orbits being able to accelerate over shorter distances than in previous cases by the parallel electric field. The peak of the pitch angle distributions appears to shift further from $180\dg$ with scattering model, but only in the case of largest resistivity: in Figure~\ref{dist2h} the $\kappa = \eta_{sp}/\eta_a$ distribution peaks at $\sim140\dg$, while the $\kappa = 10^{-8}$ distribution peaks closer to $160\dg$.

Reducing the resistivity in Case 3 appears to make the different scattering models converge, with all yielding similar distributions in the top two rows of Figure~\ref{dist2}. The effects of weaker parallel electric field appear to play more of a role, while the choice of scattering model appears to make less difference to the final distribution.  In the case of the pitch angle distribution, the population of particles with final pitch angle $\theta > 170\dg$ drops to $14\%$ in the absence of scattering, and to $4.6\%$ and $6.8\%$ for scattering with $\kappa = 10^{-8}$ and $\kappa = \eta_{sp}/\eta_a$ models respectively. In the case where $\kappa = \eta_{sp}/\eta_a$, the lowered resistivity decreases the scattering rate sufficiently to slightly increase the population of particle orbits with higher parallel velocities. The energy spectra show minimal differences for simulations with different scattering rates at the same resistivity, apart from a few orbits achieving energies above 100\unit{keV} in the $\kappa = 10^{-8}$ case (red line in Figure \ref{dist2d}). 

There are also small differences in the orbit duration distributions (see Figures~\ref{dist2c}, \ref{dist2f} and~\ref{dist2i}) with the most orbits in the range of $0.02\unit{s}$ to $0.09\unit{s}$ occurring for simulations including scattering. However, for the unscattered model more orbits achieve lifetimes $>0.1\unit{s}$, resulting in all simulations having approximately $77\%$ of the orbit durations $>0.01\unit{s}$.

\subsection{Case 4: Initial positions near top of separator}\label{extremity-particles2}

\begin{figure*}
\centering
\subfloat[KE: $\eta_a=10^{-5}$]{\label{dist4a}\resizebox{0.33\textwidth}{!}{\includegraphics[clip=true, trim=5 5 15 15]{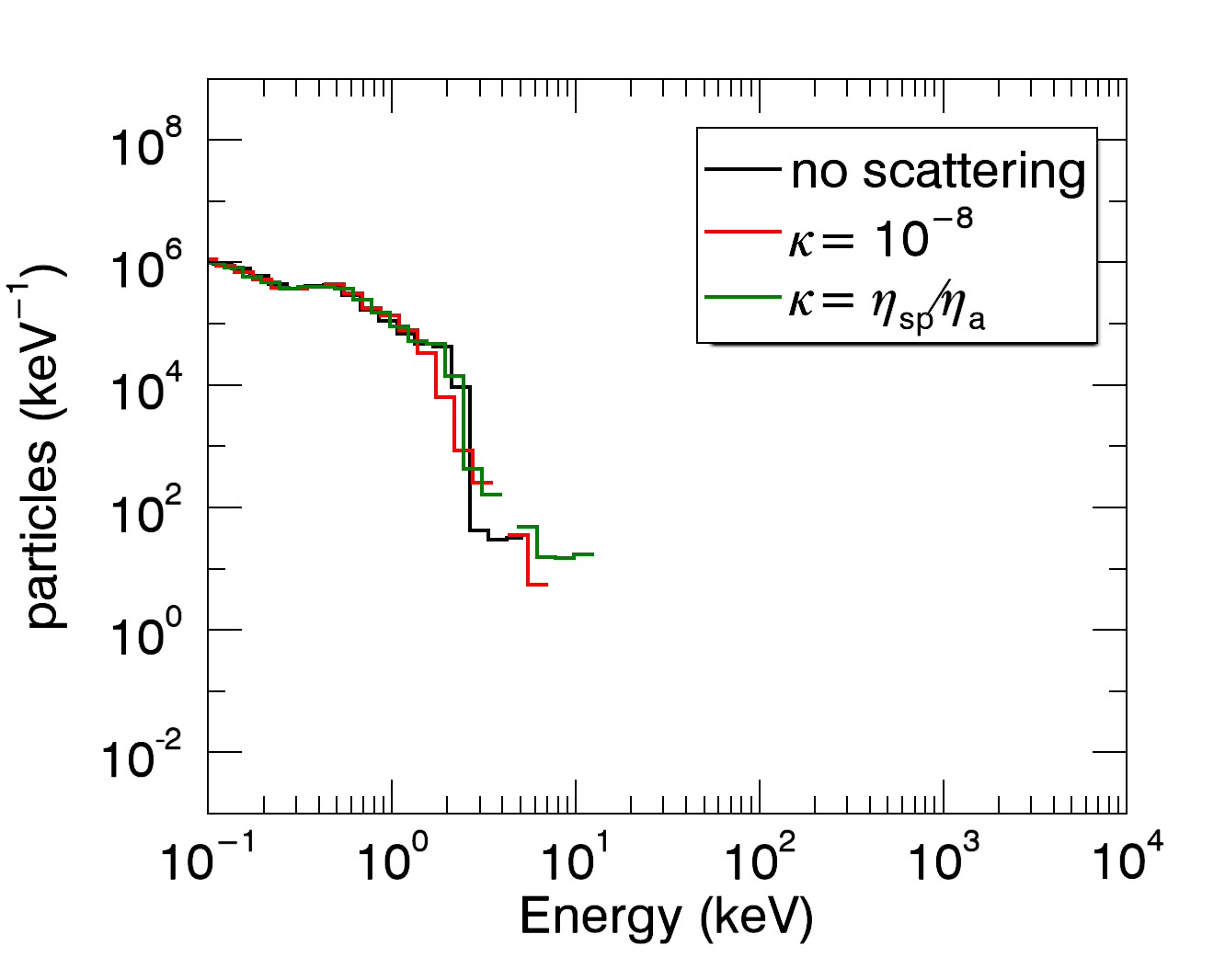}}}
\subfloat[$\theta$: $\eta_a=10^{-5}$]{\label{dist4b}\resizebox{0.33\textwidth}{!}{\includegraphics[clip=true, trim=5 5 20 20]{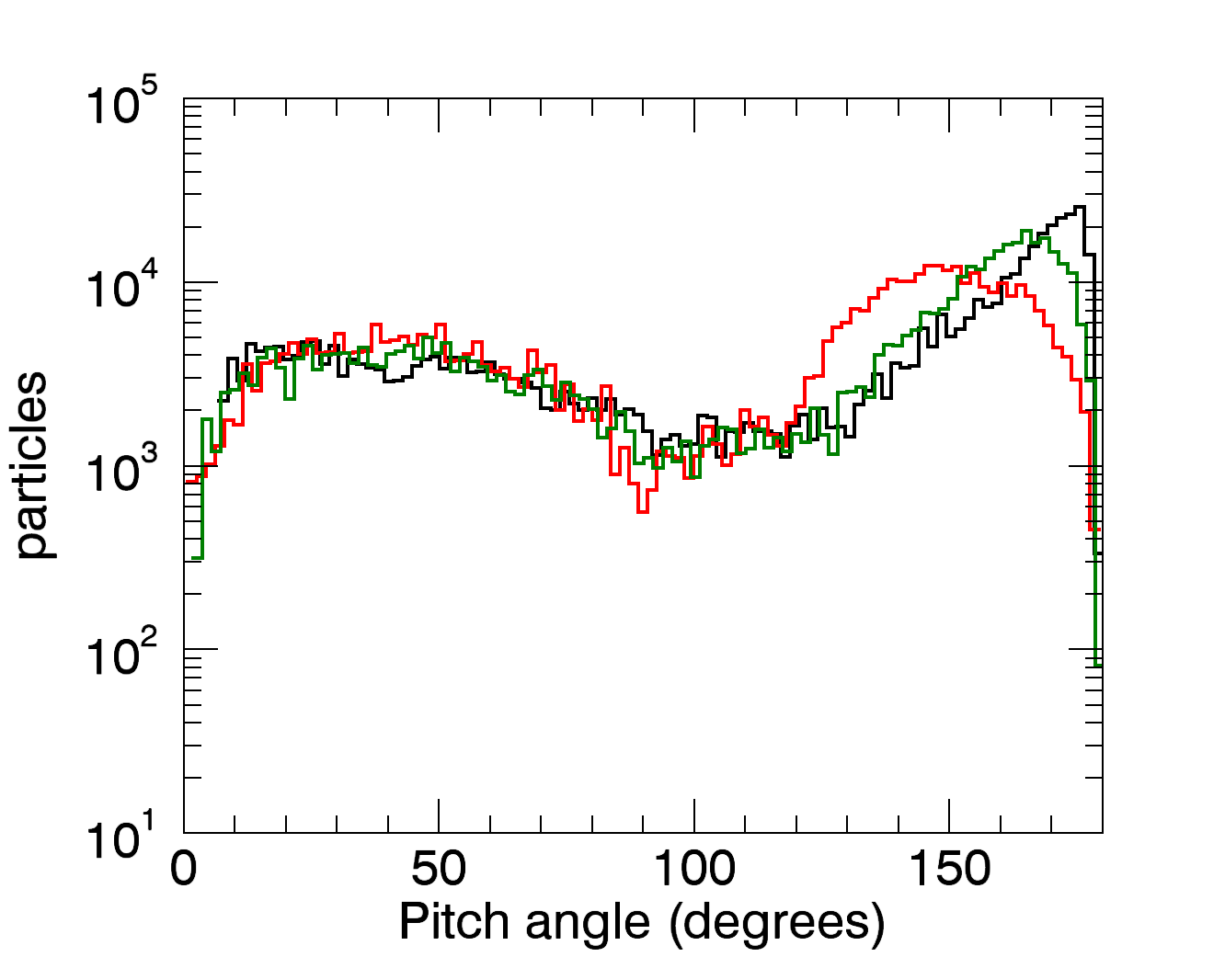}}}
\subfloat[duration: $\eta_a=10^{-5}$]{\label{dist4c}\resizebox{0.33\textwidth}{!}{\includegraphics[clip=true, trim=5 10 12 20]{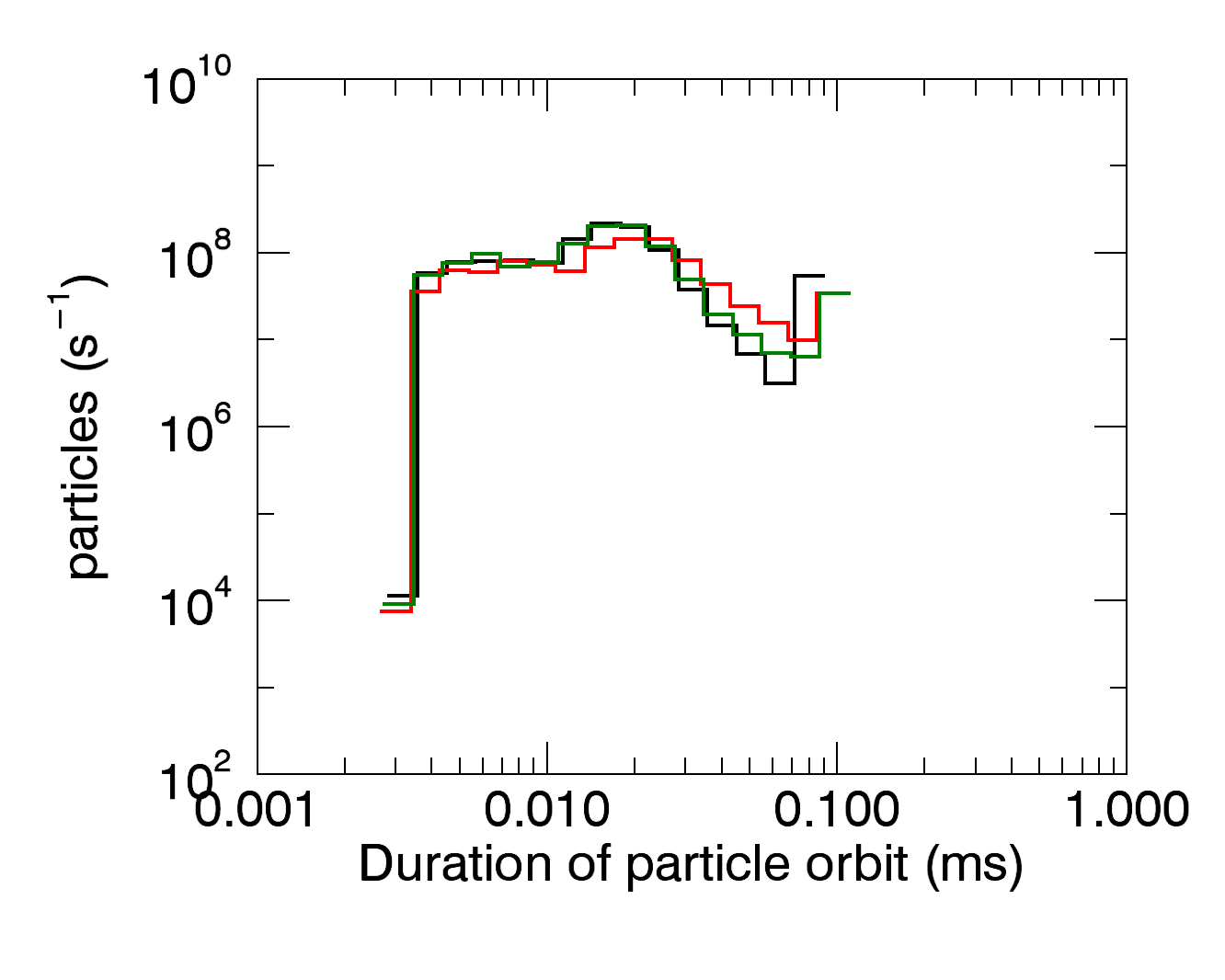}}}\\
\subfloat[KE: $\eta_a=10^{-4}$]{\label{dist4d}\resizebox{0.33\textwidth}{!}{\includegraphics[clip=true, trim=5 5 15 15]{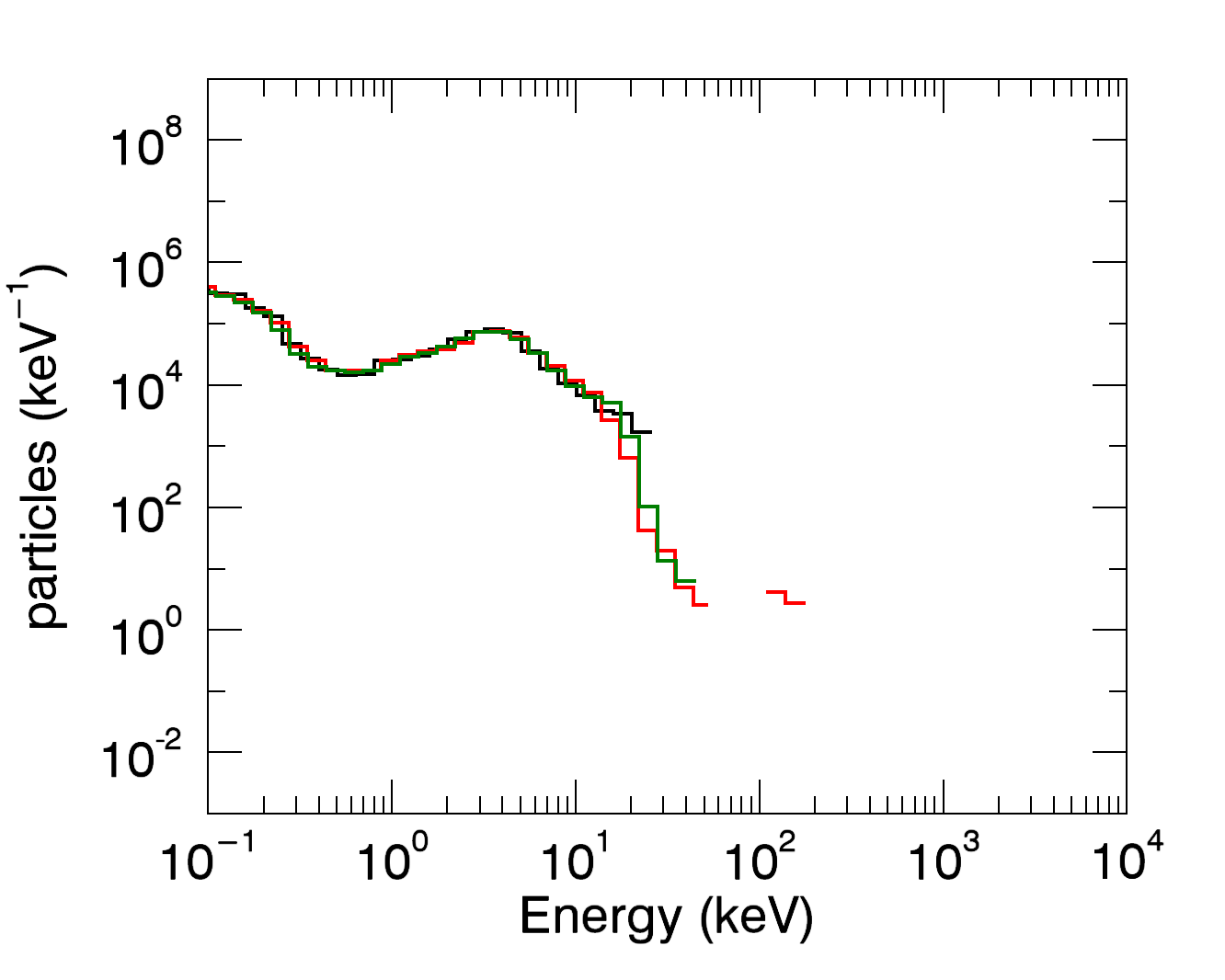}}}
\subfloat[$\theta$: $\eta_a=10^{-4}$]{\label{dist4e}\resizebox{0.33\textwidth}{!}{\includegraphics[clip=true, trim=5 5 20 20]{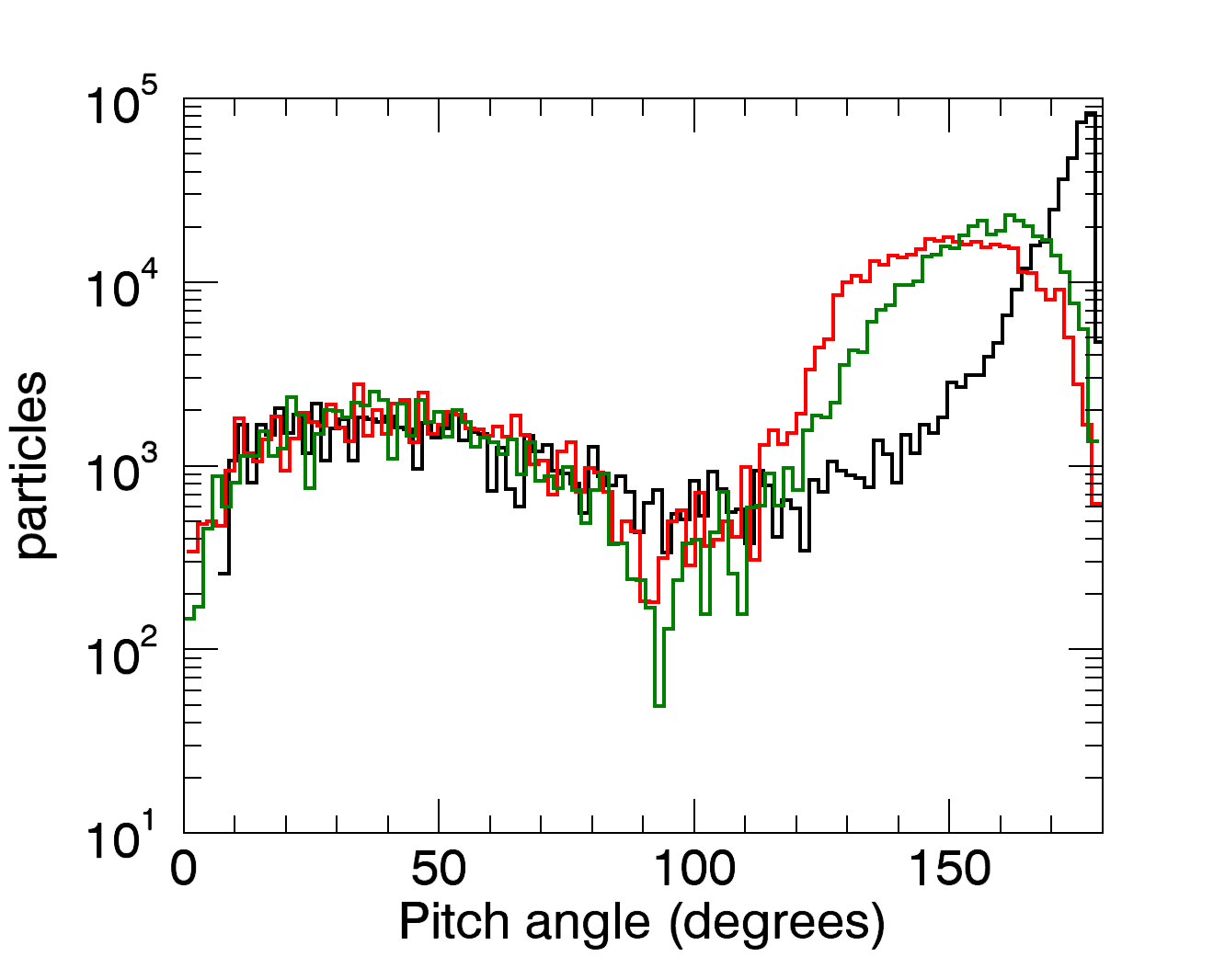}}}
\subfloat[duration: $\eta_a=10^{-4}$]{\label{dist4f}\resizebox{0.33\textwidth}{!}{\includegraphics[clip=true, trim=5 10 12 20]{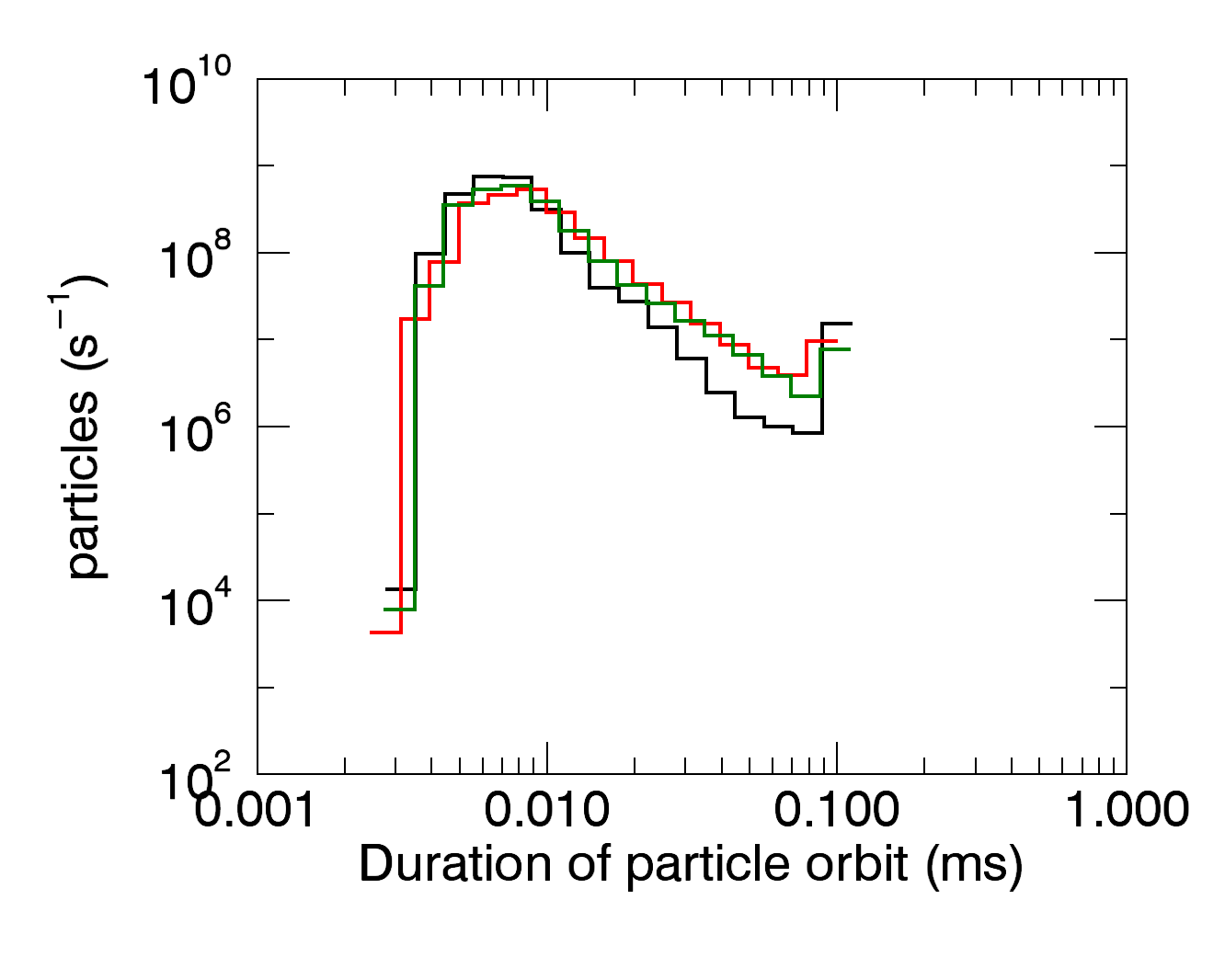}}}\\
\subfloat[KE: $\eta_a=10^{-3}$]{\label{dist4g}\resizebox{0.33\textwidth}{!}{\includegraphics[clip=true, trim=5 5 15 15]{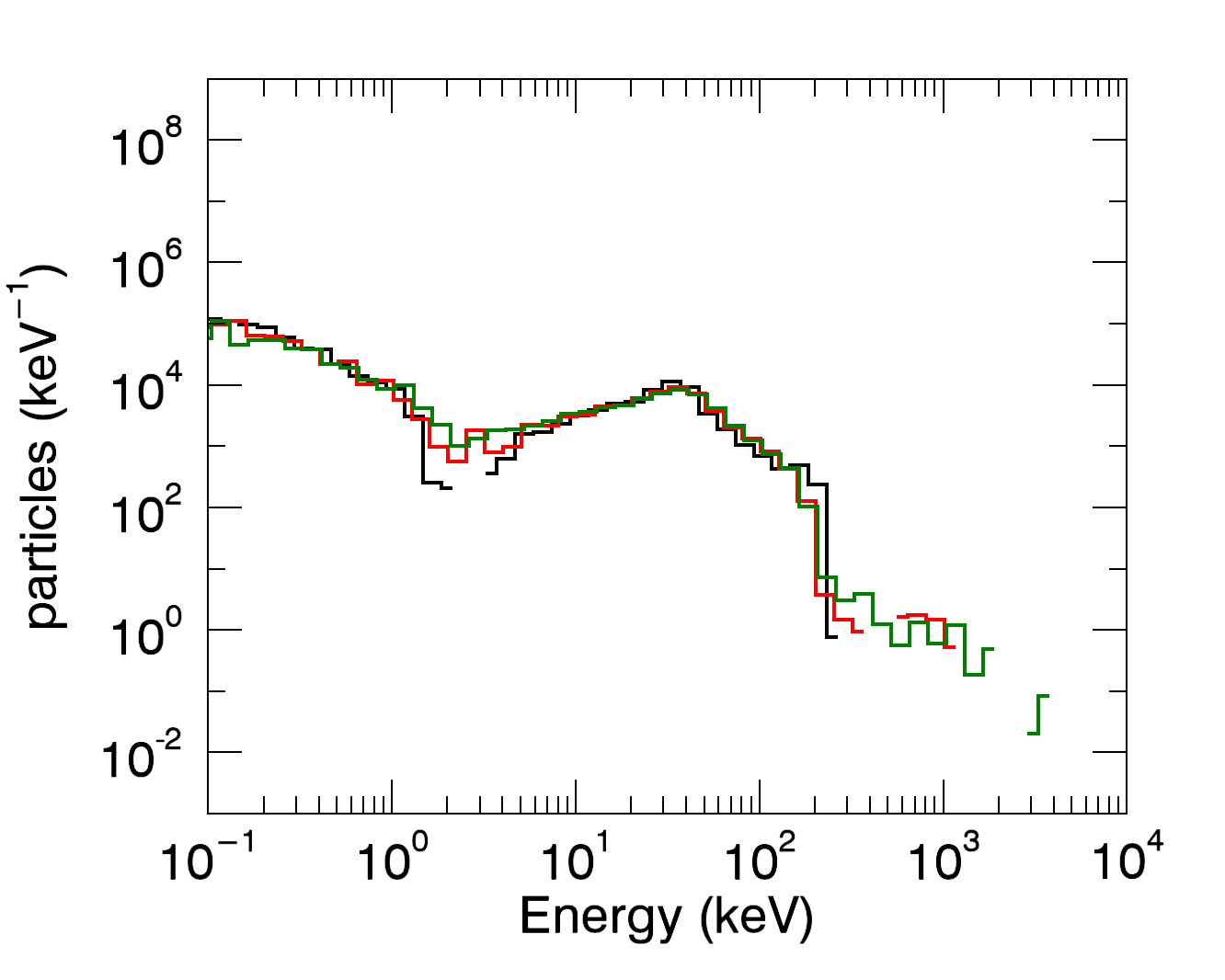}}}
\subfloat[$\theta$: $\eta_a=10^{-3}$]{\label{dist4h}\resizebox{0.33\textwidth}{!}{\includegraphics[clip=true, trim=5 5 20 20]{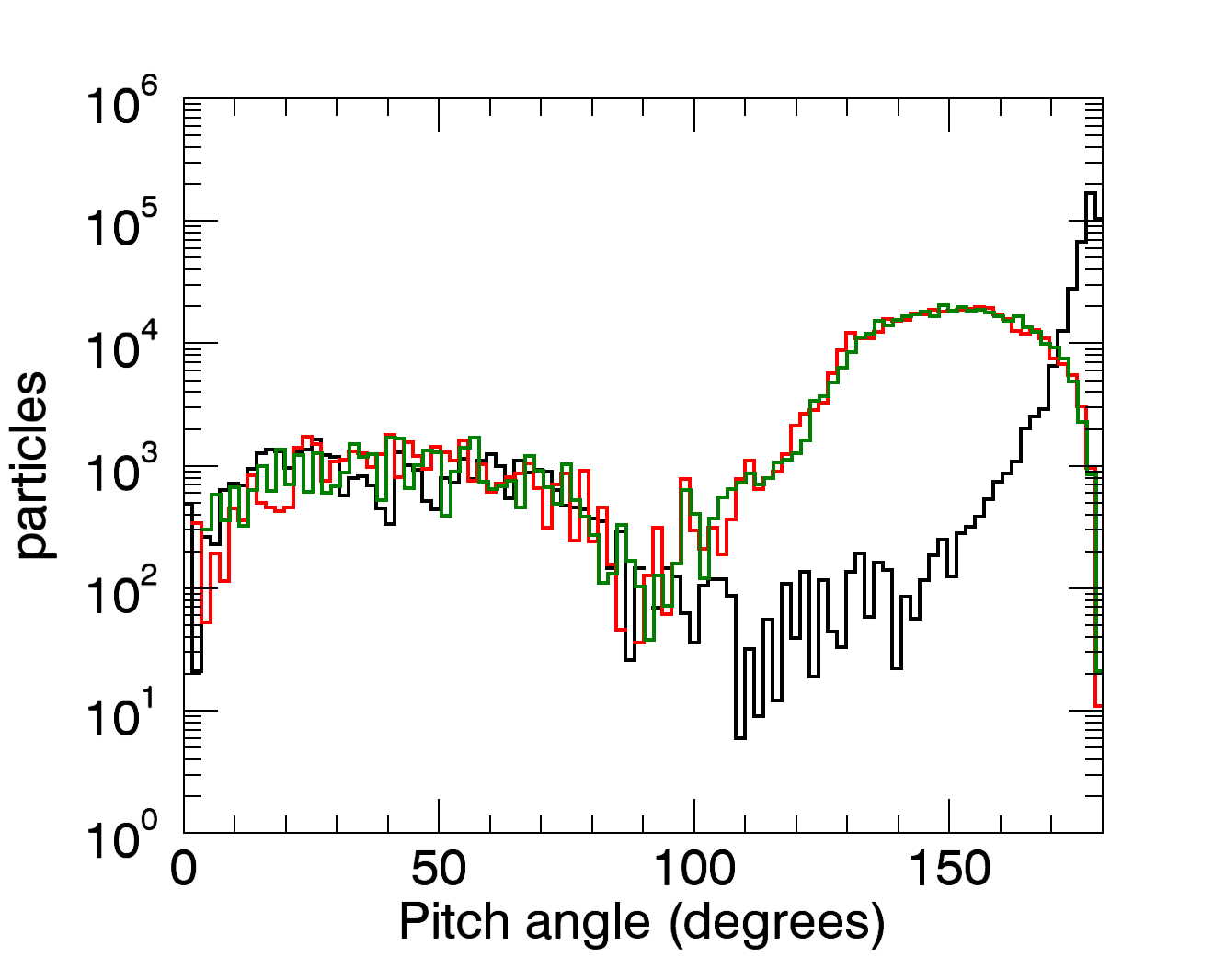}}}
\subfloat[duration: $\eta_a=10^{-3}$]{\label{dist4i}\resizebox{0.33\textwidth}{!}{\includegraphics[clip=true, trim=5 10 12 20]{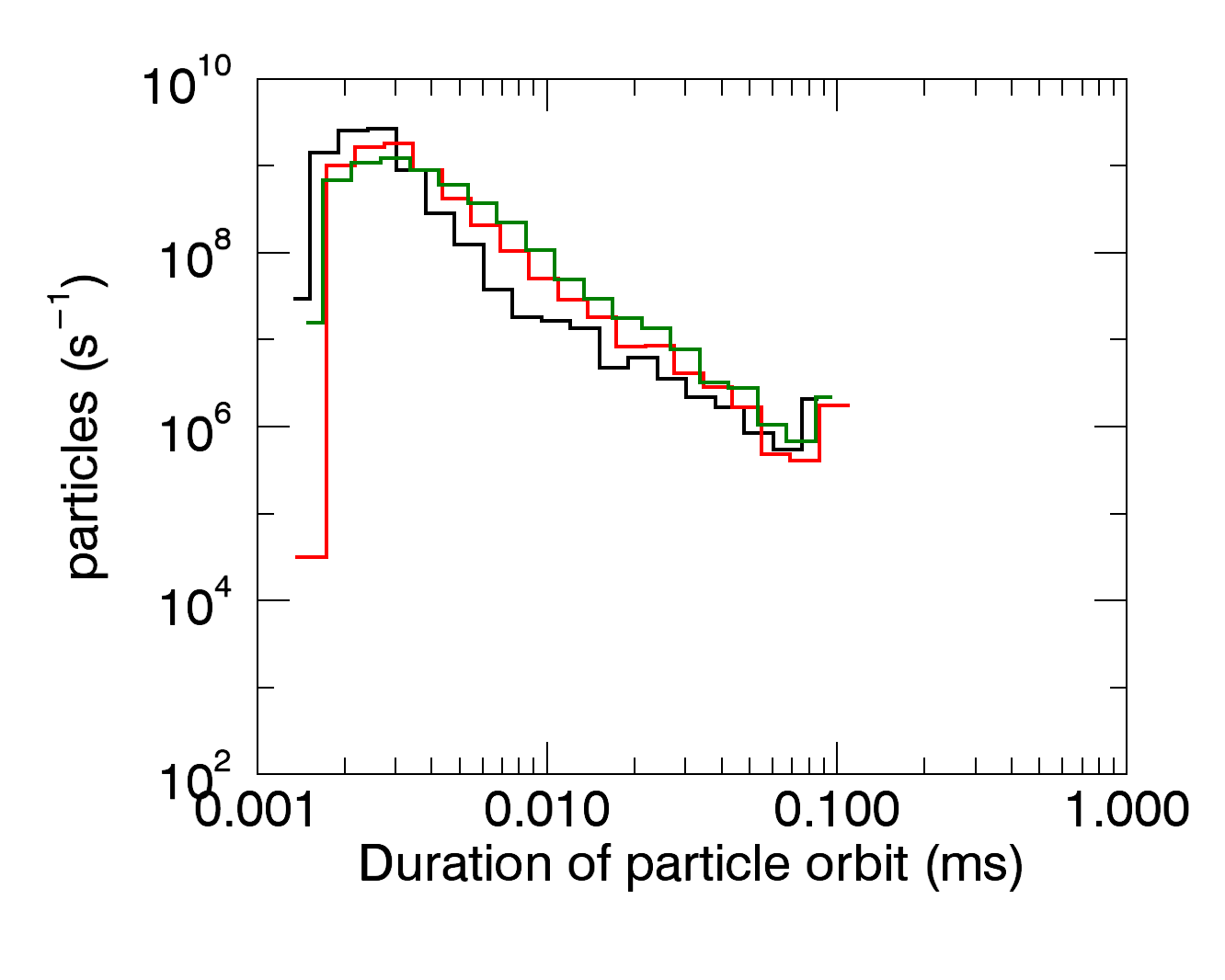}}}
\caption{Particle energy spectra, pitch angle, and orbit duration distributions for Case 4 orbits: the colours represent orbits without scattering, with scattering at a rate $\kappa = 10^{-8}$, and $\kappa = \eta_{sp}/\eta_a$ respectively (see legend), based upon initial energies $\in[10,320] \unit{eV}$, initial pitch angles $\in[10,170]\dg$ and originating from $x,y \in [-1,1]\unit m, z \in [200,299]\unit m$. Each row uses different MHD resistivity: top row (Figures~\protect\subref{dist4a}-\protect\subref{dist4c}) uses $\eta_a=10^{-5}$, middle row (Figures~\protect\subref{dist4d}-\protect\subref{dist4f}) uses $\eta_a=10^{-4}$ and bottom row (Figures~\protect\subref{dist4g}-\protect\subref{dist4i}) uses $\eta_a=10^{-3}$.}
\label{dist4}
\end{figure*}

Our final Case concerns orbit behaviour starting near the top of the separator. Figure~\ref{dist4} displays energy spectra, pitch angle and orbit duration distributions for particles initialised near the upper null, with $x,y \in [-1,1]\unit m, z \in [200,299]\unit m$ using three different MHD resistivities. 

Qualitatively, the energy spectra differ significantly from those in Figure \ref{dist2}, particularly in Figures \ref{dist4d} and~\ref{dist4g} with peaks at approximately 4 and 40\unit{keV} respectively. Table~\ref{energy-table} suggests that more particle orbits achieve higher energies in the fields from the $\eta_a = 10^{-3}$ MHD simulation, with approximately 60\% reaching $30-100\unit{keV}$ compared to approximately 3\% reaching these energies if the initial positions are $z~\in~[1,100]\unit{m}$. Furthermore, higher maximum energies are achieved, up to $4\unit{MeV}$ when $\kappa = \eta_{sp}/\eta_a$ compared to only about $2.9\unit{MeV}$ for the equivalent Case 3 conditions and scattering rate.

Differences in pitch angle distributions between Cases 3 and 4 can be observed by comparing Figures~\ref{dist4b},~\ref{dist4e} and~\ref{dist4h} with ~\ref{dist2b},~\ref{dist2e} and~\ref{dist2h}. Again, when $\eta_a = 10^{-3}$, 90.9\% of particles have pitch angle $\theta > 145\dg$ without scattering, $62.2\%$ if $\kappa = 10^{-8}$ and $63.1\%$ if $\kappa = \eta_{sp}/\eta_a$. These proportions are similar to those obtained in Case 3, however the differences between the actual pitch angle distributions are greater, particularly between scattered and unscattered simulations (for example comparing Figures \ref{dist2h} and \ref{dist4h}). The difference in the pitch angle distributions decreases between the unscattered and $\kappa = \eta_{sp}/\eta_a$ simulation at lower resistivities. Case 3 pitch angle distributions display much more asymmetry than Case 4, with many more orbits having pitch angles $<90\dg$ when originating from the top null (Case 4) than the bottom null (Case 3). This may be because orbits near the top null have a higher chance of scattering than those which originate near the lower null; Case 3 orbits are closer to leaving the computational box to begin with, and would be less likely to have significant pitch angle modification before they do so.

Finally, considering the orbit durations, smaller proportions of particles exit the box after $0.01\unit{ms}$ in Case 4 compared to Case 3; $5.1\%$ doing so without scattering, $8.7\%$ with $\kappa = 10^{-8}$ and $11.9\%$ with $\kappa = \eta_{sp}/\eta_a$ (Figure \ref{dist4i} compared to \ref{dist2i}). 

In direct contrast to the Case 3 findings (Figure \ref{dist2}), decreasing the MHD resistivity actually yields increased differences in the pitch angle and orbit duration distributions between the different scattering models. In Case 3, the strongest resistivity used (Figure~\ref{dist2h}) showed the most variation between pitch angle distributions resulting from both scattering models; in Case 4 the weakest resistivity (Figure~\ref{dist4b}) shows the greatest difference between scattering models. In Case 3, with $\eta_a=10^{-3}$, the $\kappa = \eta_{sp}/\eta_a$ distribution peaks at smaller pitch angles than the $\kappa = 10^{-8}$ model distribution; in Case 4 the distributions have switched, such that at $\eta_a=10^{-5}$ we find the $\kappa = 10^{-8}$ distribution peaking at smaller pitch angles than the $\kappa = \eta_{sp}/\eta_a$ model distribution. However, it should also be noted that in Case 4, the $\kappa = 10^{-8}$ results remain relatively unchanged with resistivity (with the red distributions in Figures~\ref{dist4b},~\ref{dist4e} and~\ref{dist4h} appearing indistinguishable). This is unsurprising, given that the scattering rate is independent of resistivity in this model. 

Meanwhile the $\kappa = \eta_{sp}/\eta_a$ model results appear to transit from behaving like the unscattered model results in the weakly resistive case (Figure~\ref{dist4b}) to matching the $\kappa = 10^{-8}$ model results at the strongest resistivity (Figure~\ref{dist4h}). It is unclear why this contradiction between Cases 3 and 4 occurs, but we would anticipate that orbits originating close to the lower null allow for fewer scattering events to occur compared to orbits originating near the top null, implying that scattering may only play a more limited role in Case 3, regardless of scattering rate.

\section{Single scattering model comparison at different MHD resistivities}\label{eta-comparison}
In Section~\ref{many-particle}, we examine the impact of the role of initial orbit positions upon large numbers of orbit calculations using different scattering models and at different resistivities. To clarify some of our findings still further, we now focus on orbit calculations for one single scattering model rate, but across MHD simulations with different anomalous resistivities. 

\begin{figure*}
\centering
\subfloat[KE]{\label{dist5a}\resizebox{0.33\textwidth}{!}{\includegraphics[clip=true, trim=5 5 15 15]{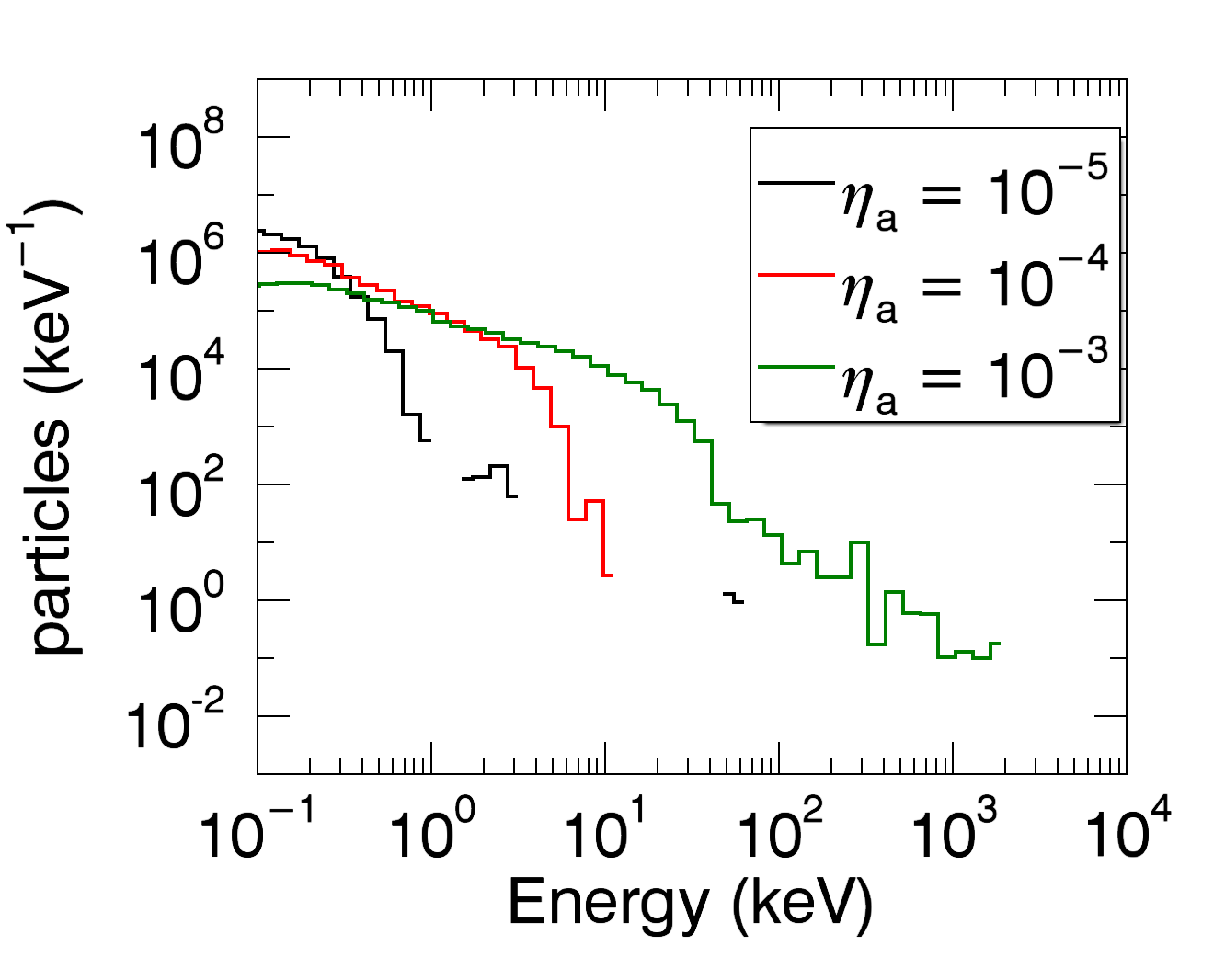}}}
\subfloat[$\theta$]{\label{dist5b}\resizebox{0.33\textwidth}{!}{\includegraphics[clip=true, trim=15 5 20 20]{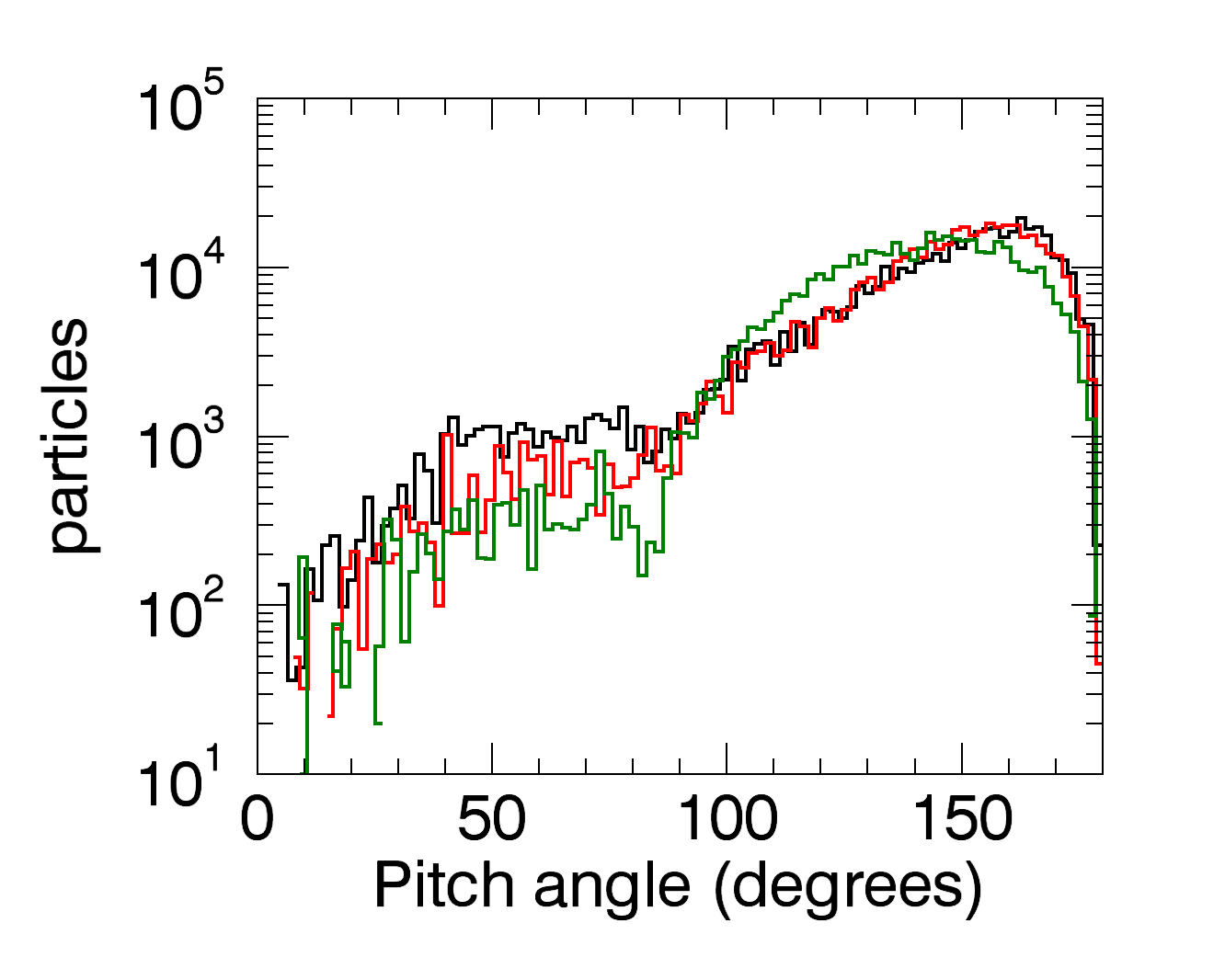}}}
\subfloat[duration]{\label{dist5c}\resizebox{0.33\textwidth}{!}{\includegraphics[clip=true, trim=15 5 20 20]{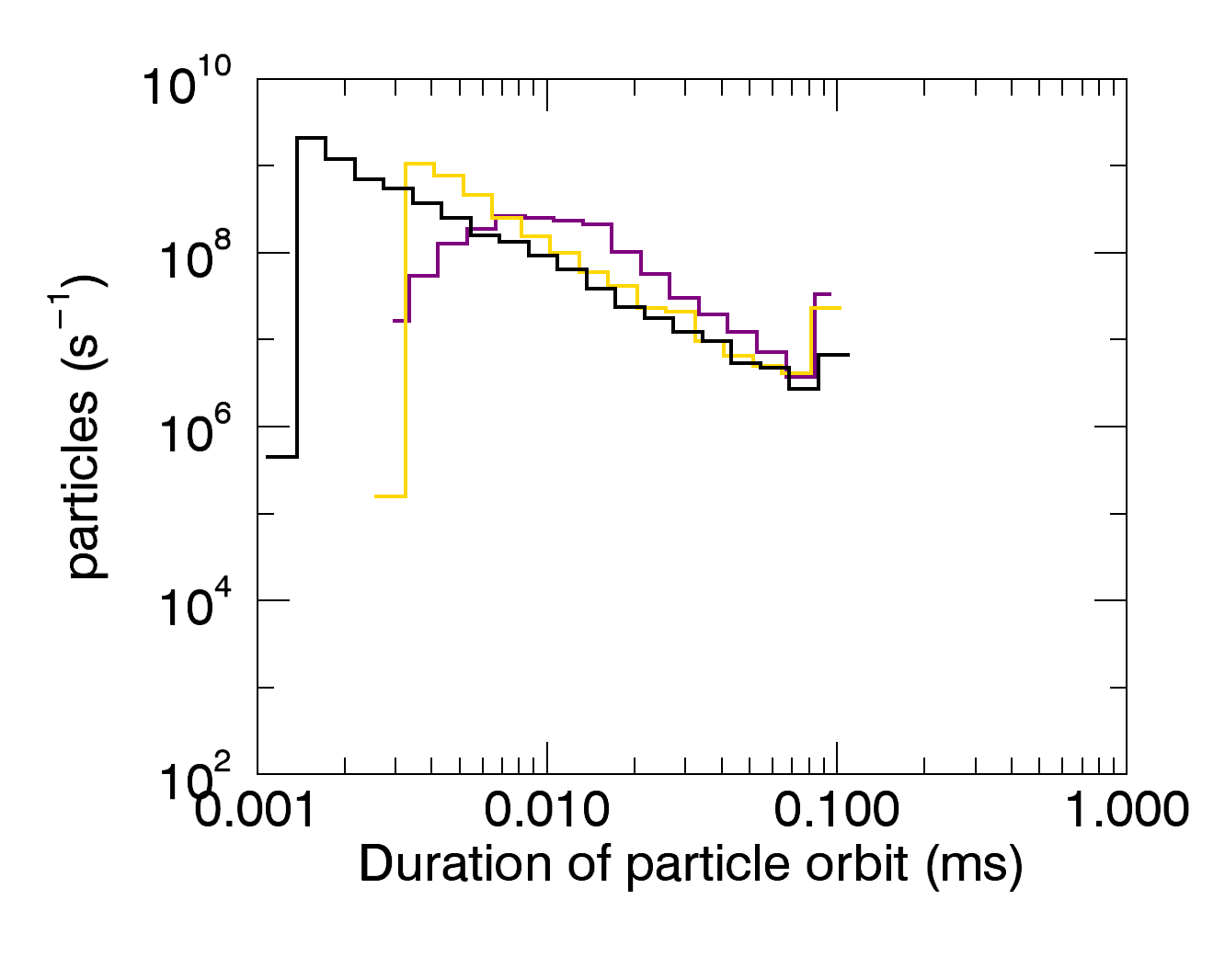}}}
\caption{Particle energy spectra, pitch angle and orbit duration distributions for orbits originating near the lower null ($x,y \in [-1,1]\unit m, z \in [1,100]\unit m$) for only a single scattering model ($\kappa = \eta_{sp}/\eta_a$).}
\label{dist5}
\end{figure*}

\begin{figure*}
\centering
\subfloat[KE]{\label{dist6a}\resizebox{0.33\textwidth}{!}{\includegraphics[clip=true, trim=5 5 15 15]{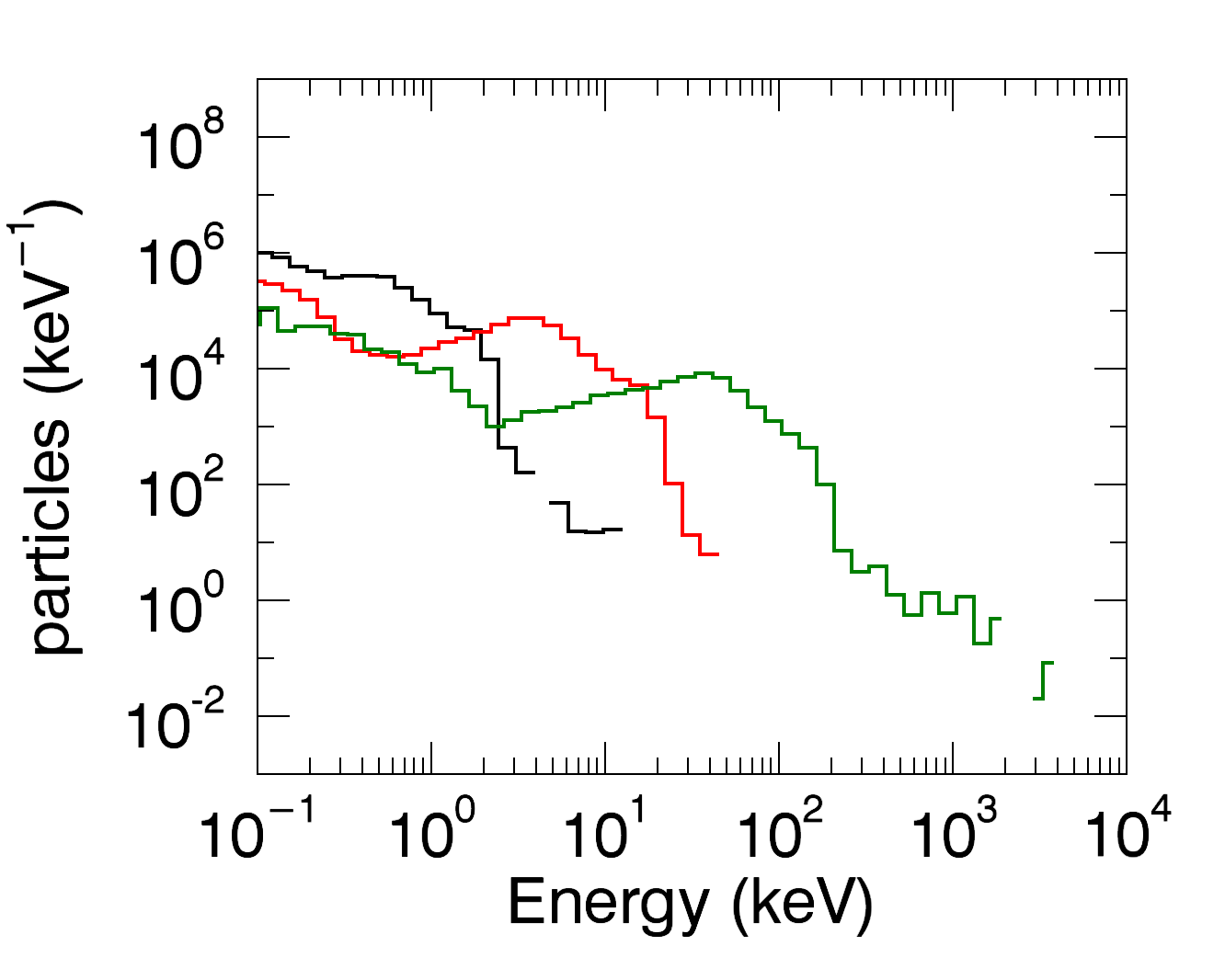}}}
\subfloat[$\theta$]{\label{dist6b}\resizebox{0.33\textwidth}{!}{\includegraphics[clip=true, trim=15 5 20 20]{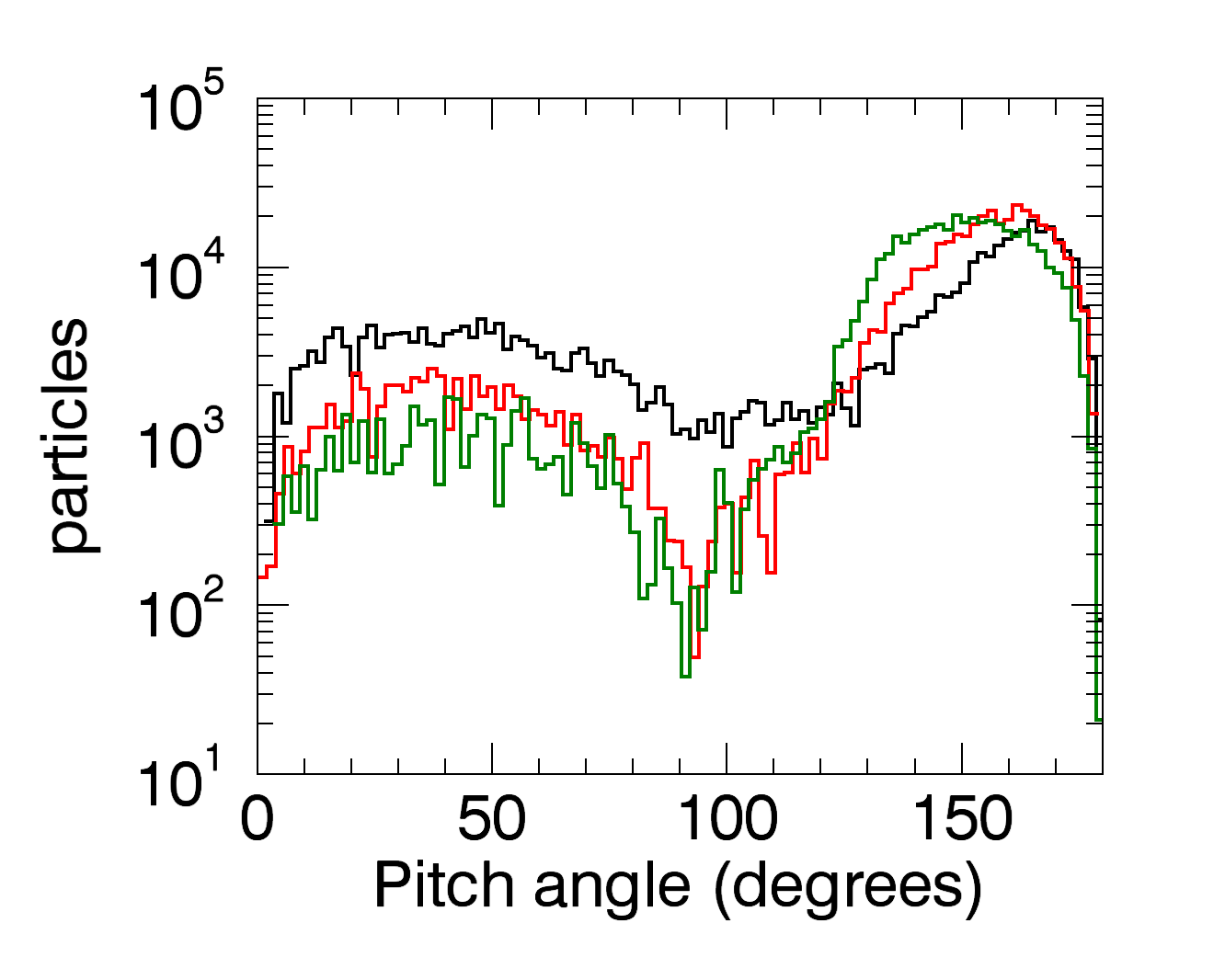}}}
\subfloat[duration]{\label{dist6c}\resizebox{0.33\textwidth}{!}{\includegraphics[clip=true, trim=15 5 20 20]{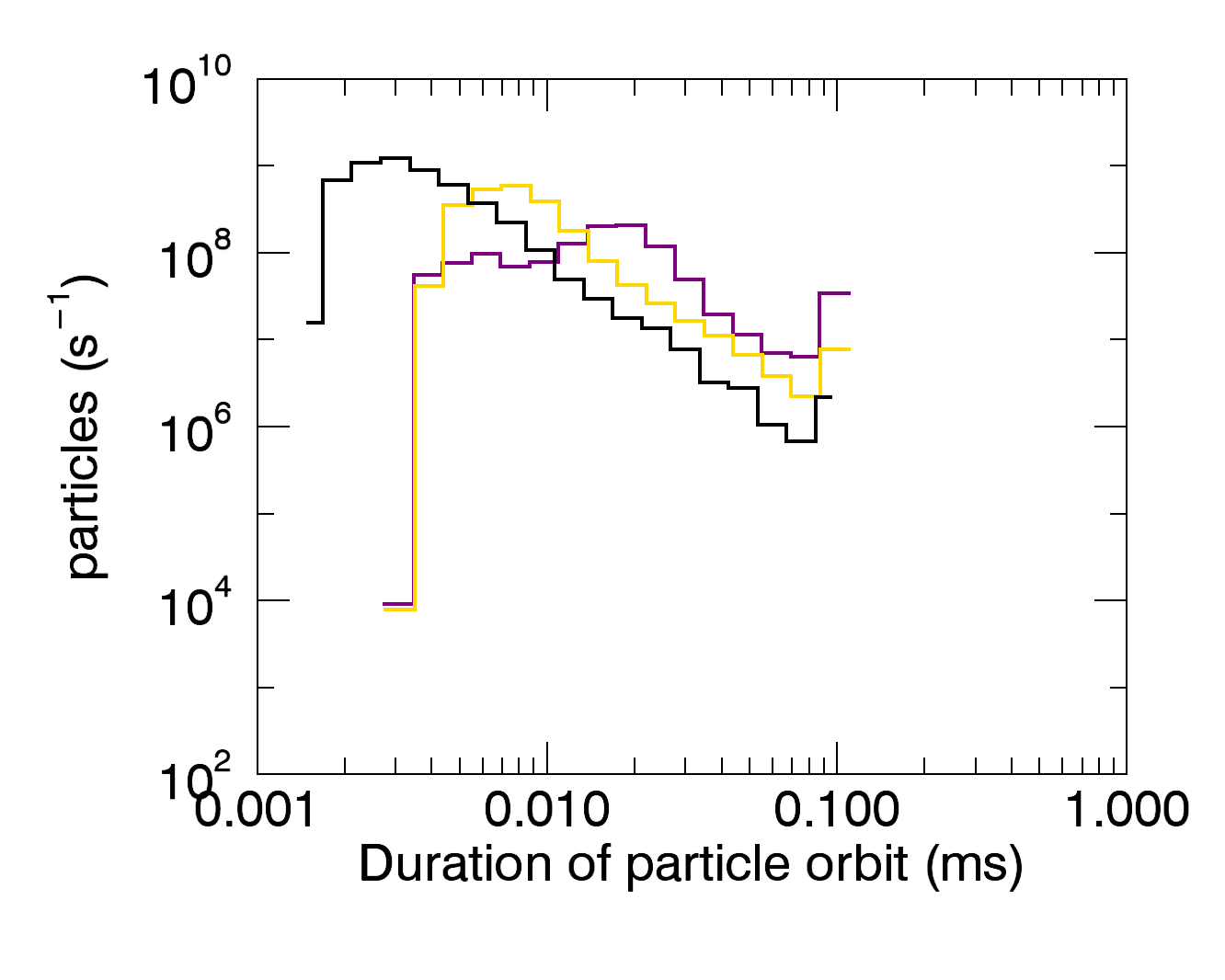}}}
\caption{Particle energy spectra, pitch angle and orbit duration distributions for orbits originating near the upper null ($x,y \in [-1,1]\unit m, z \in [200,299]\unit m$) for one specific scattering model ($\kappa = \eta_{sp}/\eta_a$).}
\label{dist6}
\end{figure*}

Figure \ref{dist5} presents an alternative comparison of the particle energy spectra, pitch angle, and orbit duration distributions for orbit calculations examined in Section \ref{extremity-particles}, where orbits were placed near the lower null. The results displayed are all carried out at the resistivity dependent scattering rate (i.e. $\kappa = \eta_{sp}/\eta_a$). Finally, Figure~\ref{dist6} compares the same scattering model rate and resistivities, but for orbits originating near the upper null (previously seen in Section \ref{extremity-particles2}).


The most obvious effect of varying resistivity is shown by the energy spectra (seen in Figures \ref{dist5a} and \ref{dist6a}), which extend to higher energies for higher resistivities, irrespective of initial position. Increasing the MHD resistivity causes a stronger parallel electric field to accelerate orbits to higher energies. Considering the pitch angle distribution (seen in Figures \ref{dist5b} and \ref{dist6b}), the effect of scattering is more subtle, particularly for orbits originating near the upper null. Here, for low resistivities (purple line) the distribution peaks close to $\theta = 170\dg$. Increasing the resistivity tends to lower the peak location and broaden the distribution (indicative of stronger scattering). This effect is particularly prevalent in Figure~\ref{dist6b}. Increasing the resistivity also reduces the number of orbits found below $\theta=90\dg$. All distributions below $90\dg$ are much more uniform: many of these orbits do not enter the reconnection region, and experience little/no acceleration or scattering. Finally, in the orbit duration distribution we see that increased resistivity leads to shorter orbit durations (Figures \ref{dist5c} and \ref{dist6c}). This indicates that in the higher resistivity simulations, larger parallel electric fields are able to accelerate particles out of the simulation domain faster, despite the increased scattering associated with the higher resistivity.


\section{Discussion}\label{discussion}

By examining the effects of pitch angle scattering on individual test particle orbits, it is evident that pitch angle scattering associated with anomalous resistivity can play a significant role in test particle orbit trajectories, durations, pitch angle evolution, and to a lesser extent, energy gain. We have, in particular, investigated particle orbits in a reconnecting 3D separator configuration. The strength of the effect, however, is significantly influenced by the test particle's location in relation to the separator, with particles experiencing the strongest effects of pitch angle scattering only if they originate very close to the separator within the reconnection region and far away from the null from which they eventually exit. This can be seen in orbit calculations whose initial positions lie outside of the reconnection region (Figures \ref{dist1a}-\ref{dist1c}), where we see no differences in the energy spectra, and orbit duration distribution, between simulations with and without scattering (although some small differences are seen in the pitch angle distributions). By considering a population of particle orbits which starts much closer to the separator (completely within the reconnection region) we see small differences in the particle energy spectra, and larger differences in the orbit duration and especially pitch angle distributions (Figures \ref{dist1d}-\ref{dist11i}). We anticipate that this behaviour results from a lack of scattering in regions where the resistivity in the MHD snapshot is zero (i.e. where the current density does not exceed a critical threshold). Thus particles which start outside of the reconnection region need to be advected into the reconnection region in order to experience any acceleration or scattering, and whether or not this happens depends on the particle's initial conditions and the strength of the $\textbf E \times \textbf B$ drift, which is determined by the background MHD fields.

There are two natural comparisons that can be made with this work. First, we may compare these results with those of other numerical simulations of test particle acceleration in the context of 3D magnetic separators, such as \cite{threlfall-et-al2016b}. That work focused on individual particle orbit trajectories and energisation in multiple separator models (including high and low beta plasmas). We recover similar particle trajectories; particles which encounter the parallel electric field near the separator are accelerated along the separator and out of the computational box near the fan plane. \citet{threlfall-et-al2016b} also examined the dynamics of protons, which exit the box in the opposite direction of electrons. Both electrons and protons gained the most energy when originating from the opposite end of the separator compared to their final positions, which is something also seen here. In contrast to the work presented here, \citet{threlfall-et-al2016b} presented orbit energy gains using a scaling relation depending on the magnetic field, length, and time scales, all of which influence the strength of the parallel electric field, and hence total particle energy gain. By including scattering, our second independent lengthscale (imposed by the mean free path of the particle) prevents us from presenting our results in the same way. However, we are able to compare how well our findings align with the scaling relations derived in \citet{threlfall-et-al2016b}. As an example, the most similar experiment in \citet{threlfall-et-al2016b} derived a peak orbit energy gain given by 
\begin{equation}
KE=1.55{\left[{\frac{B}{10G}}\right]}{\left[\frac{L}{1Mm}\right]}^{2}{\left[\frac{t}{20s}\right]}^{-1}\unit{MeV}.
\label{eqJT2016}
\end{equation}
Using our normalising quantities ($B=0.12$\unit{T}, $L=100\unit{m}$, $t=3.8\times10^{-4}\unit{s}$), we find that this would result in a peak kinetic energy gain of approximately $97\unit{keV}$, a much lower value than some of the energy gains recovered here. This discrepancy can be explained by the differences in experimental setup and by clarifying the purpose of the scaling relations: each scaling relation in \citet{threlfall-et-al2016b} represented the peak kinetic energy of orbits in a specific MHD configuration, subject to specific normalising scales. The configuration used here is different, particularly in regard to the size of the reconnection region, which is broader and more extended than the configurations used in \citet{threlfall-et-al2016b}. Such scaling relations allow estimates of the peak kinetic energies in that particular configuration at different normalising lengths, times, or field strengths. In fact, it is encouraging that despite our use of dramatically different normalising scales, the scaling relations of \citet{threlfall-et-al2016b} yield large peak orbit energies; this implies that similar reconnection processes are taking place at widely different scales.

Pitch angle evolution was not examined in \cite{threlfall-et-al2016b}, however, based on the large energy gains for reasonably sized separators they observed, it would be reasonable to guess that strongly accelerated particles would tend to have pitch angles aligned or anti-aligned with the magnetic field (based on whether they are protons or electrons), which would be broadly similar to our results. 

The other obvious comparison to be made is with the effects of pitch angle scattering due to anomalous resistivity in 2D magnetic reconnection \citep{borissov-et-al2017}. In that case significant differences in the energy spectra were observed for different scattering rates, attributed to multiple crossings of the reconnection region. Although this was also observed to a small degree in the present paper, particles tended to lose much of the energy they gained when they were scattered up the separator, hence differences in the energy spectra between scattered and unscattered simulations were much smaller, or non-existent, compared to the 2D results of \cite{borissov-et-al2017}. Furthermore, significant differences in the positions where particles exited the computational box were observed, with larger numbers of particles exiting in the reconnection outflow region in the presence of strong scattering. In contrast, in this study particle orbits tended to exit near one of the fan planes. Particle orbit durations exhibit a similar trend between the 2D simulations in \cite{borissov-et-al2017} and the 3D simulations presented here, with increased scattering tending to result in longer orbit durations. 

One significant difference between \cite{borissov-et-al2017} and this work is that here it was possible to examine the interaction between particle acceleration and scattering associated with anomalous resistivity in simulations with different resistivities. This was not possible in the 2D simulations since the reconnection rate, and hence parallel electric field strength did not depend on the resistivity. In 2D, varying the resistivity would have amounted to scaling the scattering rate if using the resistivity dependent $\kappa = \eta_{sp}/\eta_a$ scattering model. In contrast, in 3D we find that in higher resistivity simulations particles tend to reach higher energies, while orbit durations tend to be shorter, mostly due to the stronger parallel electric field caused by the higher resistivity. The effect of pitch angle scattering was primarily seen in the pitch angle distributions, with higher resistivities resulting in more particles with pitch angles that are closer to $90\dg$. Signatures of the particle pitch angle scattering may be seen in analytic and numerical models of microwave emission relevant to solar flare conditions \citep[e.g.][]{fleishman-melnikov2003,2018A&A...610A...6M}, so given that there are differences in the pitch angle scattering mechanisms it may be possible to infer the strength of the anomalous resistivity for a given model of pitch angle scattering. However, further investigation is needed in order to determine the variations in pitch angle scattering at different energies and the energy dependency of pitch angle scattering. Indeed, the MHD anomalous resistivity is controlled by pitch angle scattering of thermal electrons, the scattering of 10-100 keV electrons is relevant for hard X-ray observations, while the scattering of near relativistic electrons is essential for gyrosynchrotron emission radio emission, so the combined analysis \citep[e.g.][]{2018A&A...610A...6M} is needed. 

The work presented here should be considered as a starting point for future investigation due to the somewhat idealised nature of the underlying magnetic fields used. It is nevertheless important to begin with a sufficiently simple model in order to be able to examine in detail the effects of scattering on particle acceleration in reconnection regions. Furthermore, the local nature of reconnection means that our analysis of particle dynamics would apply to each of the individual reconnecting separators in more complex models of the Corona with multiple reconnecting separators. It would be worthwhile to examine whether differences in pitch angle distribution are reproducible in more complex field geometries \citep[such as in reconnecting magnetic flux tubes, for example,][]{gordovskyy-et-al2014}, where other effects may impact pitch angle distributions; however, particular care needs to be taken to accurately resolve the scattering processes. As already mentioned, particle scattering introduces an extra characteristic length scale into the system that is independent of MHD length scales. For high scattering rates, which were found to have the most impact on particle dynamics, this could become a limiting factor on the size of the region being modelled. Hence, a good understanding of the effect of scattering on particle acceleration in individual reconnection regions is important. In addition, studying more complex scattering models, for instance, those that include the scattering rate controlled by plasma instabilities, is necessary for a better understanding of the relationship between resistivity dependent scattering and distributions of observable quantities. 


\section{Conclusions}\label{conclusions}

We show that resistivity-dependent pitch angle scattering can play a significant role in test particle orbit evolution in the context of 3D separator reconnection. Individual test particles may be scattered into the reconnection region multiple times causing increased energy gain, however, this effect is not enough to produce significant differences in the test particle energy spectra when large numbers of particle orbits are evaluated using a variety of initial conditions. In contrast to the energy spectra, the pitch angle and orbit duration distributions show significant differences between orbit calculations both without scattering, as well as with scattering at different rates. Increased scattering tends to result in pitch angle distributions which are peaked closer to $\theta = 90\dg$, and in orbit durations which are longer. We also compared test particle orbits with pitch angle scattering at a resistivity dependent rate using MHD fields from simulations with different anomalous resistivities. In this case, the dominant factor in particle energisation was the strength of the parallel electric field, with higher resistivity leading to a higher electric field and hence energy spectra extending to higher energies. The differences in the pitch angle distributions were the opposite of what might be expected due to a simple increase in the parallel electric field, as increased resistivity resulted in the peak of the pitch angle distribution moving closer to $\theta = 90\dg$ (whereas if there was no pitch angle scattering, the peak of the distribution would approach $\theta = 0\dg,180\dg$ due to the stronger electric field). 

The model presented in this paper is an idealised case of configurations relevant to solar flares both in terms of field geometry and in particle dynamics (specifically, the scattering model). As such, it would be worthwhile investigating whether this relationship between pitch angle distribution and resistivity extends to more complex field geometries reminiscent of solar flares, and in more realistic scattering models. 

 \begin{acknowledgements}
A.B. would like to thank the University of St Andrews for financial support from the 7th Century Scholarship and the Scottish Government for support from the Saltire Scholarship. T.N., J.T. and C.E.P. gratefully acknowledge the support of the UK STFC (consolidated grants SN/N000609/1 and ST/S000402/1). E.P.K. acknowledges the financial support from the STFC consolidated 
grant ST/P000533/1.
\end{acknowledgements}

\bibliographystyle{aa}
\bibliography{references}

\end{document}